\documentclass[11pt,a4paper,preprint]{article}
\usepackage{geometry}
\usepackage{graphicx}
\usepackage{epstopdf, epsfig}
\usepackage[round]{natbib}
\usepackage{bm}%
\usepackage[colorlinks,
            linkcolor=black,
            anchorcolor=blue,
            citecolor=blue
            ]{hyperref}
          
\usepackage{subfigure}
\usepackage{multirow}
\usepackage{float}
\usepackage{amsmath}

\usepackage{txfonts}

\usepackage{color}

\begin{document}

\begin{center}
{\LARGE \textbf{Decomposition of the mean friction drag on a NACA4412 airfoil under uniform blowing/suction }}
\\[6pt]

{\small
Yitong Fan$^1$, Marco Atzori$^2$, Ricardo Vinuesa$^2$, Davide Gatti$^3$, Philipp Schlatter$^2$, and Weipeng Li$^1$\footnote{Corresponding author: liweipeng@sjtu.edu.cn}}
\\[3pt]

{\footnotesize
$^1$ School of Aeronautics and Astronautics, Shanghai Jiao Tong University, Shanghai, China, 200240

$^2$ SimEx/FLOW, Engineering Mechanics, KTH Royal Institute of Technology, Stockholm, Sweden, SE-100 44

$^3$ Institute of Fluid Mechanics, Karlsruhe Institute of Technology (KIT), Karlsruhe, Germany, D-76131

}
\end{center}

\begin{abstract}
The application of drag-control strategies on  canonical wall-bounded turbulence, such as periodic channel and zero- or adverse-pressure-gradient boundary layers, raises the question of how to describe control effects consistently for different reference cases. 
We employ the RD identity (Renard \& Deck, \textit{J. Fluid Mech.}, \textbf{790}, 2016, pp. 339--367) to decompose the mean friction drag and investigate the control effects of uniform blowing and suction applied to a NACA4412 airfoil at chord Reynolds numbers $Re_c$=$200,000$ and $400,000$.
The connection of the drag reduction/increase by using blowing/suction with the turbulence statistics (including viscous dissipation, turbulence-kinetic-energy production, and spatial growth of the flow) across the boundary layer, subjected to adverse or favorable pressure gradients,  are examined.
We found that the peaks of the statistics associated with the friction-drag generation exhibit good scaling in either inner or outer units throughout the boundary layer. They are also independent of the Reynolds number, control scheme, and intensity of the blowing/suction.
The small- and large-scale structures are separated with  an adaptive scale-decomposition method, \textit{i.e.} empirical mode decomposition (EMD), aiming to analyze the scale-specific contribution of turbulent motions to friction-drag generation.
Results unveil that blowing on the suction side of the airfoil is able to enhance the contribution of large-scale motions and to suppress that of small-scales; on the other hand, suction behaves contrarily. The contributions related to cross-scale interactions remain almost unchanged with different control strategies. 

\end{abstract}

\section{Introduction}

The friction drag (mostly associated with turbulent boundary layers) accounts for approximately $50\%$ of the total aerodynamic drag  in long-range commercial aircraft \citep{Gad-el-Hak1994}. It provides great potential for drag reduction and energy saving. Among diverse control strategies for turbulent boundary layers, such as addition of long-chain polymers, oscillating walls, superhydrophobic surfaces, and riblets \citep{White2008,Touber2012,Rastegari2015,Li2020,Ran2021}, mass blowing and suction is a promising method to control the friction drag or flow transition in wall-bounded turbulence \citep{Kim2002,Kametani2011}.

Experiments have shown that uniform blowing from smooth perforated surfaces can reduce the turbulent friction drag with net-energy saving \citep{hwan97,hwan04}. Given the proper blowing intensity, porosity, and effective roughness, the net-energy saving holds for a wide range of conditions, including both zero- and adverse-pressure-gradient (ZPG and APG) turbulent boundary layers (TBLs) \citep{welc01}.
On the other hand, uniform suction will increase the turbulent friction drag but can be employed for separation control. 
Direct numerical simulations (DNS) and large-eddy simulations (LES) of uniform blowing and suction applied to turbulent boundary layers have also been performed \citep{Park1999,Kim2002,Kametani2011,Kametani2015,bobk2016}. In most cases, these simulations were carried out in zero-pressure-gradient conditions to investigate the effects of blowing and suction on dynamics of wall-bounded turbulence. For instance, \cite{Stroh2016} compared uniform blowing and suction with body-force damping (as a model of opposition control) and introduced the concept of virtual origin to describe blowing and suction effects in the downstream of the control region. 

Recently, researchers have taken interests in the blowing and suction control of TBLs on wing sections.
Using mass suction at the leading edge of a Clark-Z airfoil to provide pressurized air for blowing, \cite{korn17} studied uniform blowing on the pressure side of the airfoil at Reynolds number $Re_c$=$U_\infty c / \nu$=$840,000$, where $U_{\infty}$ is the incoming flow velocity, $c$ is the chord length, and $\nu$ is the fluid kinematic viscosity. \cite{Eto2019} studied the effects of both active and passive blowing on the suction side of a Clark-Y airfoil at $Re_c$=$1,500,000$. \cite{korn19} employed blowing on the pressure side and suction on the suction side of a NACA0012 airfoil, and later they provided an estimation of the control energy cost under the same conditions \citep{korn20}. 
\cite{Mahfoze2019} used Bayesian optimization to discuss how to benefit from downstream effects of blowing when the control region is separated into individual areas. The first high-fidelity numerical simulation of a wing section with uniform blowing was reported by \cite{Vinuesa2017}, although at a low Reynolds number ($Re_c$=$100,000$).  
Soon after, \cite{Atzori2020} presented a dataset of highly-resolved LES of a NACA4412 airfoil at $Re_c$=$200,000$ and $400,000$ with various configurations of uniform blowing and suction, using as a reference the simulation carried out by \cite{Vinuesa2018}. This dataset has been later employed by \cite{fahl21} to validate Reynolds-Averaged Navier--Stokes (RANS) simulations, and it is also considered in the present paper.

The key objective of this study is to investigate the control effects on  mean friction drag on a wing section with uniform blowing and suction.
Although the mean friction drag is a wall property, as can be directly calculated from the normal gradient of the mean tangential velocity at the wall, it is connected to the statistical turbulence quantities across the wall layer and can be further decomposed into various physics-informed components according to different mathematical derivations and physical interpretations \citep{Li2019,Fan2019a,Fan2019}.
So far, there have been three kinds of friction-drag decomposition methods, derived from the momentum, vorticity, and energy balance, respectively. The first one is the so-called Fukagata-Iwamoto-Kasagi (FIK) identity \citep{Fukagata2002}, in which a triple integration is performed on the mean momentum balance equation and gives a direct relationship between the skin-friction coefficient and the Reynolds-shear-stress profile. The FIK identity has been widely used and extended for more complex situations over the years, \textit{e.g.} \cite{Mehdi2011,Mehdi2014,Modesti2018,Peet2009,Bannier2015}, to name a few. \cite{Kametani2015} and \cite{stro15} applied FIK identity to quantify the variation of skin-friction coefficients caused by blowing and suction in ZPG-TBLs.
Inspired by the mathematical derivation of FIK identity, \cite{Yoon2016} derived a vorticity-based formula relating the mean friction-drag generation with the motion of vortical structures, by performing a triple integration on the mean spanwise vorticity transport equation. They later used this method to analyze the contribution of outer large-scale motions to the friction-drag generation in a moderate APG-TBL \citep{Yoon2018}.
Finally, an energy-based decomposition method was proposed by \cite{Renard2016} from the perspective of streamwise kinetic energy balance. Under an absolute reference frame where the wall is moving, the friction drag develops a non-zero power, which is characterized as the energy transfered from the wall to the fluid, by means of molecular viscosity dissipation, turbulence-kinetic-energy production, and spatial growth of the flow. This method is referred to as RD identity hereafter. The RD identity has been used to analyze the friction-drag generation in channel flows, ZPG/APG-TBLs, turbulent square-duct flows, and pipe flows \citep{Fan2019a,Fan2019, Fan2020,Fan2020a,Wei2018}. 
\cite{Li2019} and \cite{Fan2019} generalized the RD identity to a compressible form to quantify the compressibility effects on the friction-drag generation. 
All these three methods \citep{Fukagata2002,Yoon2016, Renard2016} are mathematically correct and have been widely validated.
In the present study, we only adopt the RD identity, considering that the momentum- and vorticity-based method (by three successive integrations) are lack of physics-informed interpretations, thus  their decomposed constituents hardly carry the causal relationship for the friction-drag generation \citep{Deck2014,Renard2016,Fan2019a}.

The turbulent boundary layers on the suction/pressure sides of the wing section are subjected to adverse/favorable pressure gradients. The pressure gradients have significant impacts on the scales of coherent structures across the wall layer. 
For instance, inner-outer scale separation is more evident in APG-TBLs than in ZPG-TBLs, even at relatively low Reynolds numbers, due to the enhancement of outer-scale motions \citep{Tanarro2020}. Additionally, a significant increment of small-scale energy was found in the outer region, as the vertical motion induced by the APG transports small scales from the near-wall region to the outer layer \citep{Vinuesa2018,Tanarro2020}.
In the present study, we also aim to quantify the contribution of structures with different scales to the generation of friction drag on the wing section with/without blowing and suction. 
Studies on such cases are likely to promote new drag control strategies.
To this end, an appropriate approach to separate the multi-scale coherent structures is in need. 
Typically, Fourier analysis might be a tempting tool to decompose the raw signals into modes with given wavelengths, yet it relies on {\it a-priori} definition of cutoff wavelength and suffers from inflexibility for complex and transient signals \citep{Cheng2019}.
Another frequently-used method is proper orthogonal decomposition (POD)~\citep{Lumley1967,Wu2010}, which sorts the contribution of velocity fluctuations to the turbulence kinetic energy. However, \cite{Wang2018, Wang2019} pointed out that, the energy-ranking spatial modes cannot fully recover the dynamics of turbulent motions in different length scales. 
In contrast, empirical mode decomposition (EMD), proposed by \cite{Huang1998}, provides an adaptive, data-driven, and \textit{a-posteriori} technique to delineate the transient and local characteristics of signals. It is in principle free from pre-established basis functions and represents the original signal as a superposition of several mono-components and a residual, with the characteristic wavelengths of the signals automatically determined. 
With EMD, \cite{Huang2008} studied the scaling properties and intermittency of homogeneous turbulence, and \cite{Ansell2017} analyzed the features of large-scale vortical structures in a turbulent mixing layer.
\cite{Agostini2014,Agostini2016} used bidimensional empirical mode decomposition (BEMD) to analyze the modulation of large-scale motions on the small-scale eddies in the near-wall region, and later they discussed the scale-specific contributions of large- and small-scale structures to the friction-drag generation  by means of FIK and RD identity \citep{Agostini2019} in channel flows. 
\cite{doga19} have used EMD to characterize the inner-outer interaction based on the modulation coefficient.
\cite{Cheng2019} adopted BEMD to identify attached eddies in turbulent channel flows and quantify their relationship with the friction-drag generation.
However, no relevant study has been found in the open literature to analyze the scale-specific contribution of turbulent motions to friction-drag generation on the wing section with/without blowing and suction.

This paper is organized as follows. In Sec.\ref{method}, we introduce the friction-drag decomposition method and the database of flow over a NACA4412 airfoil.  The decomposition results on the suction and pressure side of the NACA4412 are discussed in Sec. \ref{suction} and \ref{pressure}, respectively. Concluding remarks are given in Sec. \ref{conclusion}.

\section{Friction-drag decomposition method and the database of flow over the NACA4412 airfoil} \label{method}

With the energy-based RD identity \citep{Renard2016}, the skin-friction coefficient $C_f$ of a turbulent boundary layer can be decomposed as:
\begin{eqnarray}\label{apg_formula}
C_{f}=\underbrace{\frac{2}{U_e^3}\int_{0}^{\infty}\nu \left(\frac{\partial \left<u\right>}{\partial y}\right)^2 {\rm d}y}_{C_{f,V}}+\underbrace{\frac{2}{U_e^3}\int_{0}^{\infty}-\left<u'v'\right>\frac{\partial \left<u\right>}{\partial y} {\rm d}y}_{C_{f,T}}\nonumber\\
+\underbrace{\frac{2}{U_e^3}\int_{0}^{\infty}\left(\left<u\right>-U_e\right)\frac{\partial}{\partial y}\left(\nu\frac{\partial \left<u\right>}{\partial y}-\left<u'v'\right>\right) {\rm d}y}_{C_{f,G}},
\end{eqnarray}
where $\left<\cdot\right>$ is the Reynolds averaging operator, the prime $'$ denotes fluctuations with respect to the Reynolds averages, $U_e$ is the velocity at the boundary-layer edge $\delta_{99}$, $x$ and $y$ represent the directions tangential and normal to the wall surface respectively, and $u$ and $v$ are the corresponding velocity components. The derivation of the RD identity can be retrieved in \cite{Renard2016}.

Three contributive friction constituents are obtained in equation (\ref{apg_formula}). Thereinto, (i) $C_{f,V}$ represents the direct molecular viscous dissipation, transforming the mechanical power into heat; (ii) $C_{f,T}$ represents the power spent for turbulence-kinetic-energy production; (iii) $C_{f,G}$ accounts for the spatial growth of the flow, which is also interpreted as the rate of gain of mean streamwise kinetic energy by the fluid in the absolute frame. Note that the integrand in $C_{f,G}$ has been substituted with local information which only depends on the well-documented wall-normal profiles~\citep{Renard2016}. This is especially applicable for the cases where the accurate calculation of explicit streamwise derivatives is unfeasible.

For adverse-/favorable-pressure-gradient turbulent boundary layers around an airfoil, the roles of the wall-normal convection and pressure gradient are of particular importance and should be individually discussed, thus a further decomposition of $C_{f,G}$ is carried out, \textit{viz.}

\begin{eqnarray}\label{cf3}
C_{f,G}=
\underbrace{\frac{2}{U_e^3}\int_{0}^{\infty}\left(\left<u\right>-U_e\right)\left(\left<v\right>\frac{\partial\left<u\right>}{\partial y}\right) {\rm d}y}_{C_{f,C}}
+\underbrace{\frac{2}{U_e^3}\int_{0}^{\infty}\left(\left<u\right>-U_e\right)I_x{\rm d}y}_{C_{f,D}}\nonumber\\
+\underbrace{\frac{2}{U_e^3}\int_{0}^{\infty}\left(\left<u\right>-U_e\right)\left(\frac{{\rm d}  p/\rho}{{\rm d}  x}\right) {\rm d}y}_{C_{f,P}},
\end{eqnarray}
where $I_x={\partial\left<u'u'\right>}/{\partial x}+\left<u\right>{\partial\left<u\right>}/{\partial x}-\nu{\partial^2\left<u\right>}/{\partial x^2}$, $p$ is the static pressure, and $\rho$ is the density. Contributions of the mean wall-normal convection ($C_{f,C}$), streamwise development ($C_{f,D}$), and the pressure gradient ($C_{f,P}$) are separated in equation (\ref{cf3}).

We consider a set of well-resolved LESs of a NACA4412 airfoil at angle of attack of $5^\circ$ at two chord Reynolds numbers, \textit{i.e.} $Re_c$=$200,000$ and $400,000$. The simulations were performed with the spectral-element code \textit{Nek5000}, developed by \cite{fisc08}. The spatial derivatives in the incompressible Navier--Stokes equations are discretized employing a Garlerkin method, following the $P_N-P_{N-2}$ formulation by \cite{pate84} and the solution is expressed within each spectral element in terms of a nodal-base of Legendre polynomials on the Gauss--Lobatto--Legendre (GLL) quadrature points. The discretization of the time derivatives is explicit for the non-linear terms and implicit for the viscous term, employing an extrapolation and a backward differentiation scheme, respectively, both of the third order. In order to trigger transition to turbulence, we employed tripping through a volume force, implemented as proposed by \cite{schl12}, at $x/c=0.1$ on both suction and pressure sides.

The cases with/without control, listed in Table~\ref{tab:dataset}, include various configurations of uniform blowing and suction applied on the suction side and uniform blowing applied on the pressure side. 
\begin{table}
\begin{center}
\def~{\hphantom{0}}
\resizebox{\textwidth}{!}{
    \begin{tabular}{cccccccc}
    \hline
        Case & Control strategies & Intensity ($V_{wall}$)  &$\Delta(c_l)$ & $\Delta(c_d)$  & $\Delta(L/D)$ & $Re_\tau$ & $\beta$\\
	   $Re200k, ss, ref$ & -- & --  &  -- &  -- & -- & $[132,224]$ & $[0.16,11.07]$\\
       $Re200k, ss, blw1$ & blowing, suction s. ($0.25<x/c<0.86$) & $0.1\%U_\infty$  & $-4\%$ & $+3\%$ & $-7\%$ & $[126,209]$ & $[0.14,19.62]$\\
       $Re200k, ss, blw2$ & blowing, suction s. ($0.25<x/c<0.86$) & $0.2\%U_\infty$  &  $-8\%$ & $+8\%$ & $-15\%$ & $[118,193]$ & $[0.12,36.54]$\\
       $Re200k, ss, sct1$ & suction, suction s. ($0.25<x/c<0.86$) & $0.1\%U_\infty$  &  $+4\%$ & $-2\%$ & $+6\%$ & $[139,238]$ & $[0.17,6.97]$\\      
       $Re200k, ss, sct2$ & suction, suction s. ($0.25<x/c<0.86$) & $0.2\%U_\infty$  &  $+7\%$ & $-4\%$ & $+11\%$ & $[145,248]$ & $[0.18,4.71]$\\
       $Re400k, ss, ref$ & -- & --  &  --  &  -- &  -- & $[183,363]$ & $[0.15,9.16]$\\
       $Re400k, ss, blw1$ & blowing, suction s. ($0.25<x/c<0.86$) & $0.1\%U_\infty$  &  $-4\%$ & $+5\%$ & $-9\%$ & $[174,335]$ & $[0.14,17.56]$\\ 
       $Re400k, ss, sct1$ & suction, suction s. ($0.25<x/c<0.86$) & $0.1\%U_\infty$  &  $+3\%$ & $-1\%$ & $+4\%$ & $[193,390]$ & $[0.15,5.37]$\\
       $Re200k, ps, ref$ & -- & --  &  --  & --  & --  & $[96,219]$ & $[-0.30,-0.01]$\\
       $Re200k, ps, blw1$ & blowing, pressure s. ($0.2<x/c<1.00$) &  $0.1\%U_\infty$  &  $+0\%$ & $-3\%$ & $+4\%$ & $[87,232]$ & $[-0.42,-0.03]$\\
       $Re200k, ps, blw2$ & blowing, pressure s. ($0.2<x/c<1.00$) & $0.2\%U_\infty$  &  $+1\%$ & $-5\%$ & $+7\%$ & $[80,234]$ & $[-0.55,-0.04]$\\
       \hline
       \end{tabular}}
    \caption{Cases with/without control considered in the present paper and the relative changes of the total lift and drag coefficients (denoted by $\Delta(c_l)$ and $\Delta(c_d)$, respectively) and aerodynamic efficiency (denoted by $\Delta(L/D)$) in respect to the reference case. 
    Note that ``suction s.'' and ``pressure s.'' are suction and pressure sides, respectively, $Re200k$ and $Re400k$ are the chord Reynolds number ($Re_c$) of each case , $Re_\tau$ denotes the friction Reynolds number, and $\beta$ is the Rotta-Clauser pressure-gradient parameter.}
    \label{tab:dataset}
\end{center}
\end{table}
The relative proportions of pressure drag and skin-friction drag determine the control effects on the total drag, denoted by $c_d$. At these moderate Reynolds numbers, pressure drag is relatively high and uniform blowing on the suction side increases it by an amount that is high enough to overcome the skin-friction reduction, eventually leading to higher $c_d$. On the contrary, uniform suction increases skin-friction drag, but it decreases the pressure drag enough to result in lower $c_d$. At the same time, uniform blowing and suction on the suction side also decreases and increases lift ($c_l$), respectively. Uniform blowing on the pressure side has different effects on the pressure distribution around the airfoil, decreasing both skin friction and pressure drag and increasing lift. The friction Reynolds numbers ($Re_\tau=u_\tau\delta_{99}/\nu$) and the Rotta-Clauser pressure-gradient parameters \citep{Rotta1950,Clauser1954,Clauser1956} ($\beta=\delta^*/\tau_w{\rm d}Pe/{\rm d}x$) are also listed in Table \ref{tab:dataset}.
Note that $u_\tau=\sqrt{(\tau_w/\rho)}$ is the friction velocity, $\delta^*$ is the displacement thickness, $\tau_w$ is the wall shear stress, and ${\rm d}Pe/{\rm d}x$ is the streamwise pressure gradient at the boundary-layer edge.
For a more complete description of the numerical setup and the aerodynamic effects of control, we refer to \cite{Vinuesa2018} and \cite{Atzori2020}.

\section{Friction-drag decomposition on the suction side}\label{suction}

\subsection{The control effects}\label{sec:1}

\begin{figure}[h]
\centering\centering
\subfigure{\includegraphics[width = 5.5cm]{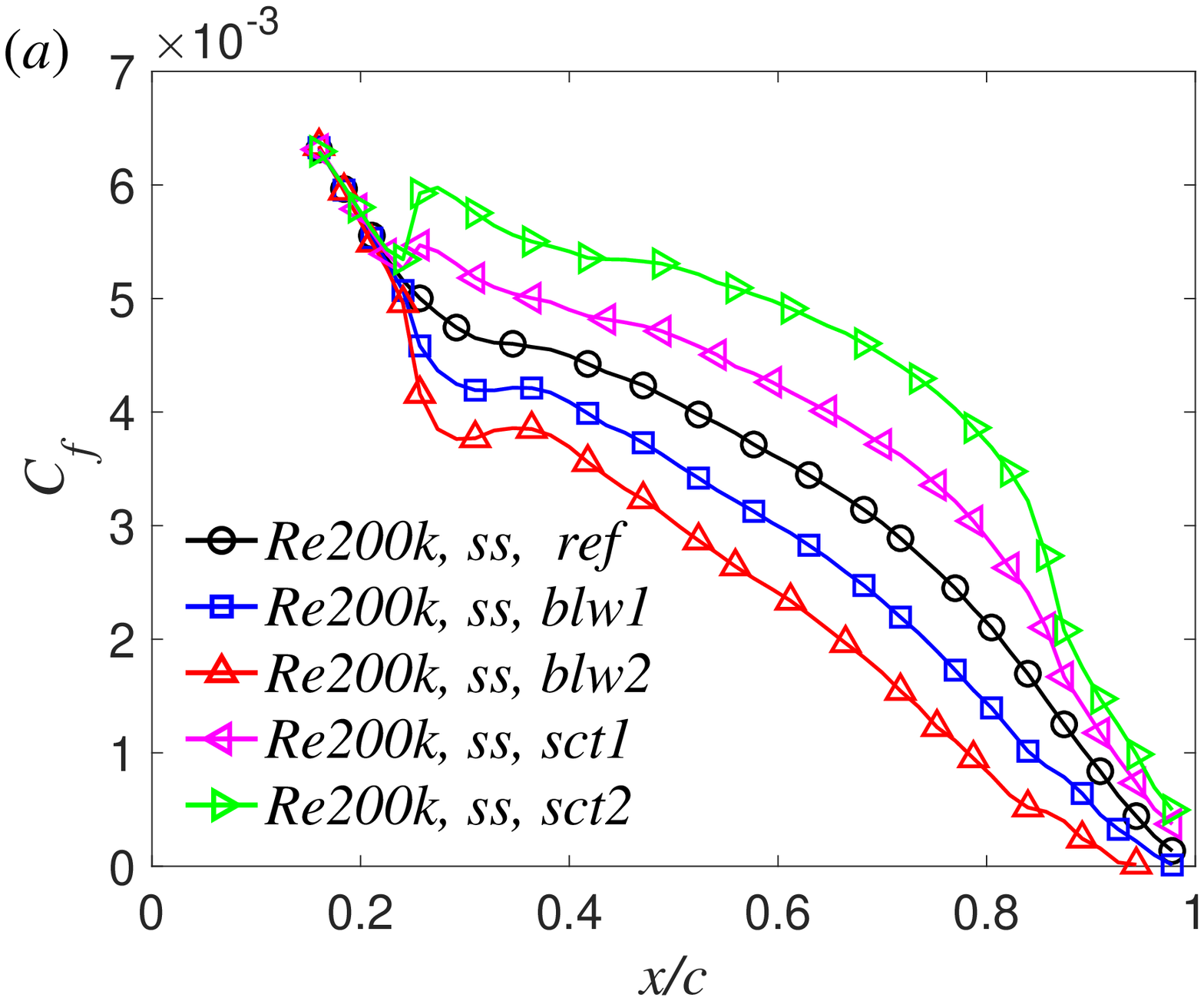}\label{topcf:a}}
\subfigure{\includegraphics[width = 5.5cm]{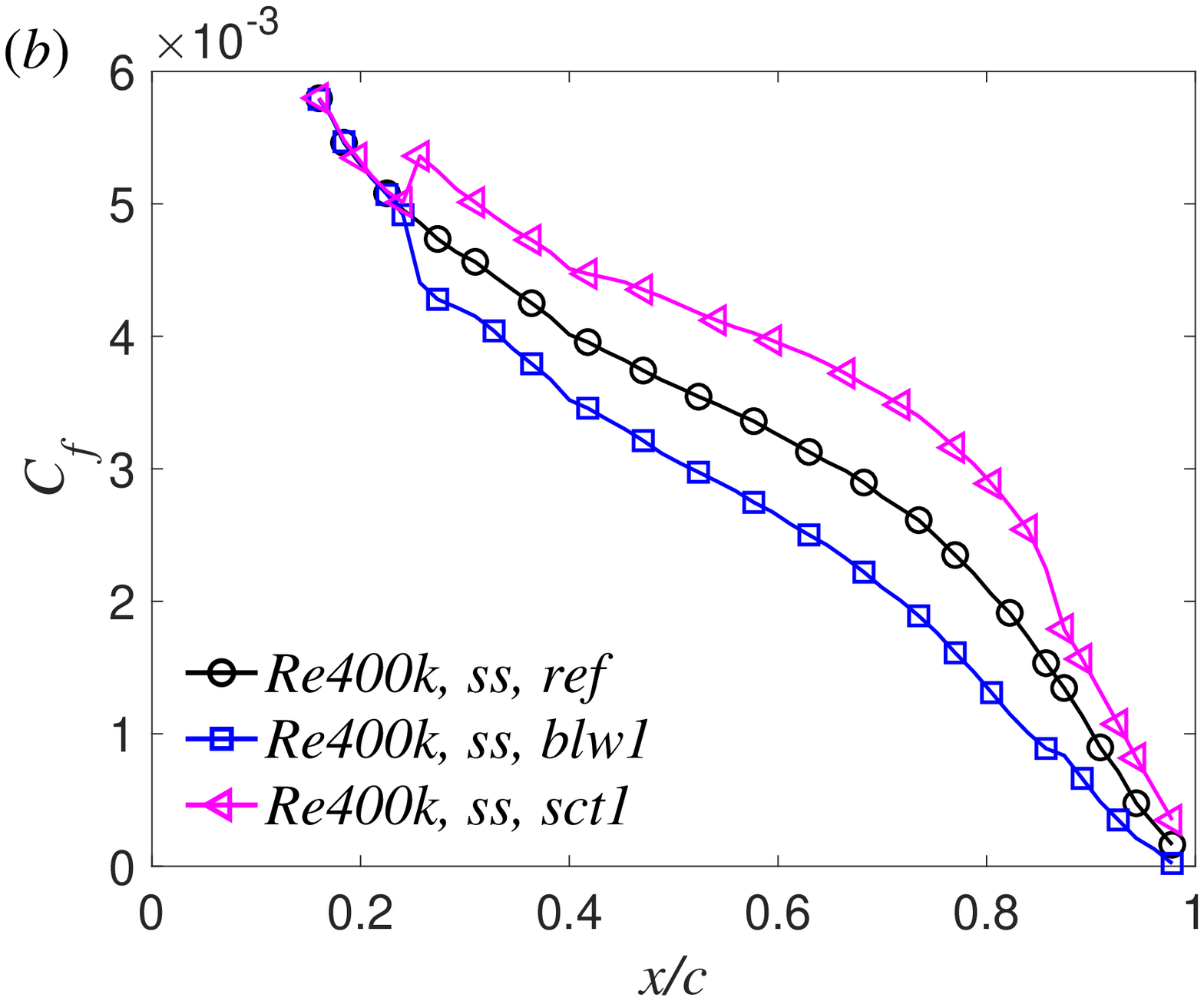}\label{topcf:b}}
\caption{Variation of skin-friction coefficients on the suction side of a NACA4412 wing section at ($a$) $Re_c=200,000$ and ($b$) $Re_c=400,000$.}
\label{topcf}
\end{figure}

Using the database, we first show the variation of skin-friction coefficients on the suction side of a NACA4412 wing section in figure \ref{topcf}.
It can be easily found that uniform blowing causes friction-drag reduction whereas suction causes friction-drag increase, regardless of the Reynolds number and streamwise position on the control surface. 
Stronger intensity of blowing/suction leads to larger drag-reduction/increase rate, as expected. Such phenomena are in consistence with previous studies~\citep{Kametani2011,Kametani2015,Atzori2020}. 
The mechanisms of the drag reduction/increase by blowing/suction are associated with the interactions between the ``cross-stream'' and the quasi-streamwise vortical structures in the near-wall region, which probably enhance or damp the behavior of sweep/ejection events and yield modifications of the mean velocity profiles~\citep{Park1999,Kim2002,hwan04,Kornilov2015}.
The variation of turbulent dynamics in the near-wall region leads to a redistribution of the turbulence kinetic energy and alteration of the turbulent momentum transport across the wall layer. The vortical structures in the outer layer will also be influenced by the near-wall blowing/suction, especially in the downstream of the control surface.
Consequently, the generation of the skin-friction drag, which is linked to the turbulence statistics across the wall layer, will be correspondingly changed.

In order to clarify such control effects, we conduct the decomposition of skin-friction coefficients on the suction side (within $0.2\le x/c \le0.85$) of NACA4412 by RD identity. Note that the relative errors, $(C_{f,V}+C_{f,T}+C_{f,G}-C_f)/C_f$, where $C_f$ is directly calculated with the normal gradient of tangential velocity at the wall surface (\textit{i.e.} $C_f=(\mu \partial\left<u\right>/\partial y)|_{wall}/(0.5\rho U_e^2)$), are well confined within {$\pm0.12\%$} for all cases considered, which confirms the reliability of the decomposition method.

\begin{figure}[h]
\centering
\subfigure{\includegraphics[width = 5.5cm]{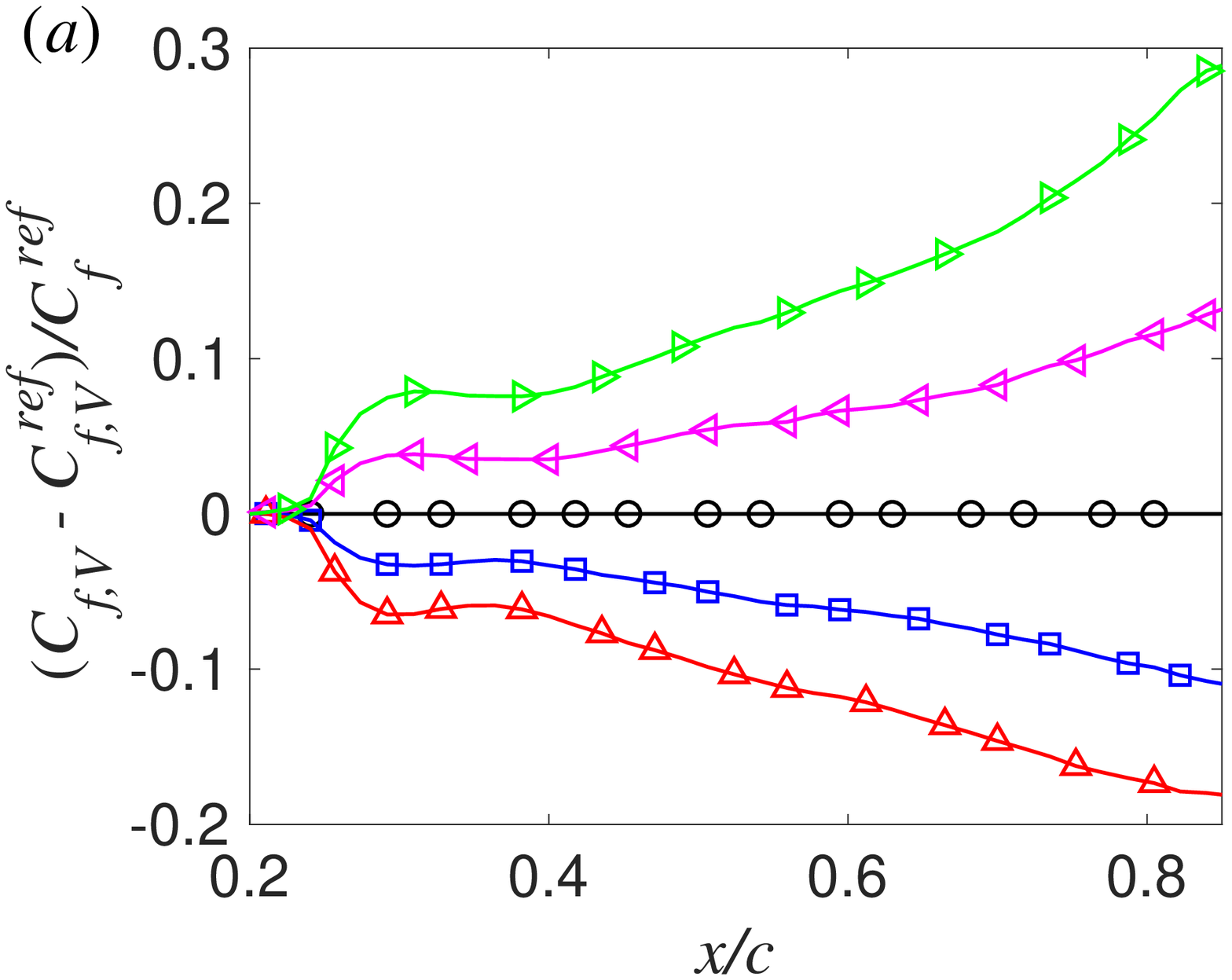}\label{top_con:a}}
\subfigure{\includegraphics[width = 5.5cm]{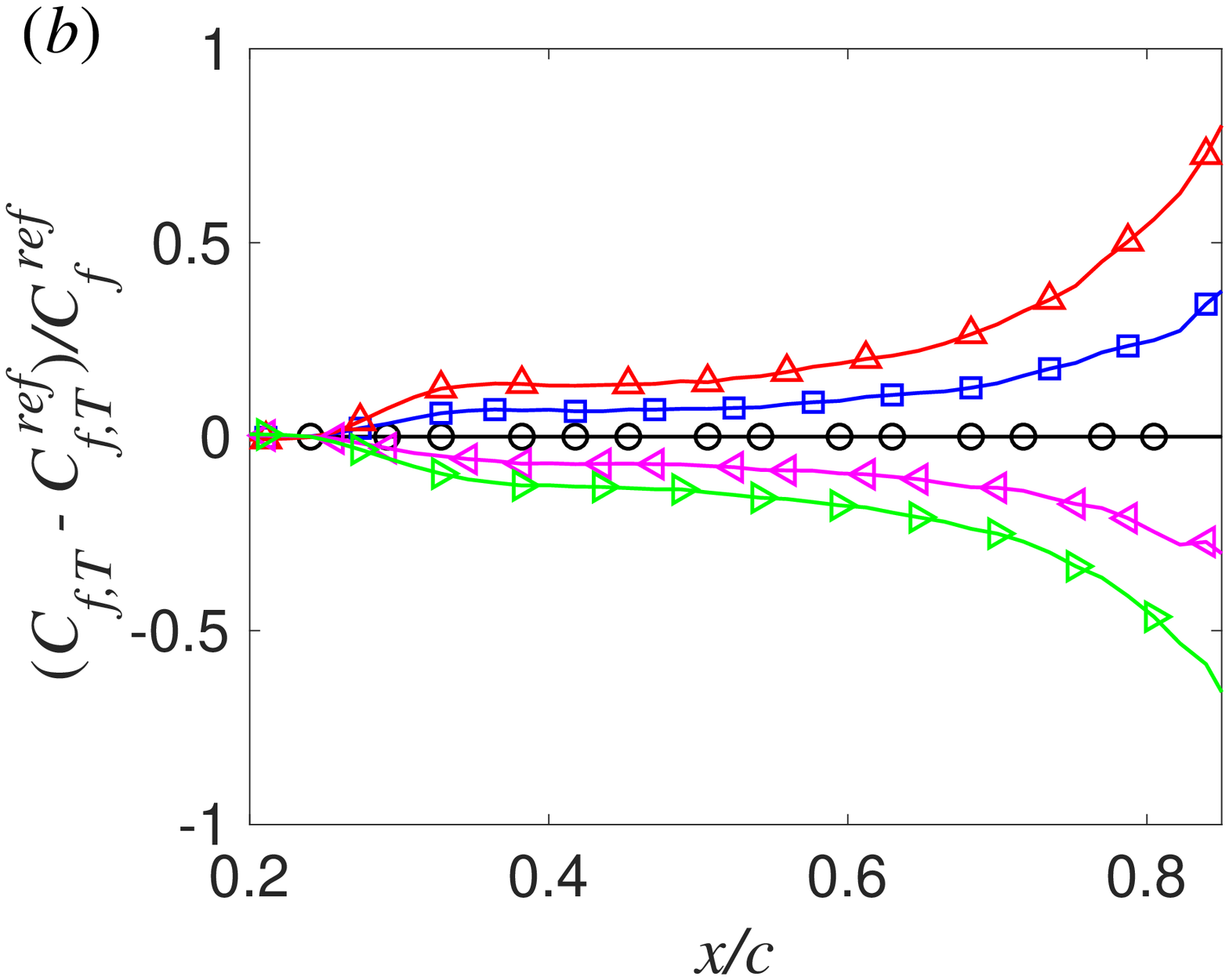}\label{top_con:b}}
\subfigure{\includegraphics[width = 5.5cm]{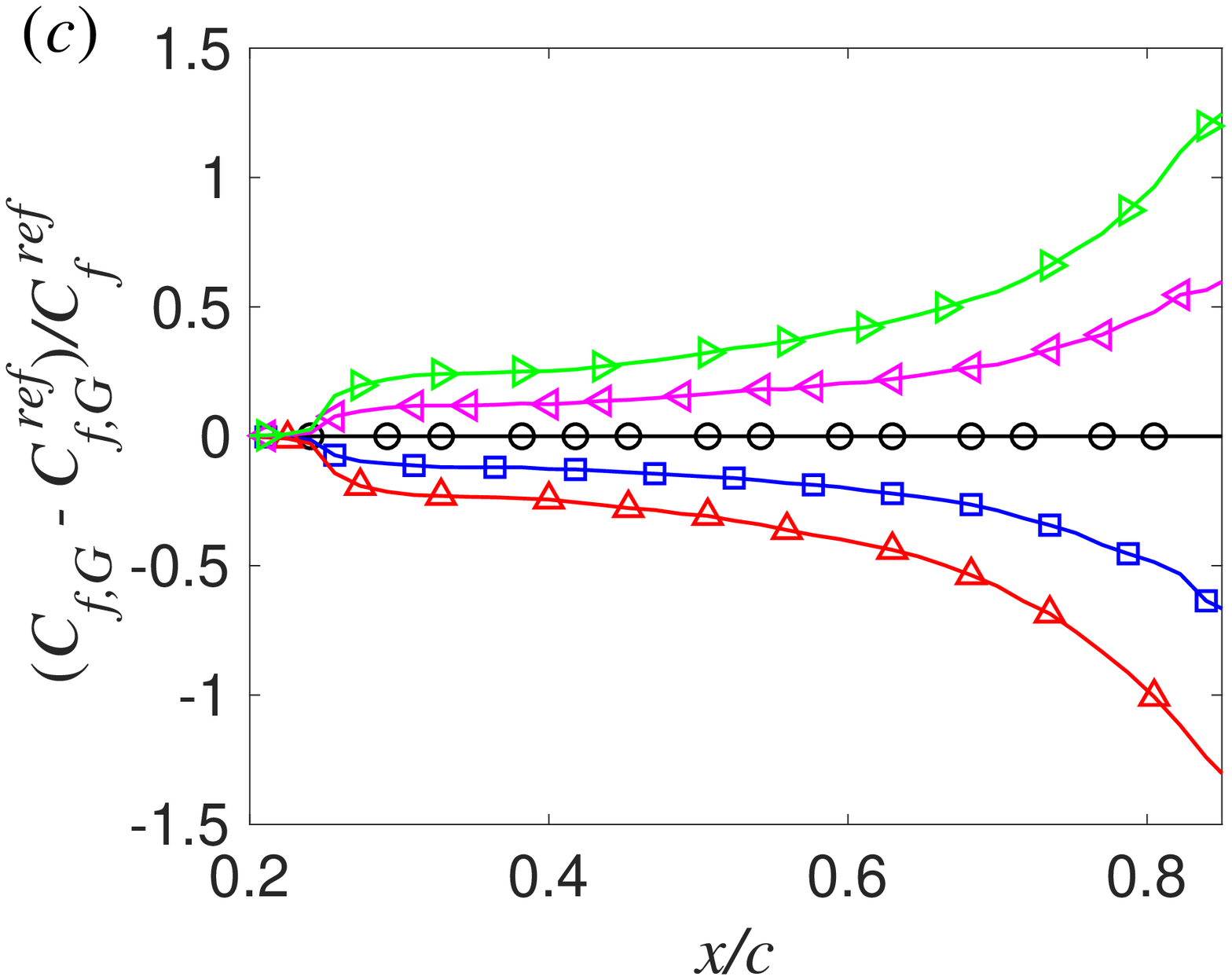}\label{top_con:c}}
\subfigure{\includegraphics[width = 5.5cm]{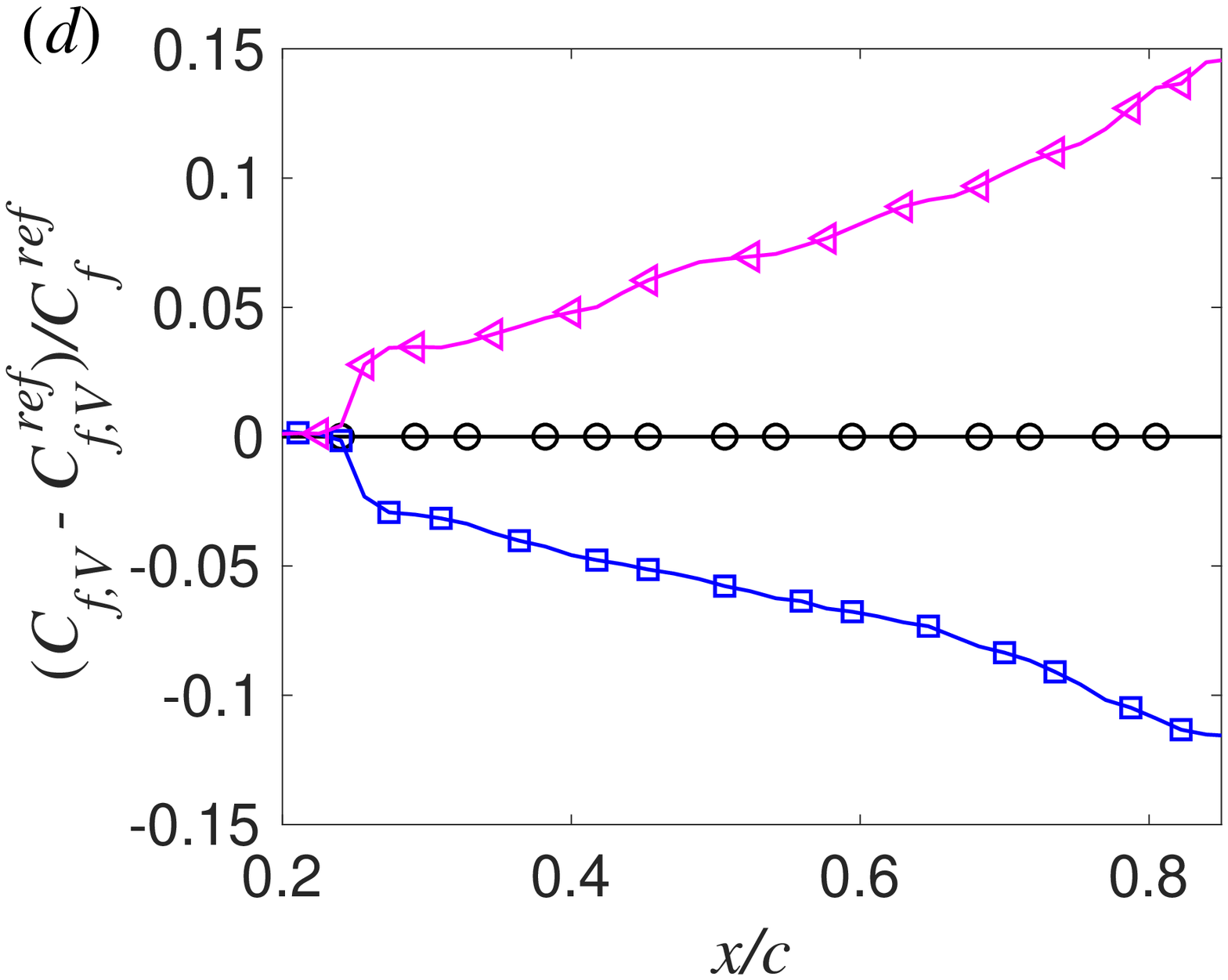}\label{top_con:d}}
\subfigure{\includegraphics[width = 5.5cm]{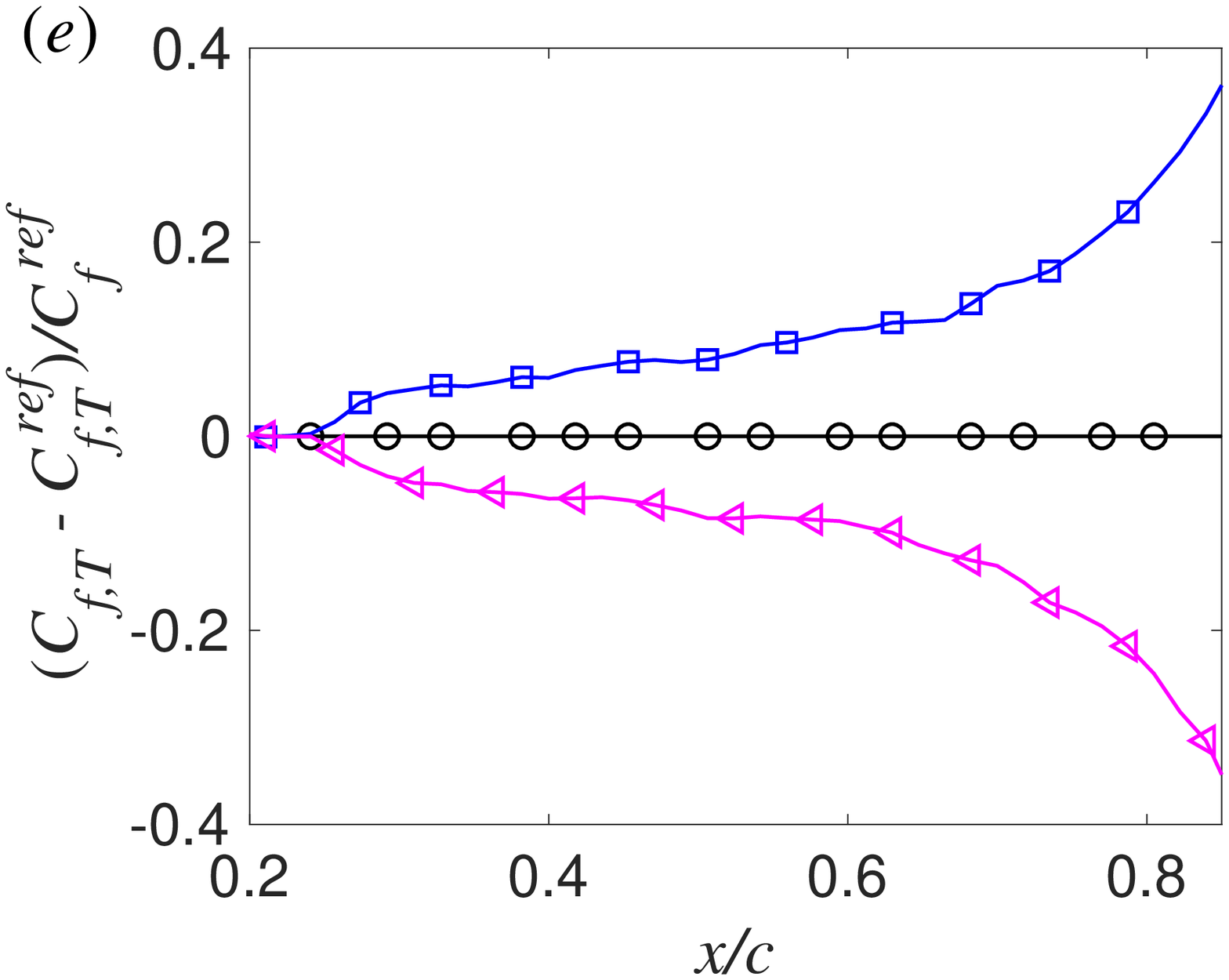}\label{top_con:e}}
\subfigure{\includegraphics[width = 5.5cm]{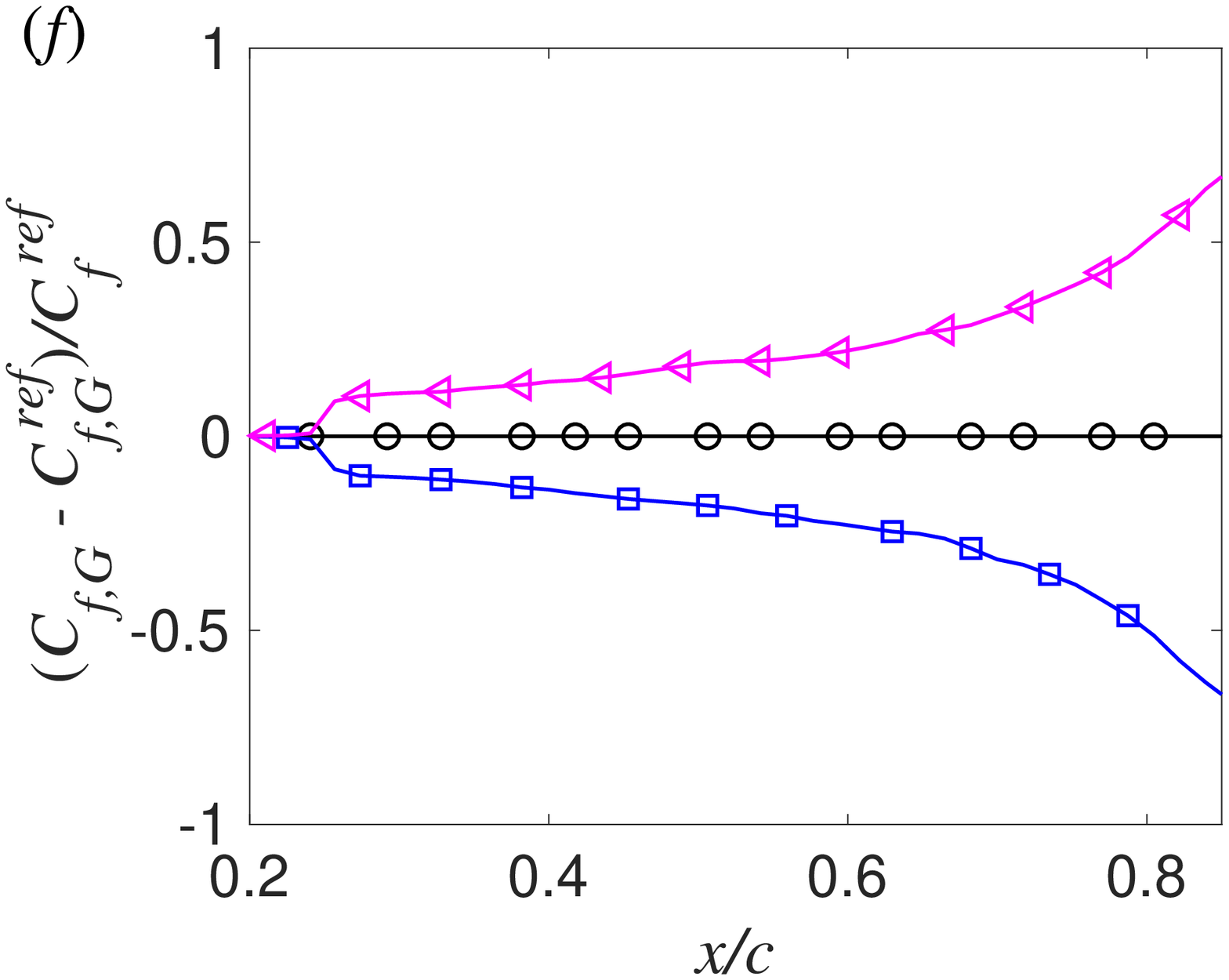}\label{top_con:f}}
\caption{Variation of ($a,d$) $C_{f,V}$, ($b,e$) $C_{f,T}$, and ($c,f$) $C_{f,G}$ with regard to the reference case on the suction side of a  NACA4412 wing section at ($a-c$) $Re_c=200,000$ and ($d-f$) $Re_c=400,000$. (The superscript of ``\textit{ref}'' represents the reference case without control. Legends in ($a-c$) refer to figure \ref{topcf:a}, while those in ($d-f$) refer to figure \ref{topcf:b}.)}
\label{top_con}
\end{figure}

Figure~\ref{top_con} shows the variations of $C_{f,V}$, $C_{f,T}$, and $C_{f,G}$ induced by uniform blowing and suction with regard to the reference case, at $Re_c$=$200,000$ (figures~\ref{top_con:a}--\ref{top_con:c}) and $Re_c$=$400,000$ (figures~\ref{top_con:d}--\ref{top_con:f}). With blowing, the friction contribution of direct viscous dissipation ($C_{f,V}$) is reduced at both Reynolds numbers. This is closely related to the suppression of the near-wall sweep events, which transport high-speed fluid towards the wall. With blowing, an increase of the generation of turbulence-kinetic-energy production ($C_{f,T}$) is found in figures \ref{top_con:b} and \ref{top_con:e}.
The variations of $C_{f,V}$ and $C_{f,T}$ are essentially associated with the influences on the wall-normal profiles of mean viscous shear stress and Reynolds shear stress across the wall layer, which will be discussed in Sec. \ref{sec:2}.

\begin{figure}[h]
\centering
\subfigure{\includegraphics[width = 5.5cm]{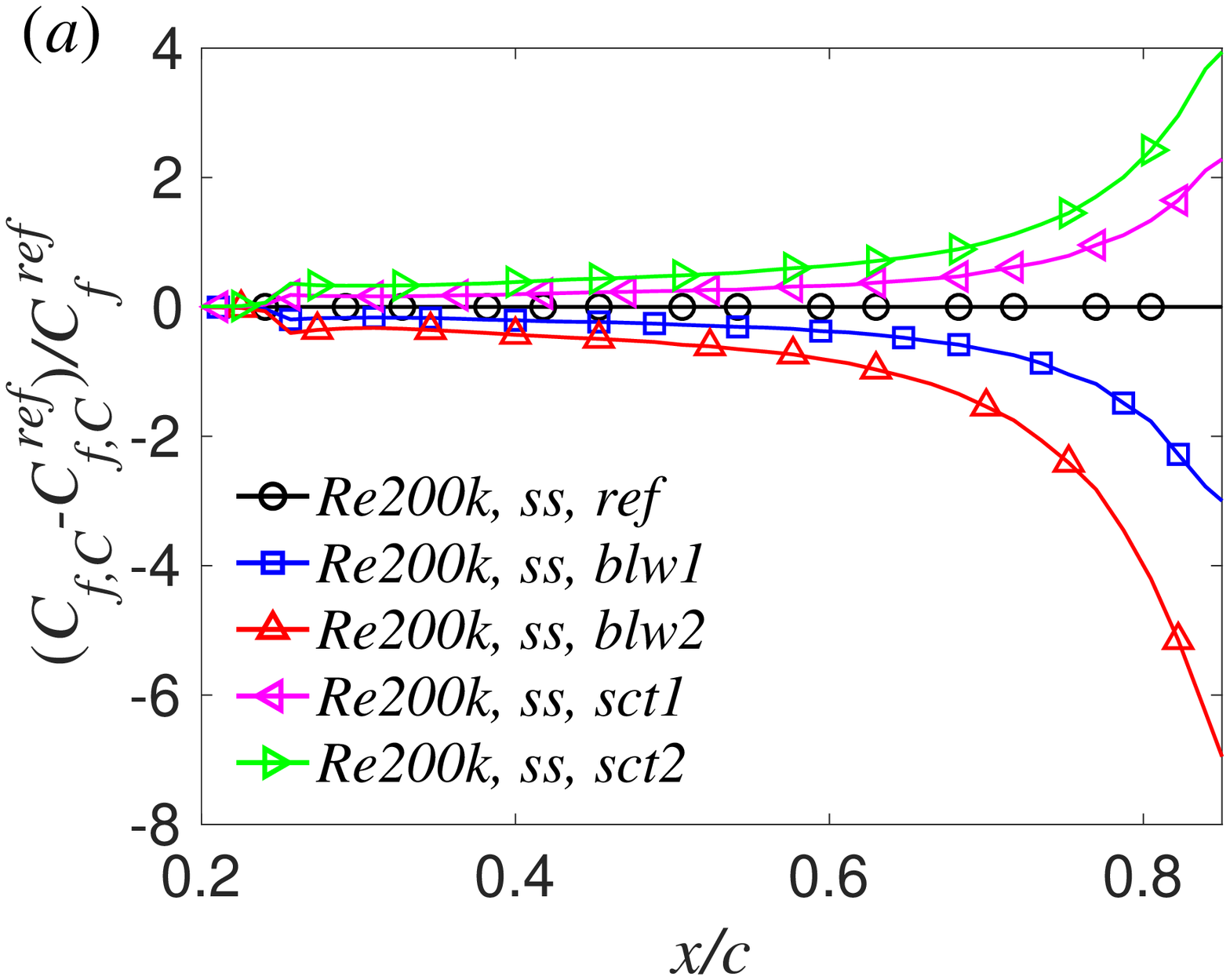}\label{top_con3:a}}
\subfigure{\includegraphics[width = 5.5cm]{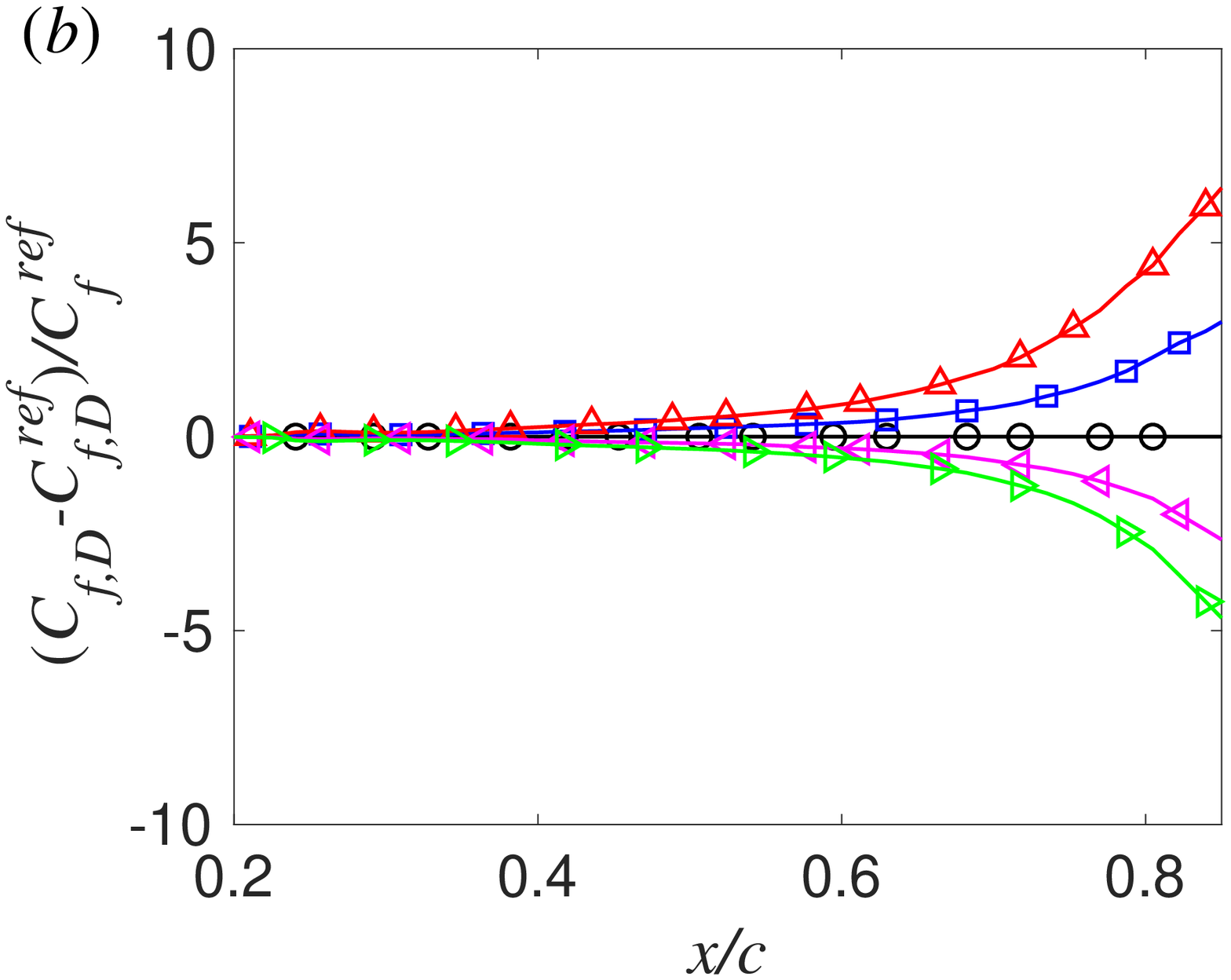}\label{top_con3:b}}
\subfigure{\includegraphics[width = 5.5cm]{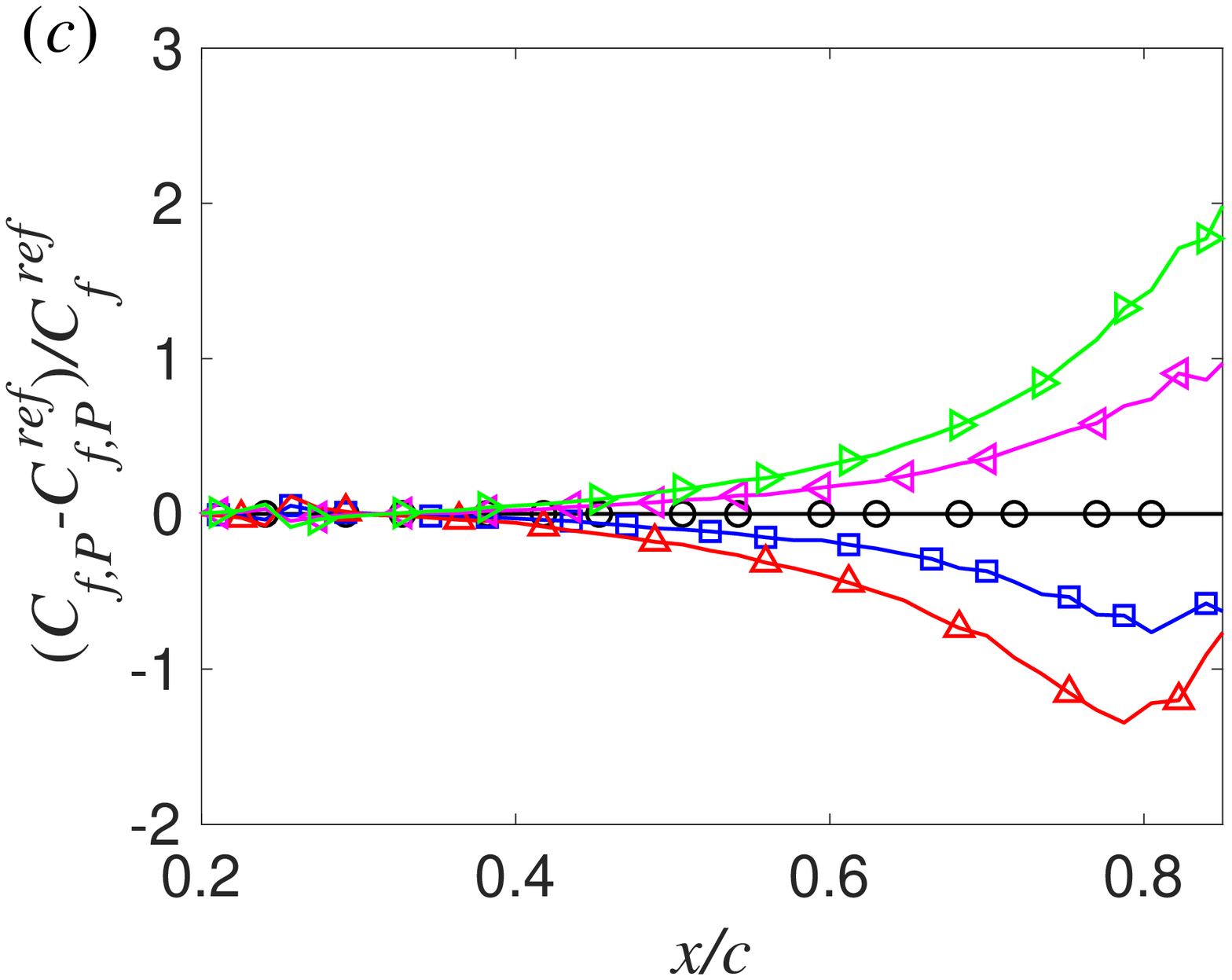}\label{top_con3:c}}
\subfigure{\includegraphics[width = 5.5cm]{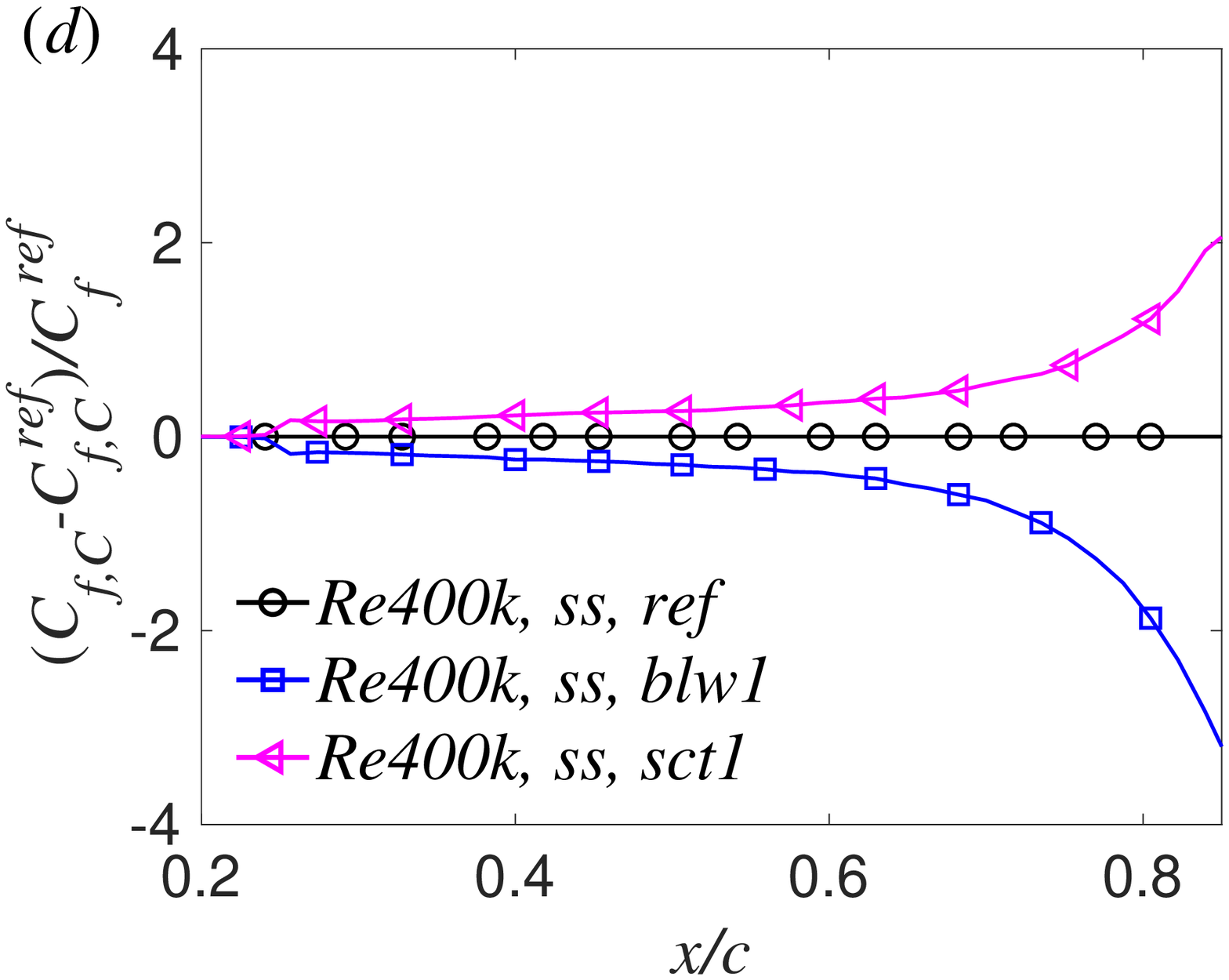}\label{top_con3:d}}		
\subfigure{\includegraphics[width = 5.5cm]{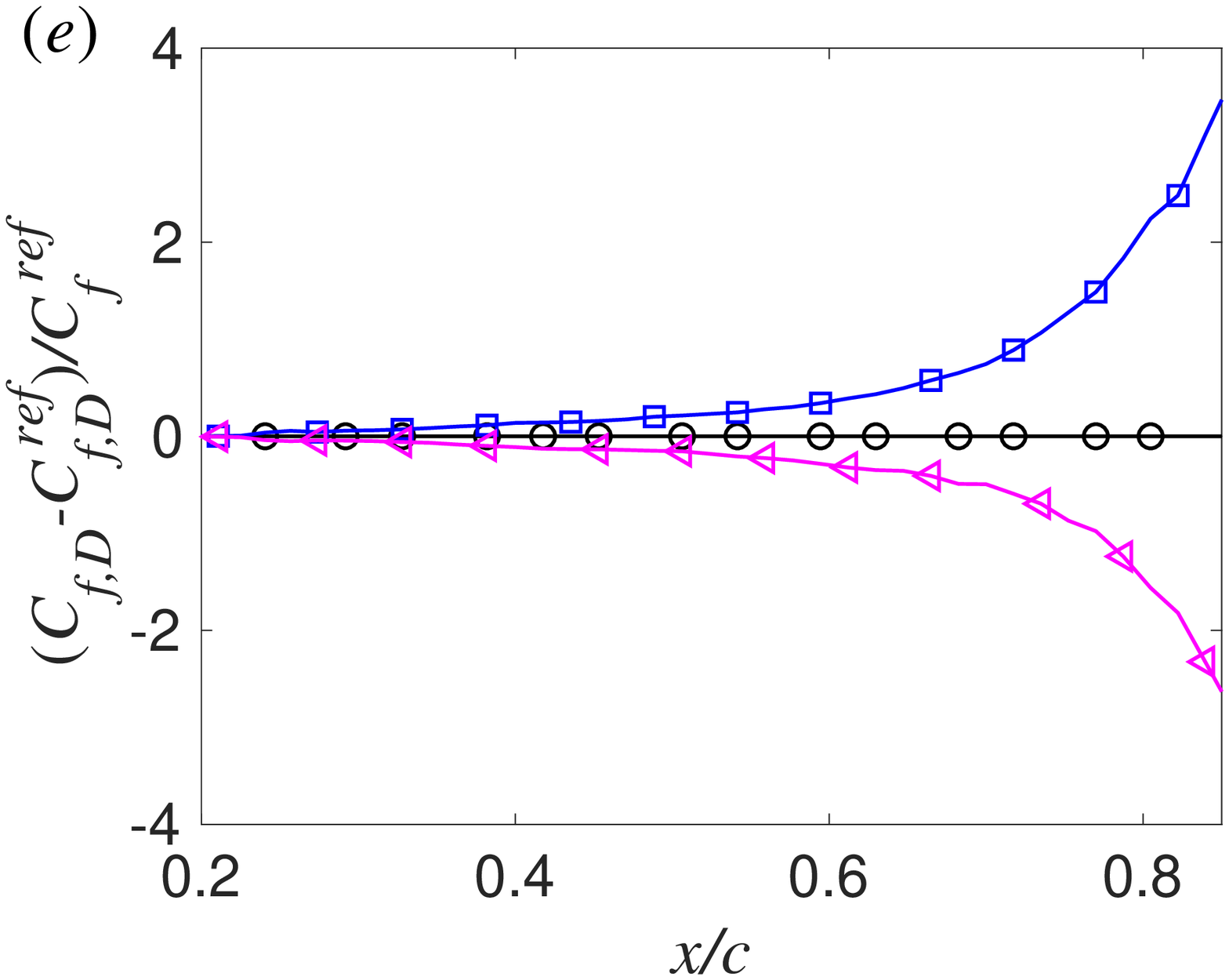}\label{top_con3:e}}
\subfigure{\includegraphics[width = 5.5cm]{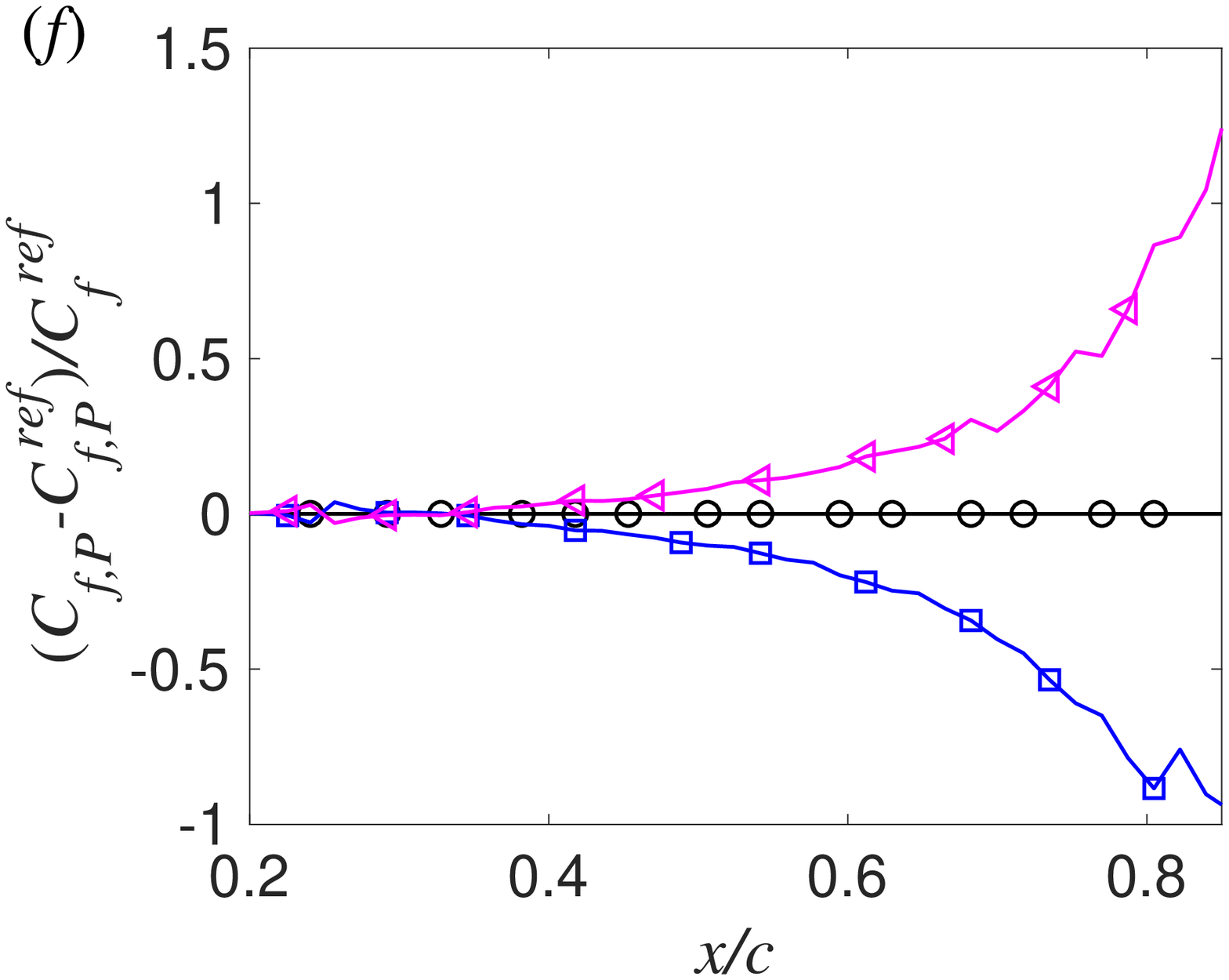}\label{top_con3:f}}
\caption{Variation of ($a,d$) $C_{f,C}$, ($b,e$) $C_{f,D}$, and ($c,f$) $C_{f,P}$ with regard to the reference case on the suction side of a NACA4412 wing section at ($a-c$) $Re_c=200,000$ and ($d-f$) $Re_c=400,000$. }
\label{top_con3}
\end{figure}

As for the generation of the spatial growth ($C_{f,G}$), it is decreased by the blowing and the influence on $C_{f,G}$ is stronger than that on $C_{f,V}$ and $C_{f,T}$.
To further clarify the cause of such variation, we trace back to its sub-constituents in equation \eqref{cf3} and plot the variations of $C_{f,C}$, $C_{f,D}$, and $C_{f,P}$ in figure \ref{top_con3}.
In the cases with blowing, the generation of wall-normal convection ($C_{f,C}$) and adverse pressure gradient ($C_{f,P}$) is decreased, while that of streamwise development ($C_{f,D}$) is increased. A slight turnup of $C_{f,P}$ is observed near the trailing edge for the blowing cases (see figure \ref{top_con3:c}), which probably relates to the fact that the boundary layer is approaching the condition of mean separation \citep{Atzori2020}. Generally, the positive variation of $C_{f,D}$  is overcome by the negative influence on $C_{f,C}$ and $C_{f,P}$, which consequently leads to the overall reduction of $C_{f,G}$ by blowing \citep{Mahfoze2019}, as shown in figures \ref{top_con:c} and \ref{top_con:f}. In Sec. \ref{sec:2}, we will further analyze the wall-normal distributions of these friction constituents, to reveal the connection of the  $C_f-$constituents with the turbulent dynamics across the boundary layer.
With mass suction, the control effects on the $C_f-$constituents  shown in figure~\ref{top_con} and \ref{top_con3} are opposite to those with blowing. 
Here we did not add more discussions on the suction cases for brevity.

\begin{figure}[h]
\centering
\subfigure{\includegraphics[width = 7.5cm]{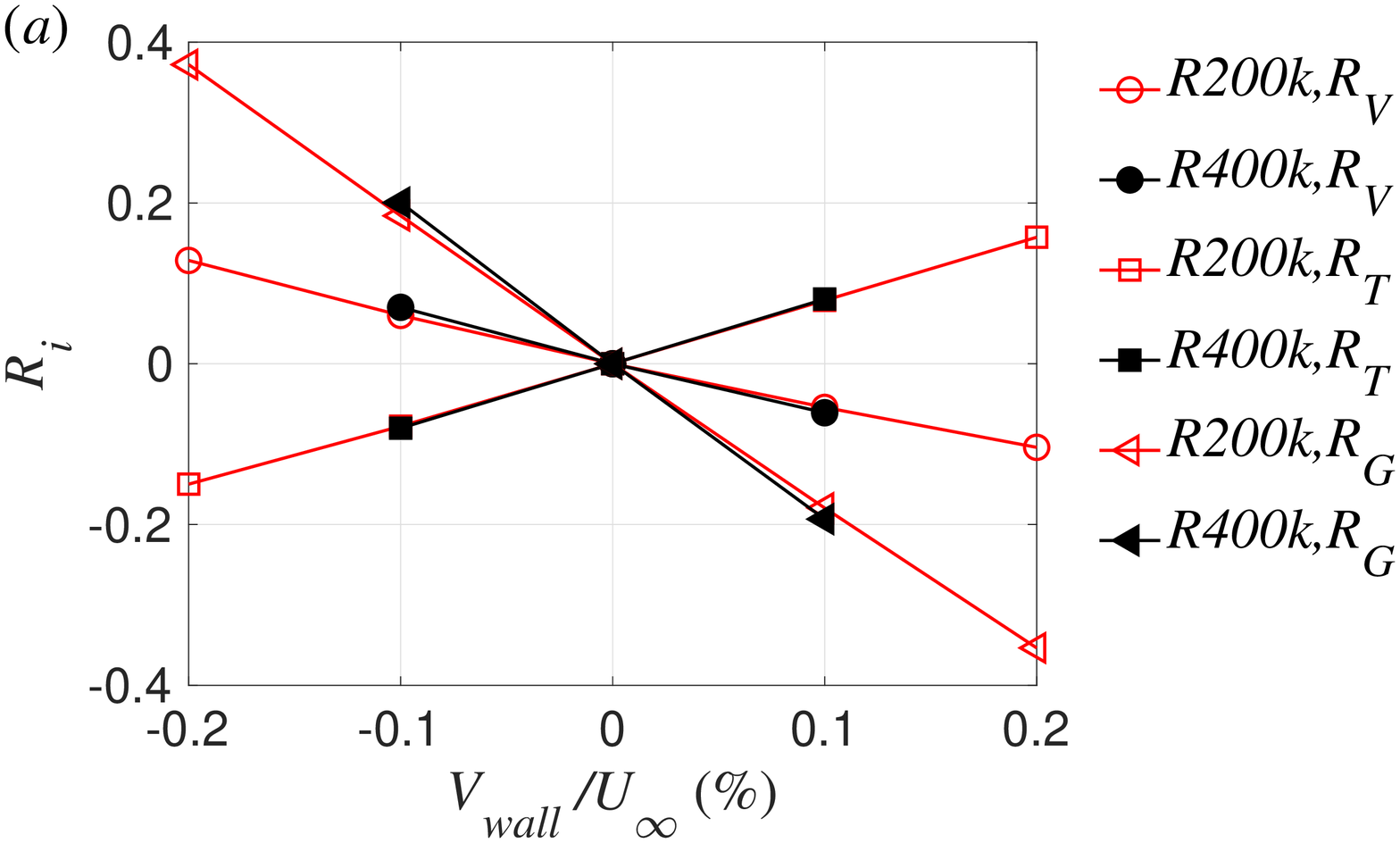}\label{rr:a}}
\subfigure{\includegraphics[width = 7.5cm]{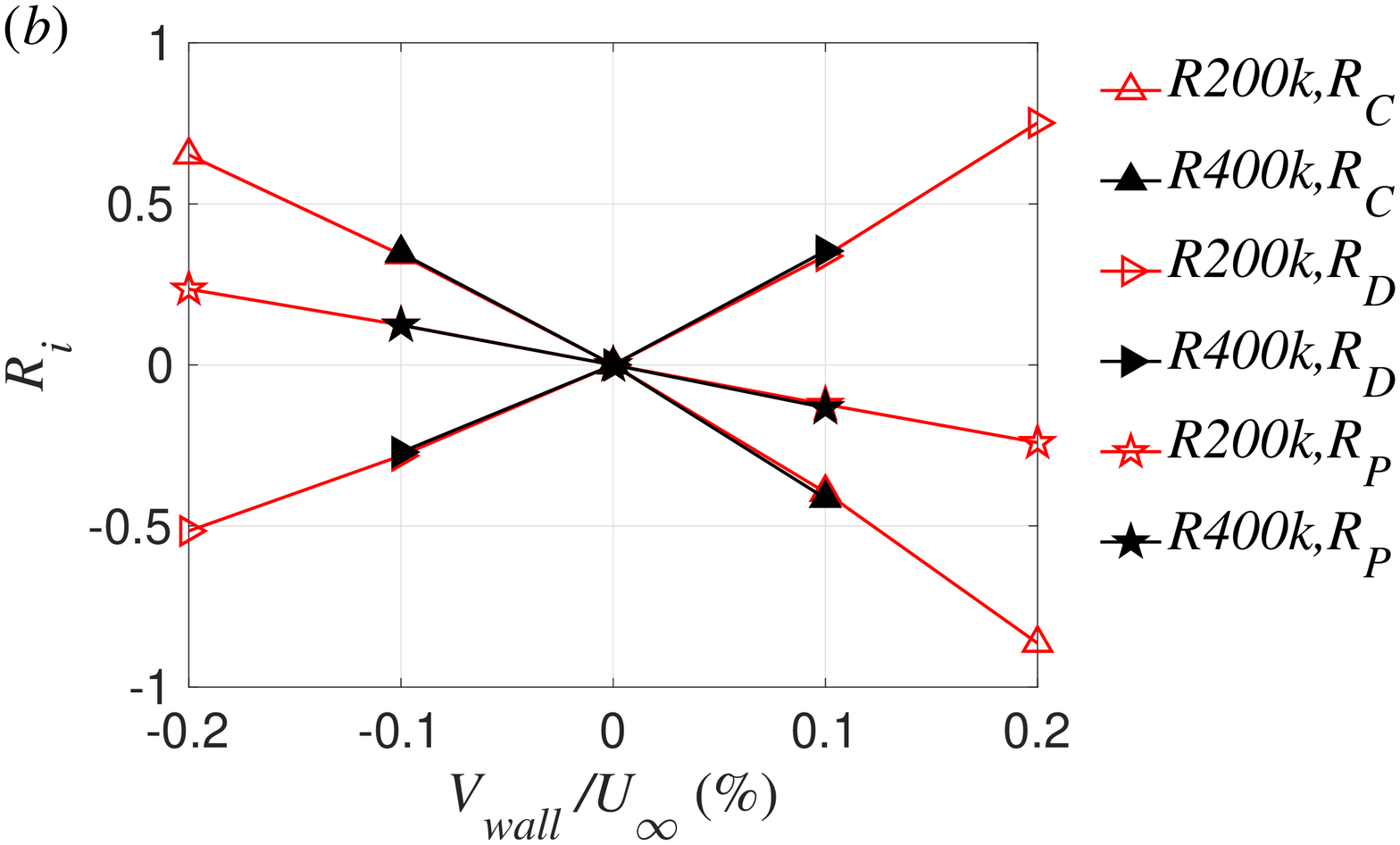}\label{rr:b}}
\caption{Total friction-drag change rate as a function of control intensity ($V_{wall}/U_\infty$) on the suction side of a NACA4412 wing section.}
\label{rr}
\end{figure}

In order to assess the control effects on the total skin-friction drag over the control surface, a parameter $\bar{D}_{f,i}$ is introduced:
\begin{equation}
\bar{D}_{f,i}={\int\limits_{\Omega_{\rm ctr}}{\tau_{w,i}}\left(\vec{t} \cdot\vec{n}\right){\rm d}\xi},
\end{equation}
where ${\tau_{w,i}}=C_{f,i}\cdot\left(0.5\rho U_e^2\right)$ is the decomposed component of wall-shear stress, with the subscript `\textit{i}' working as a label to denote each $C_f-$constituent, {\it i.e.} `\textit{V}', `\textit{T}', `\textit{G}', `\textit{C}', `\textit{D}', and `\textit{P}' as mentioned in equations \eqref{apg_formula} and \eqref{cf3}, $\vec{t}$ and $\vec{n}$ denote the unit vectors tangential to the airfoil surface and along the free-stream direction respectively, $\xi$ is the curvilinear coordinate along the airfoil surface, and $\Omega_{\rm ctr}$ represents the area of the control surface. The change rate of $\bar{D}_{f,i}$, with respect to the uncontrolled reference case, is then defined as:
\begin{equation}
R_i=\frac{\left(\bar{D}_{f,i}-\bar{D}_{f,i}^{\rm ref}\right)}{\bar{D}_{f}^{\rm ref}}.
\end{equation}

Figure \ref{rr} shows the result of $R_i$ under different control schemes at $Re_c$=$200,000$ and $400,000$.
It can be seen that $R_i$ appears to be linearly dependent on the control intensity of the blowing/suction, within $-0.2\% \leq V_{\rm wall}/U_\infty \leq 0.2\%$. However, note that this might not always be true when $V_{\rm wall}/U_\infty$ becomes much larger, and needs to be validated in the future work.
Among the decomposed constituents, the most significant control effect lies on the friction constituent of spatial growth of the flow ($C_{f,G}$), with its sub-constituents primarily correlated with the convection, streamwise growth, and pressure gradient in the outer region \citep{Fan2020}.
This reveals that the tremendously influenced outer-layer dynamics play an important role in the drag control with blowing/suction.
Moreover, weak Reynolds-number effects are found, especially for $R_V$ and $R_G$, where a stronger control effect is achieved at higher Reynolds number which to some extent validates the theoretical estimation by \cite{Kametani2011}.

\subsection{Wall-normal distributions of the $C_f-$constituents}\label{sec:2}
To answer the question that how uniform blowing/suction specifically influences the sources of skin-friction generation, the wall-normal distributions of the decomposed $C_f-$constituents across the boundary layer are investigated. 
As $C_f$ and its constituents vary along the streamwise direction on the wing surface, we only discuss the wall-normal contributions of the $C_f-$constituents at $x/c \approx 0.75$, where the friction Reynolds numbers are $Re_\tau\approx224$, $205$, $180$, $237$, $245$, $362$, $332$, and $387$ in the case  of ``$Re200k, ss, ref$'', ``$Re200k, ss, blw1$'', ``$Re200k, ss, blw2$'', ``$Re200k, ss, sct1$'', ``$Re200k, ss, sct2$'', ``$Re400k, ss, ref$'', ``$Re400k, ss, blw1$'', and ``$Re400k, ss, sct1$'' respectively. 
Similar conclusions can be drawn at other positions within $0.2\le x/c \le0.85$, and the results are not shown here for simplicity.

The $C_f-$constituents are expressed in intrinsic scales as:
\begin{eqnarray}
{C_{f,V}} &=& 2/U_e^{+3}\int_{0}^{\infty}\left(\frac{\partial \left<u\right>^+}{\partial y^+}\right)^2{\rm d}y^+,\label{cf1_cf}\\  
{C_{f,T}} &=& 2/U_e^{+3} \int_{0}^{\infty}\left<-u'v'\right>^+\frac{\partial \left<u\right>^+}{\partial y^+}{\rm d}y^+,\label{cf2_cf}\\
{C_{f,G}} &=& 2/U_e^{+3} \int_{0}^{\infty}\left(\left<u\right>^+-U_e^+\right)\frac{\partial}{\partial y^+}\left(\frac{\partial \left<u\right>^+}{\partial y^+}-\left<u'v'\right>^+\right){\rm d}y^+,\label{cf3_cf}
\end{eqnarray}
where the superscript $+$ denotes normalization by viscous units, \textit{i.e.} friction velocity $u_\tau$ and viscous length scale $\delta_\nu=\nu/u_\tau$.

\begin{figure}[h]
\centering
\subfigure{\includegraphics[width = 5.5cm]{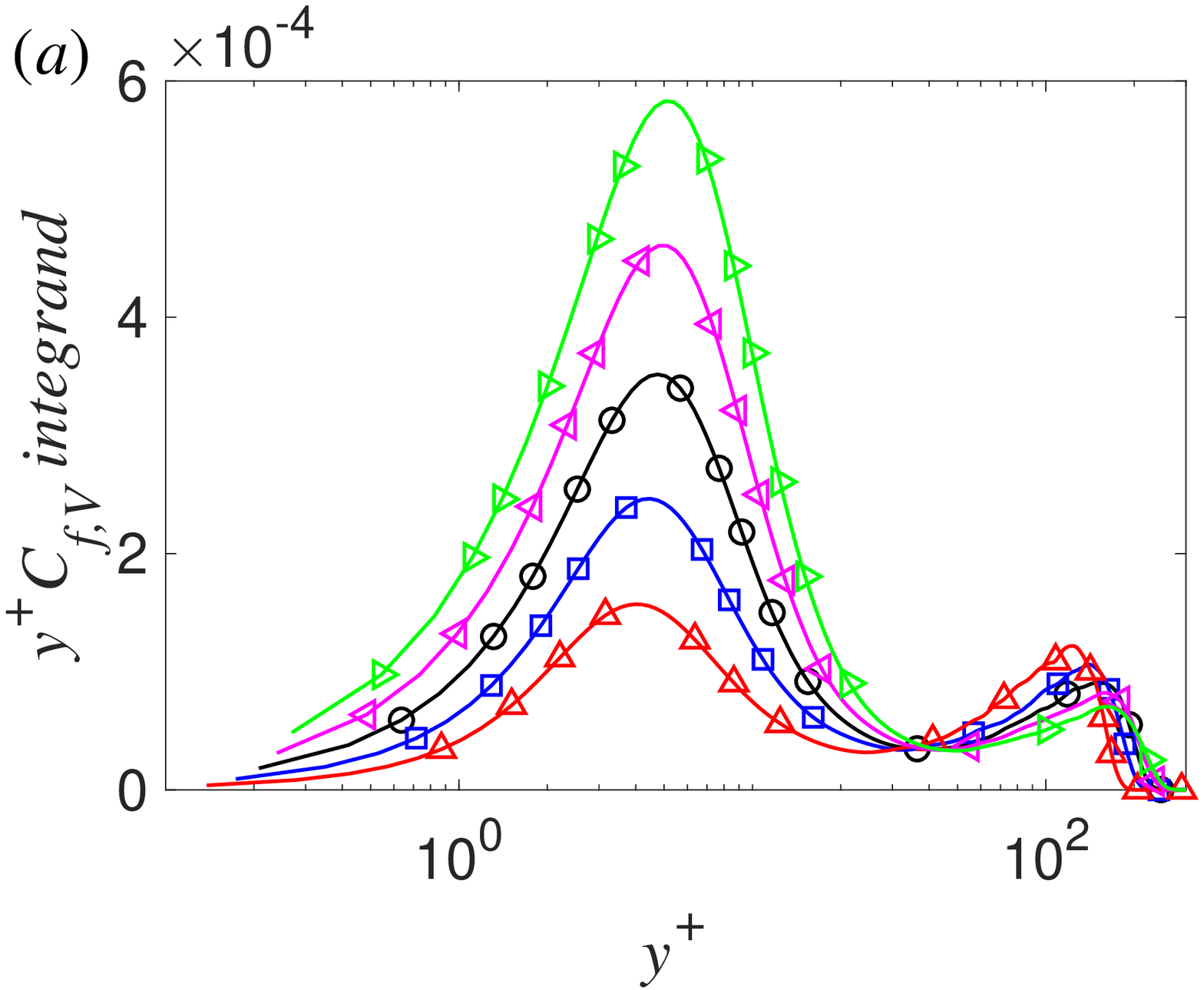}\label{topcfd:a}}
\subfigure{\includegraphics[width = 5.5cm]{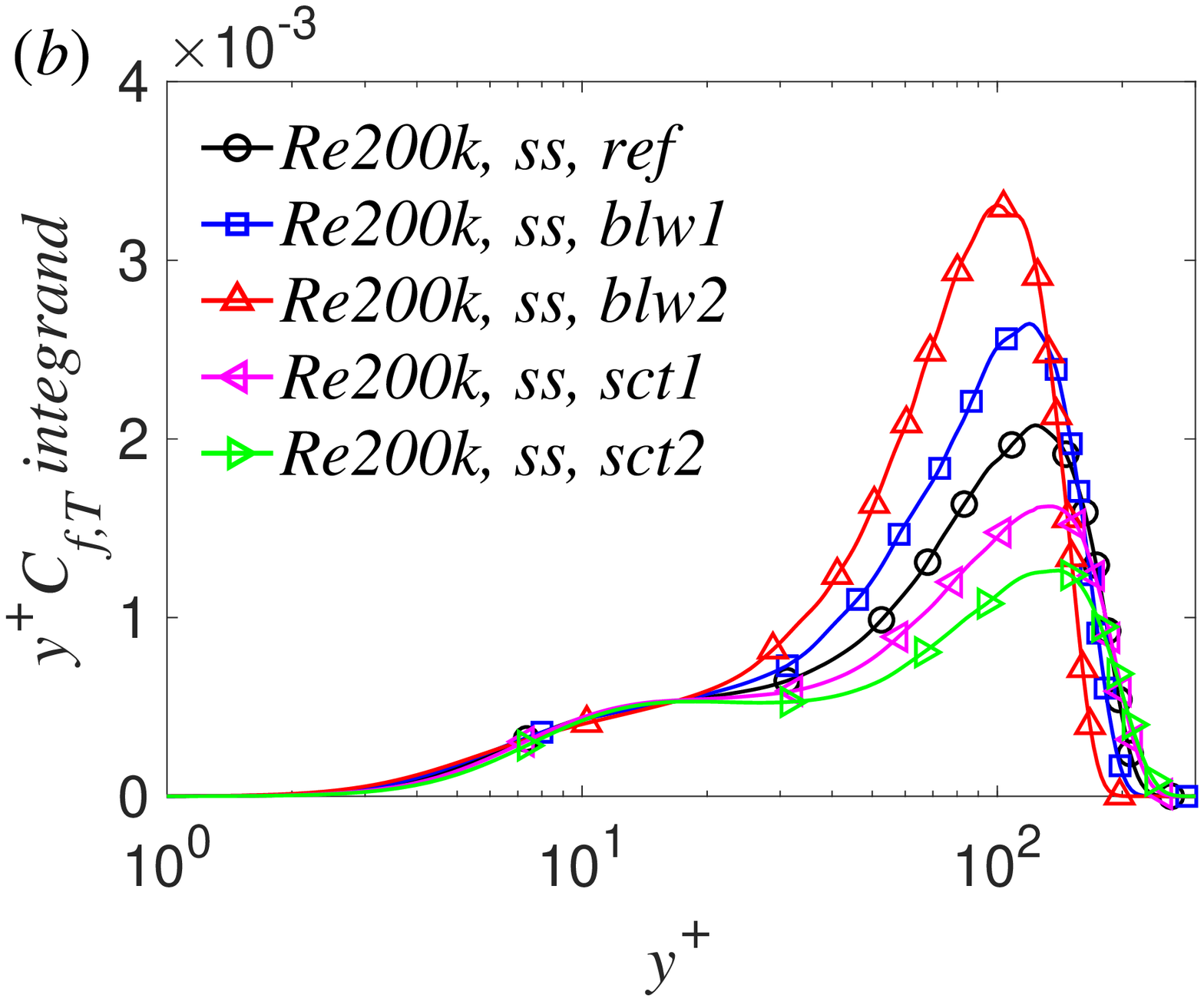}\label{topcfd:b}}
\subfigure{\includegraphics[width = 5.5cm]{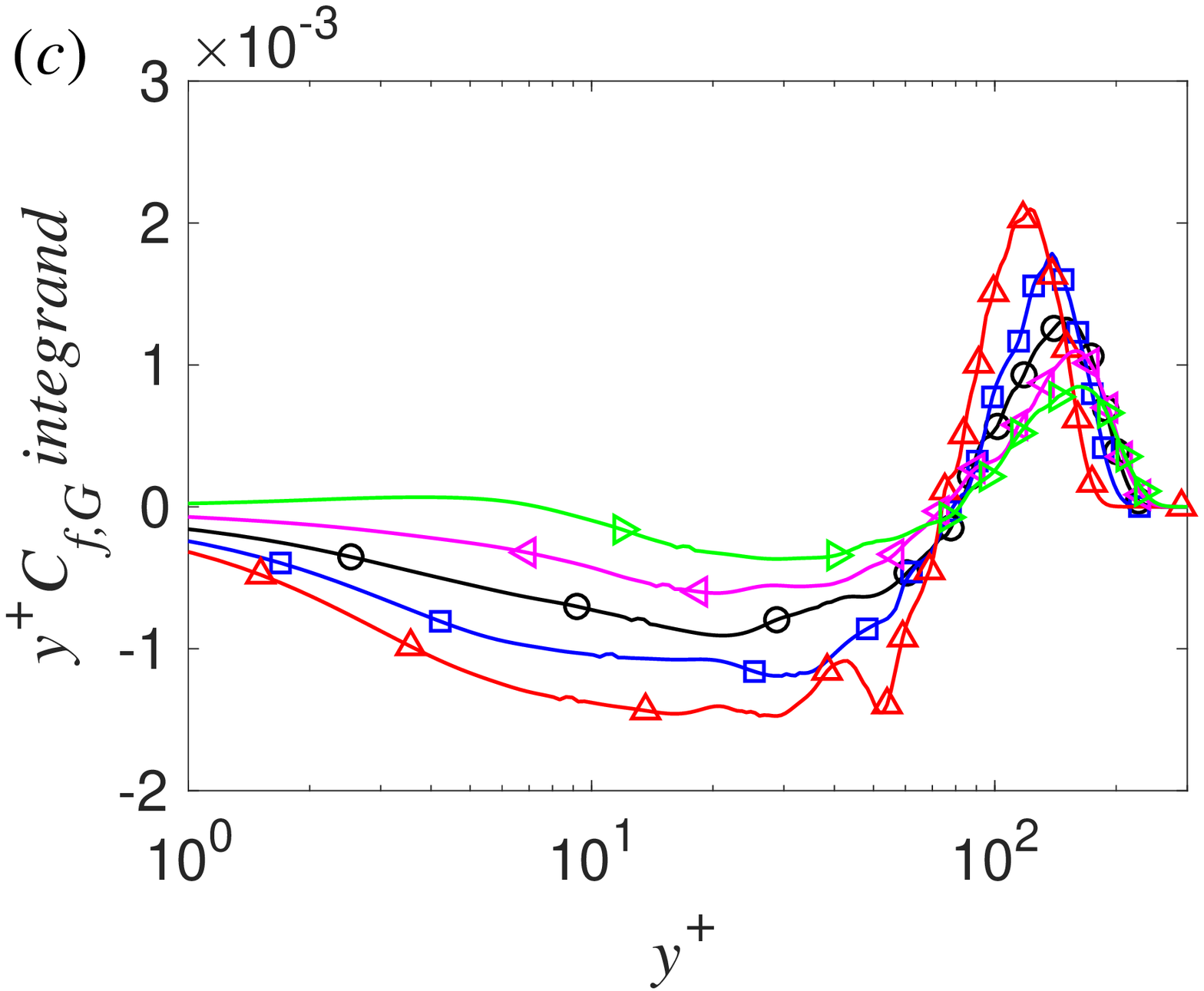}\label{topcfd:c}}
\subfigure{\includegraphics[width = 5.5cm]{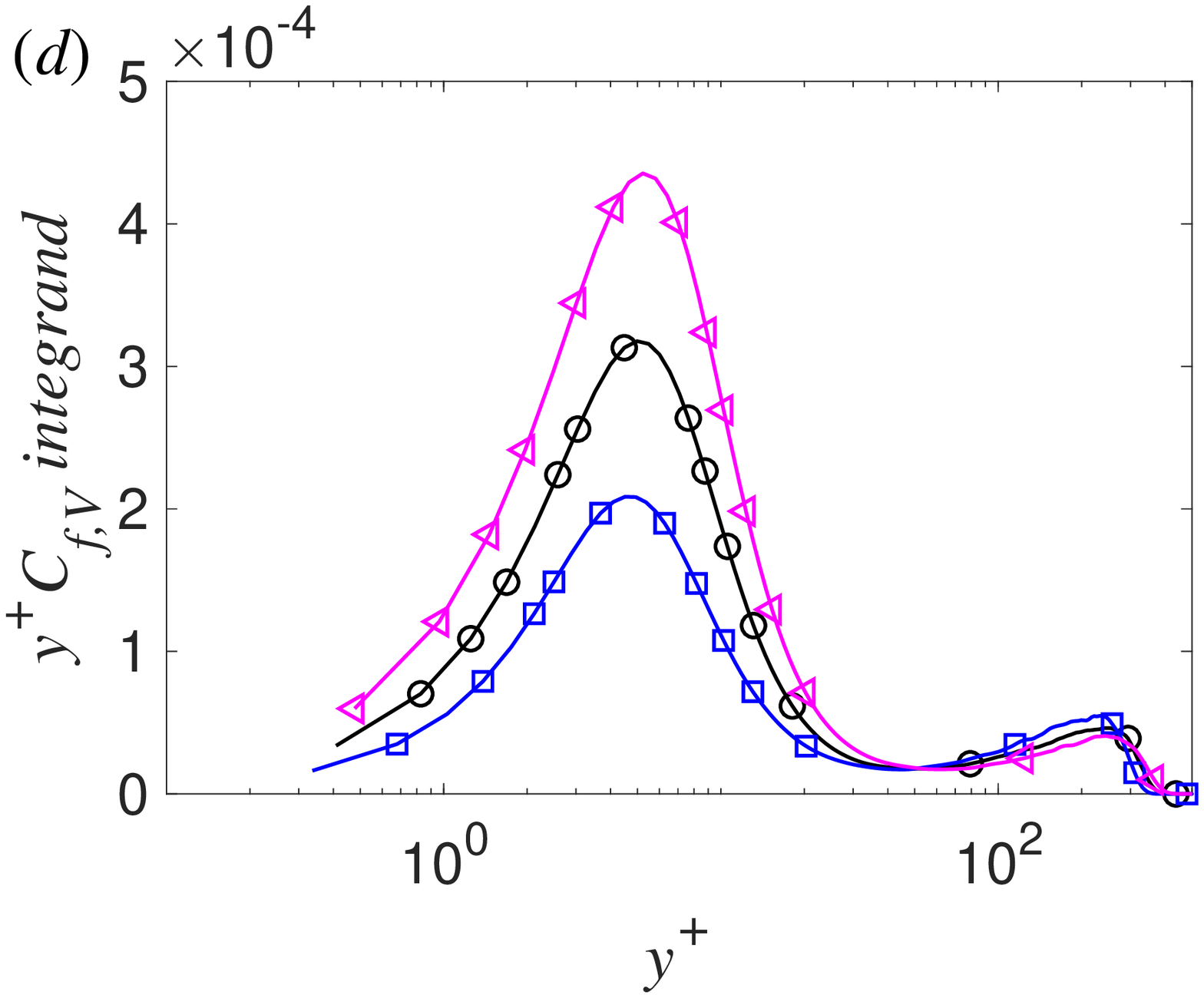}\label{topcfd:d}}
\subfigure{\includegraphics[width = 5.5cm]{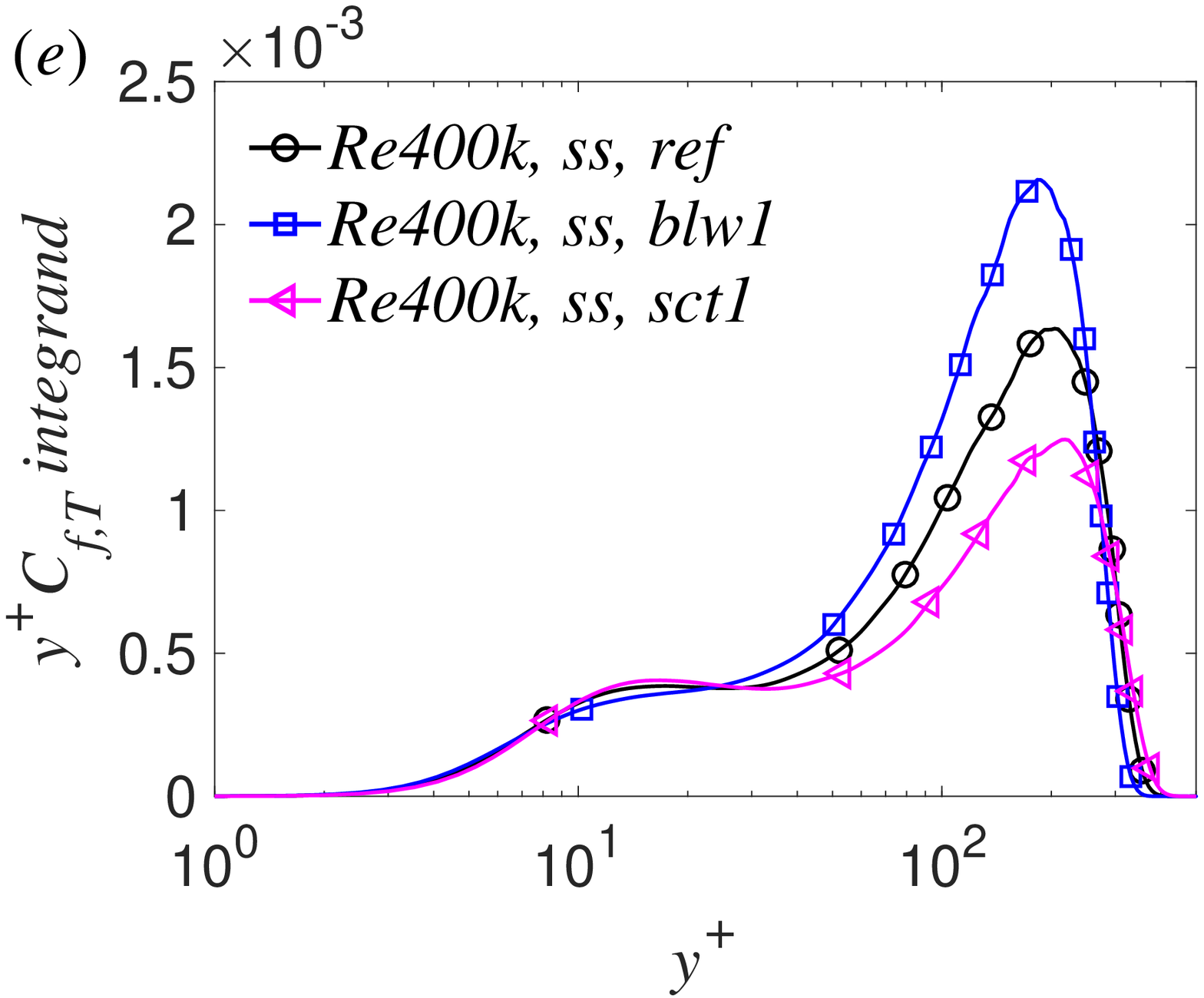}\label{topcfd:e}}
\subfigure{\includegraphics[width = 5.5cm]{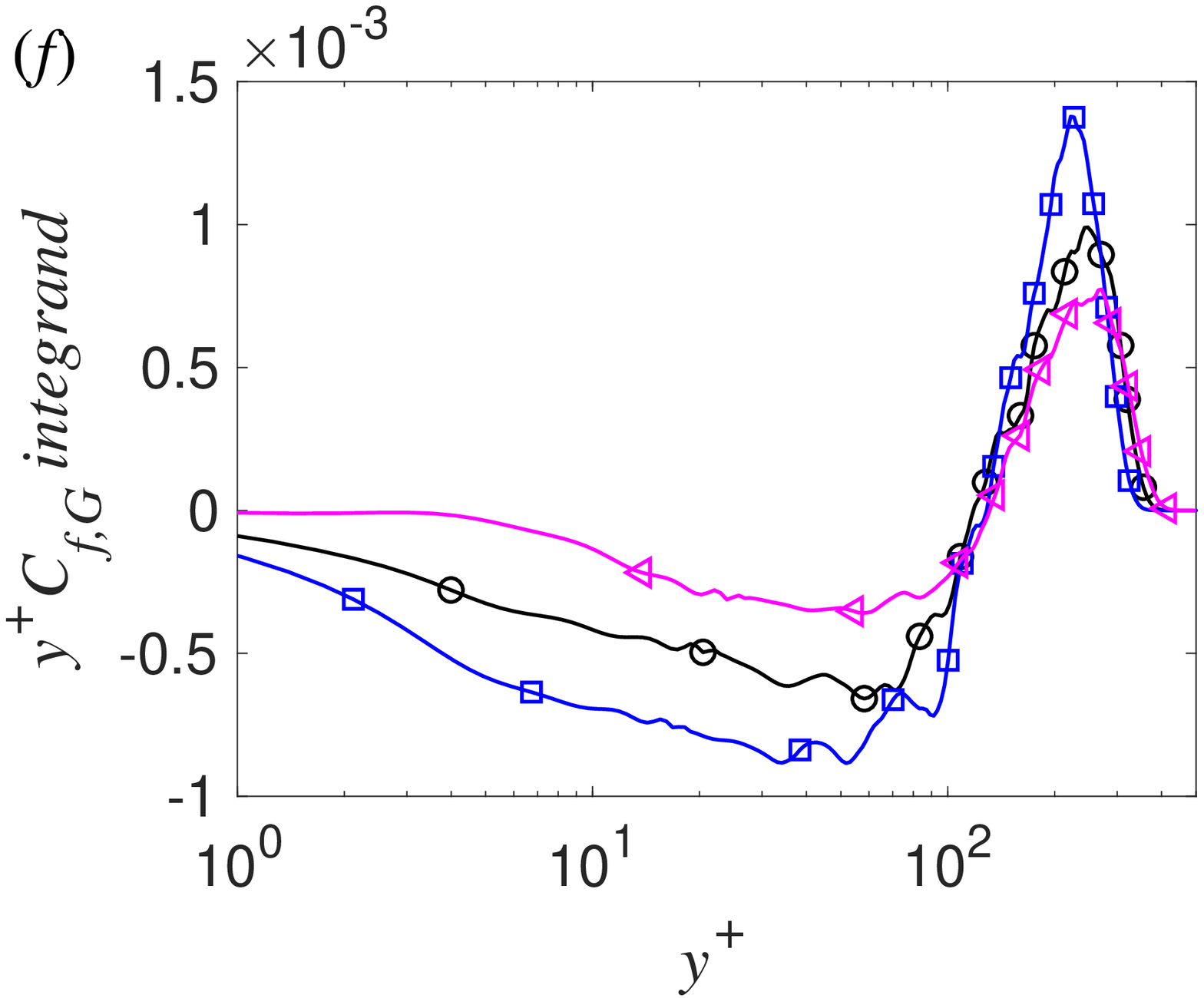}\label{topcfd:f}}
\caption{Pre-multiplied integrands of ($a,d$) $C_{f,V}$, ($b,e$) $C_{f,T}$, and ($c,f$) $C_{f,G}$ at $x/c\approx0.75$, as a function of $y^+$, on the suction side of a NACA4412 wing section at ($a-c$) $Re_c=200,000$ and ($d-f$) $Re_c=400,000$.}
\label{topcfd}
\end{figure}

\begin{figure}[h]
\centering
\subfigure{\includegraphics[width = 5.5cm]{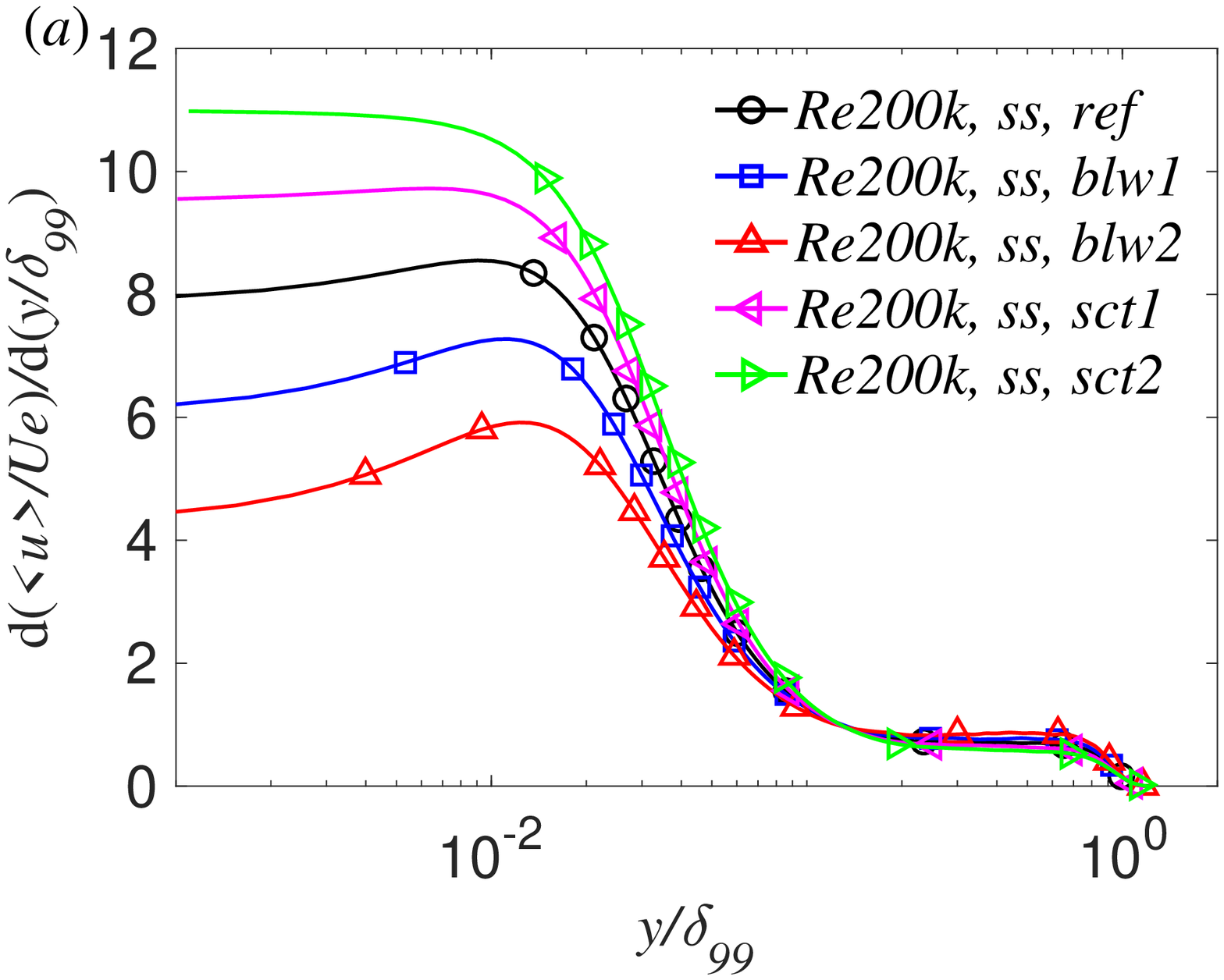}\label{topprof:a}}\subfigure{\includegraphics[width = 5.5cm]{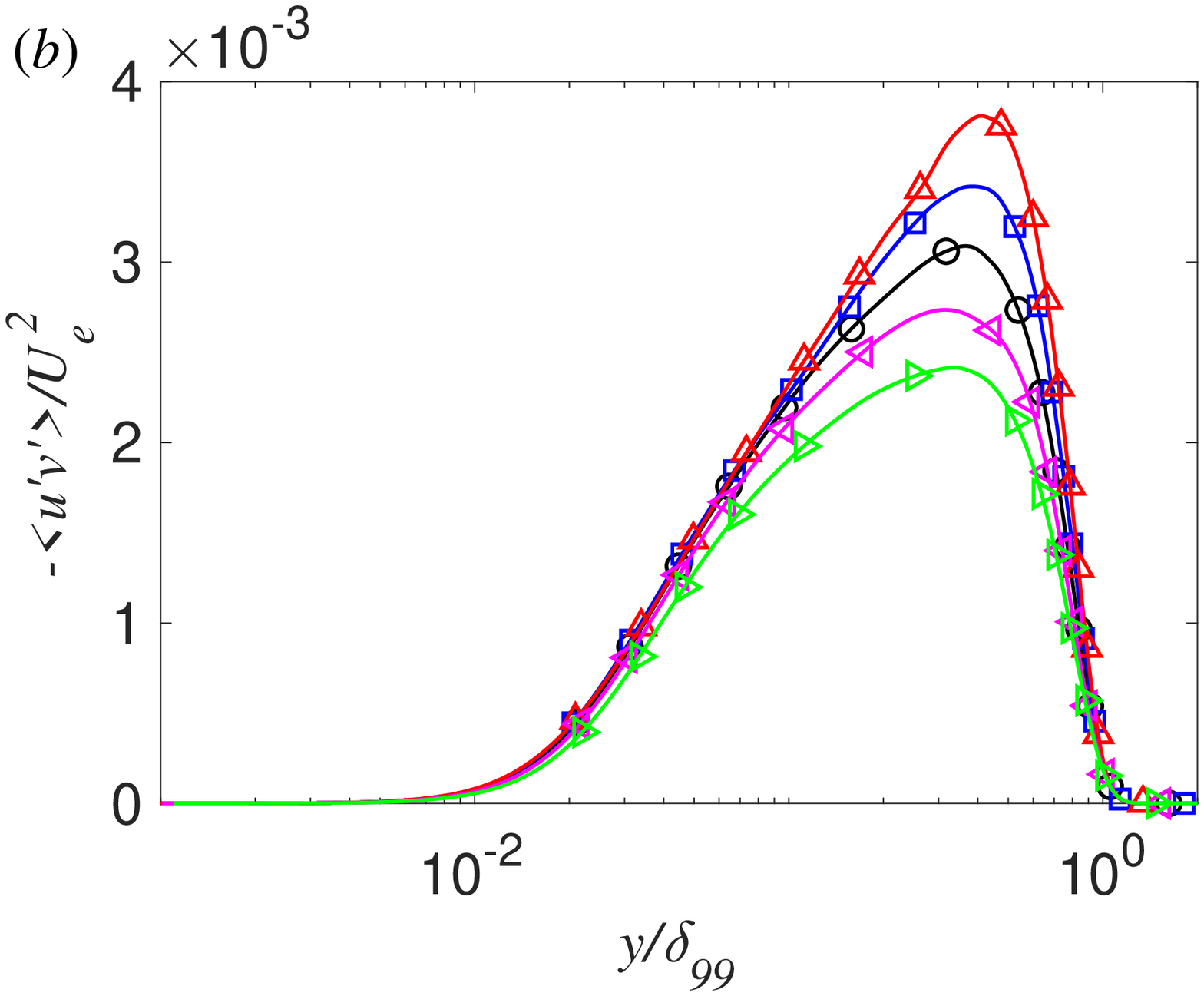}\label{topprof:b}}
\subfigure{\includegraphics[width = 5.5cm]{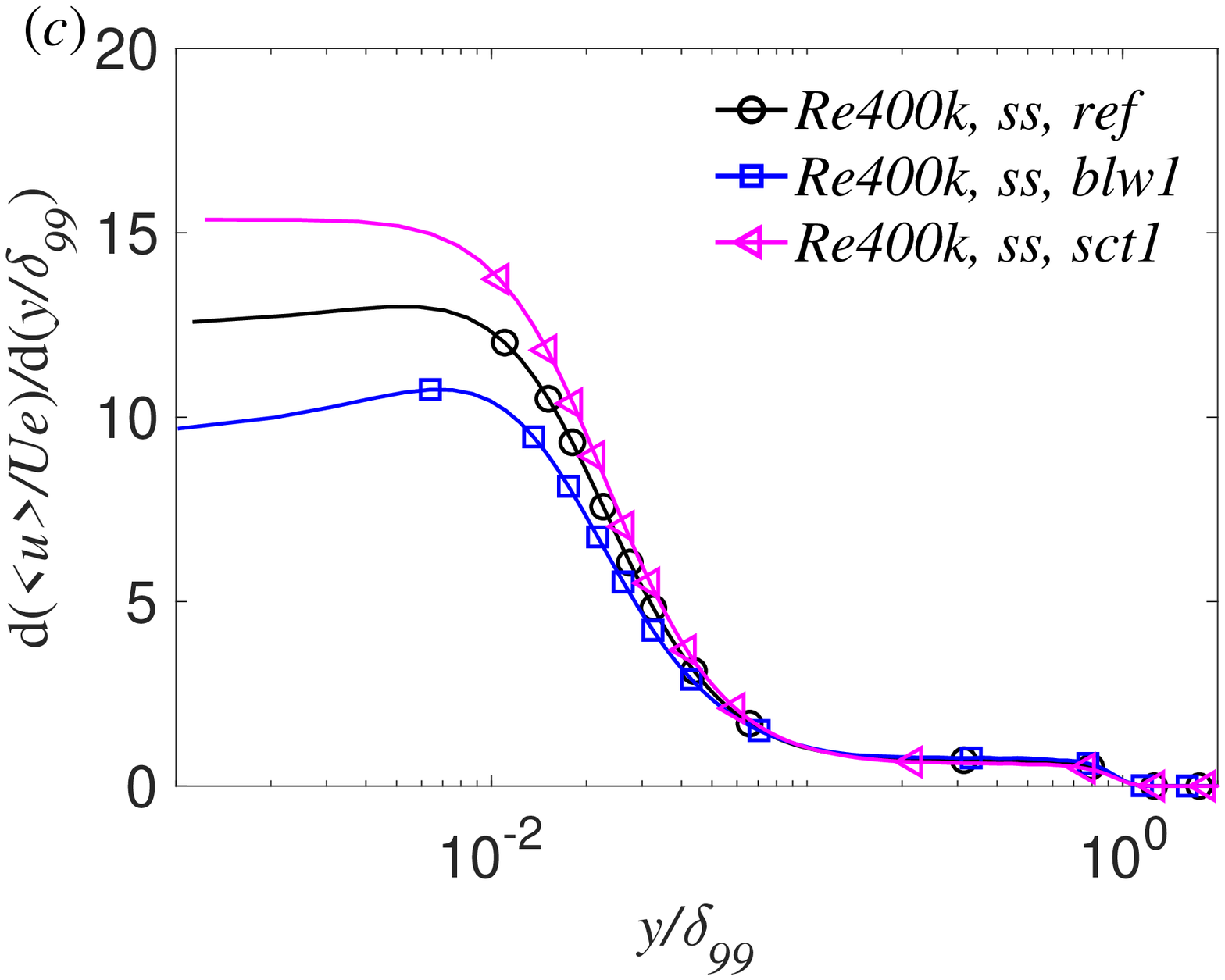}\label{topprof:c}}\subfigure{\includegraphics[width = 5.5cm]{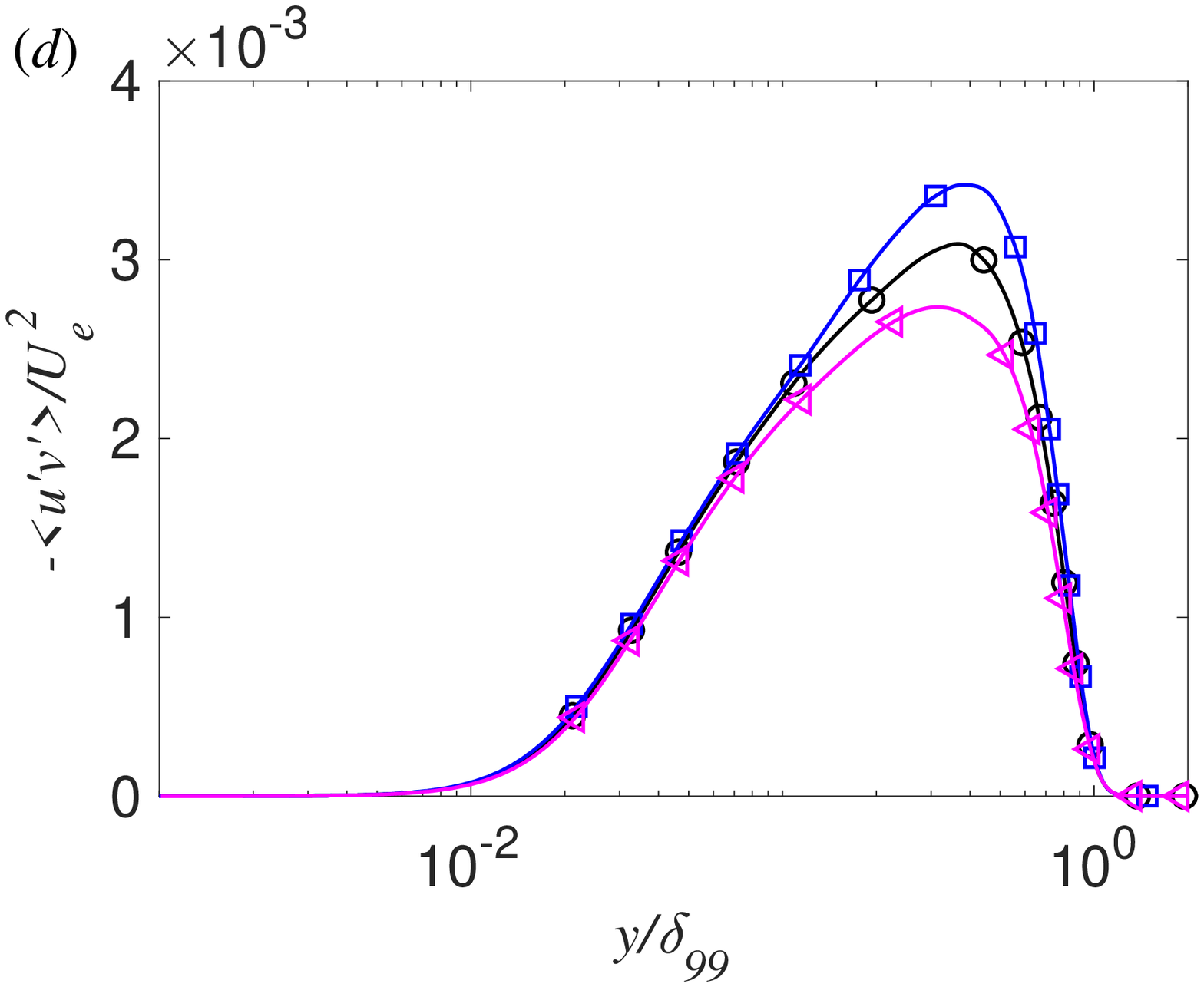}\label{topprof:d}}
\caption{Profiles of ($a,c$) wall-normal velocity gradient and ($b,d$) Reynolds shear stress, on the suction side of a NACA4412 wing section at ($a,b$) $Re_c=200,000$ and ($c,d$) $Re_c=400,000$.}
\label{topprof}
\end{figure}

Figure \ref{topcfd} shows the wall-normal distributions of the pre-multiplied integrand of $C_{f,V}$, $C_{f,T}$, and $C_{f,G}$ in equations \eqref{cf1_cf}--\eqref{cf3_cf}, as a function of $y^+$. The semi-logarithmic plots retain the advantage that the areas beneath the curves directly yield the total generation of the constituents. 
Comparisons between figures \ref{topcfd:a}--\ref{topcfd:c} for the cases at $Re_c$=$200,000$ and \ref{topcfd:d}--\ref{topcfd:f} for the cases at $Re_c$=$400,000$ confirms that the Reynolds-number variation does not change the conclusions which will be presented in the following from a qualitative perspective.

For the distribution of $C_{f,V}-$contributions, two peaks are respectively observed in the near-wall and outer region of APG-TBLs. Most of the $C_{f,V}-$contributions come from the inner region ($y^+<30$), indicating that the viscous dissipation is mostly concentrated in the near-wall region, as expected. In the meantime, a secondary peak appears in the outer region, which is probably due to the energy enhancement by APG \citep{Tanarro2020,Vila2020}. The secondary peak is absent in the ZPG-TBLs even at higher friction Reynolds number up to $Re_\tau=1270$ \citep{Fan2019}.
When uniform blowing/suction is applied, the locations of inner peaks are fixed at a wall-normal distance of $y^+\approx 5.0-6.0$, regardless of the control scheme.
In the blowing cases, the inner peak of $C_{f,V}-$contributions is reduced whereas the outer peak is increased, suggesting a lowered mean shear in the near-wall region while enhanced in the outer region due to the lifting-up of boundary layer \citep{Kornilov2015}, which is also validated by showing the wall-normal gradient of the tangential velocity in figures \ref{topprof:a} and \ref{topprof:c}. This reveals that blowing has different actions in different sub-layers, namely inhibiting the contribution of inner-layer dynamics to skin-friction generation while promoting that of outer-layer dynamics. When blowing intensity is up to 0.2\%, the outer-layer contributions seem to be comparable to the inner-layer contributions. On the other hand, suction behaves quite opposite for all constituents, which will not be repeated hereafter.

Similar inner and outer peaks are also observed in the pre-multiplied distribution of $C_{f,T}-$contributions, as shown in figures \ref{topcfd:b} and \ref{topcfd:e}, with the former well collapsed at the inner-scaled wall-normal distance $y^+\approx 16.0-17.0$.
It can be found that the outer-layer motions dominate the contributions of $C_{f,T}$, although the $Re_\tau$ is lower than $400$,  which is much different from the features in ZPG-TBLs \citep{Fan2019}. 
The prominent peak in the outer region suggests the energization of large-scale outer motions by APGs \citep{Harun2013}.
When blowing is applied, the inner peak of $C_{f,T}-$contributions is reduced, while the outer peak is increased. This phenomenon is linked to the wall-normal distributions of the wall-normal velocity gradient and Reynolds shear stress, as shown in figure \ref{topprof}.
With blowing, the wall-normal velocity gradients are suppressed significantly in the near-wall region, probably resulting from the damping of the near-wall sweep motions. On the other hand, in the outer layer, the Reynolds shear stress is amplified by the blowing. 
These two actions are consequently responsible for the changes of the inner and outer peaks in the distribution of $C_{f,T}-$contributions.

As for the  distribution of $C_{f,G}-$contributions in figures \ref{topcfd:c} and \ref{topcfd:f},  negative contributions are observed within $y^+\lesssim 80$ for the uncontrolled reference case at $Re_c$=$200,000$ ($y^+\lesssim130$ at $Re_c$=$400,000$), and positive values beyond this region. This differs from the result in ZPG-TBLs: $C_{f,G}$ always remains positive across the wall layer \citep{Fan2020}. 
Blowing enhances both the negative and positive distributions, as the strengthened adverse pressure gradient promotes a more pronounced growth of the boundary layer and a more prominent outer region \citep{Vinuesa2018}.

\begin{figure}[h]
\centering
\subfigure{\includegraphics[width = 5.5cm]{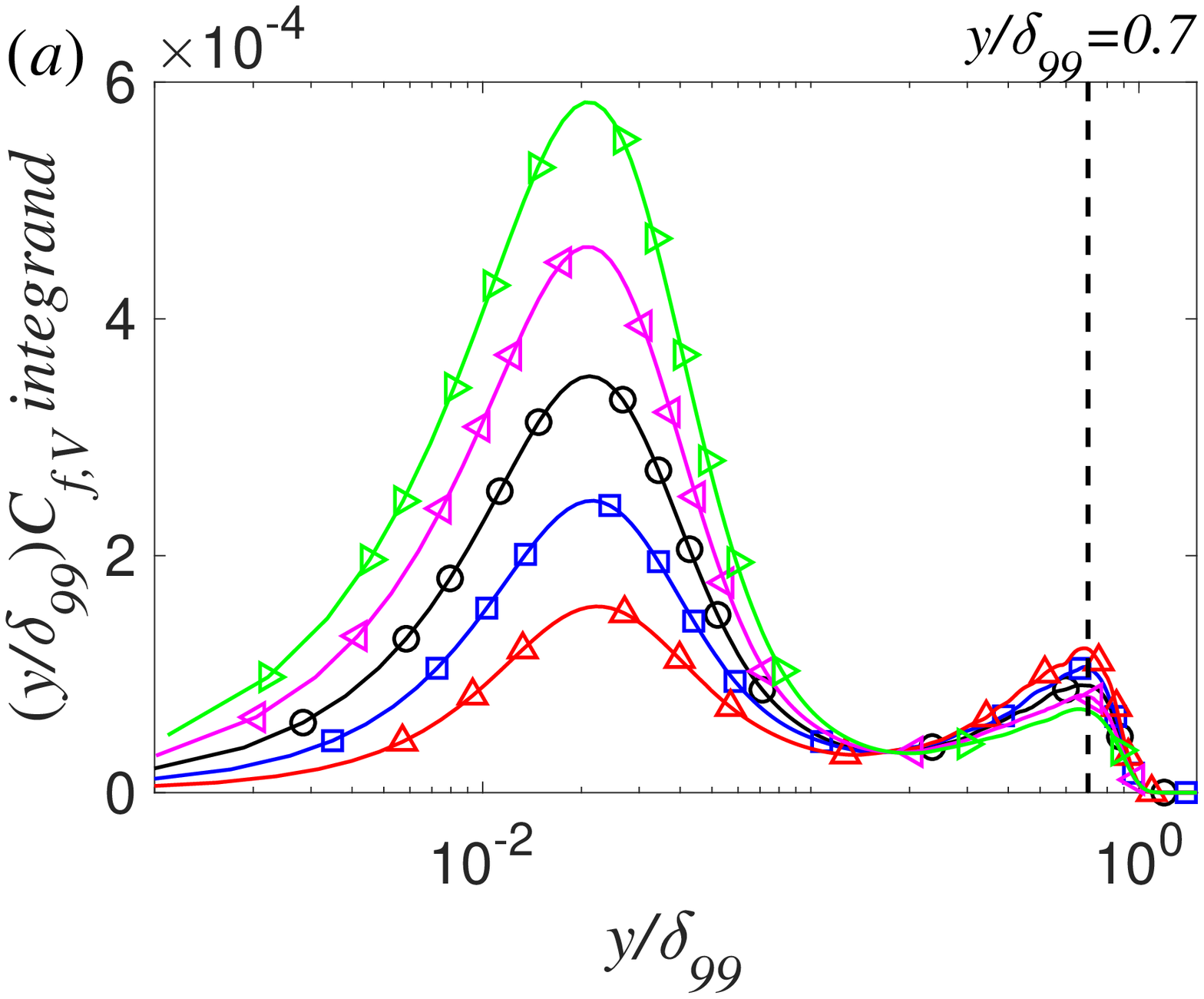}\label{topcfd2:a}}
\subfigure{\includegraphics[width = 5.5cm]{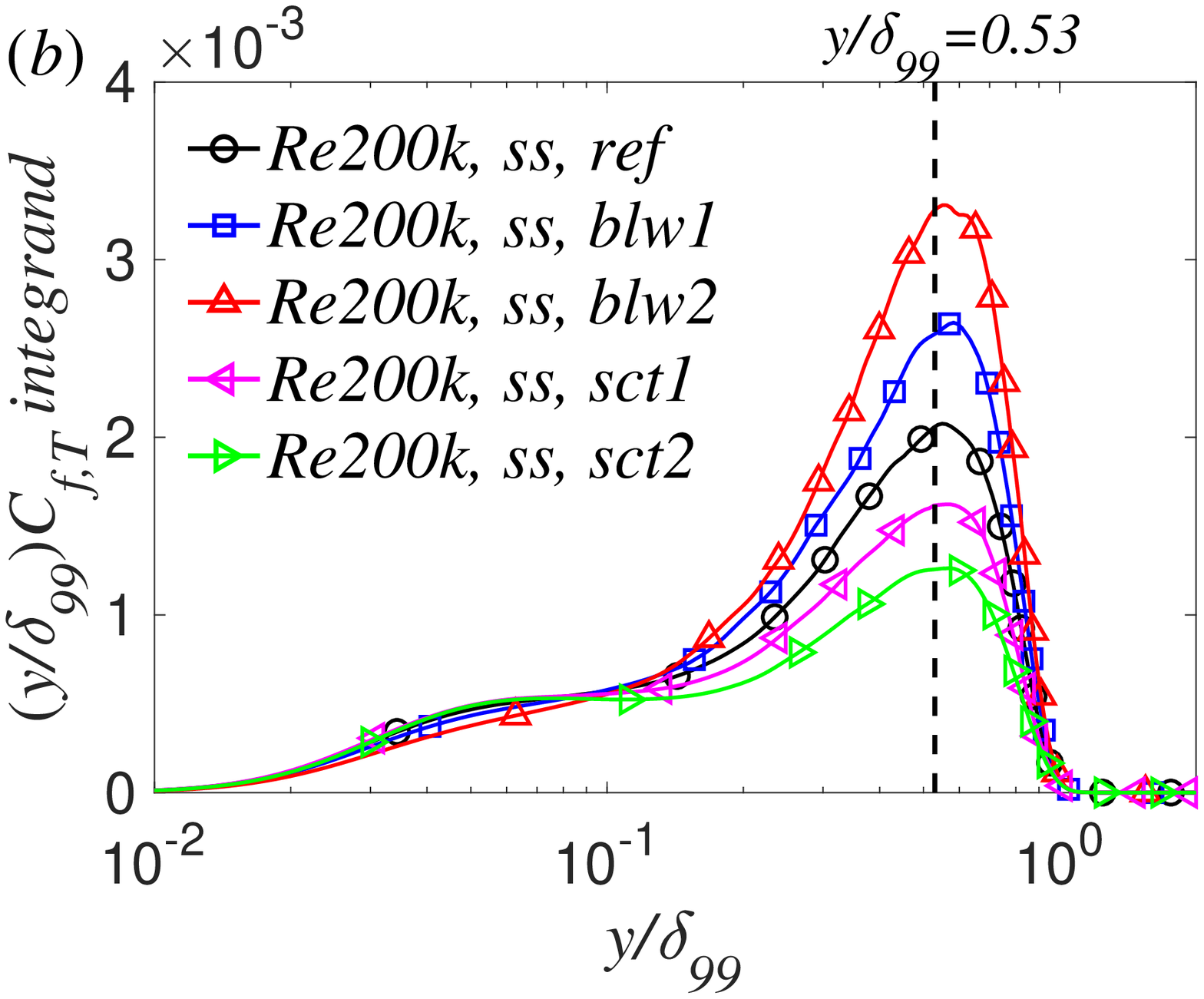}\label{topcfd2:b}}
\subfigure{\includegraphics[width = 5.5cm]{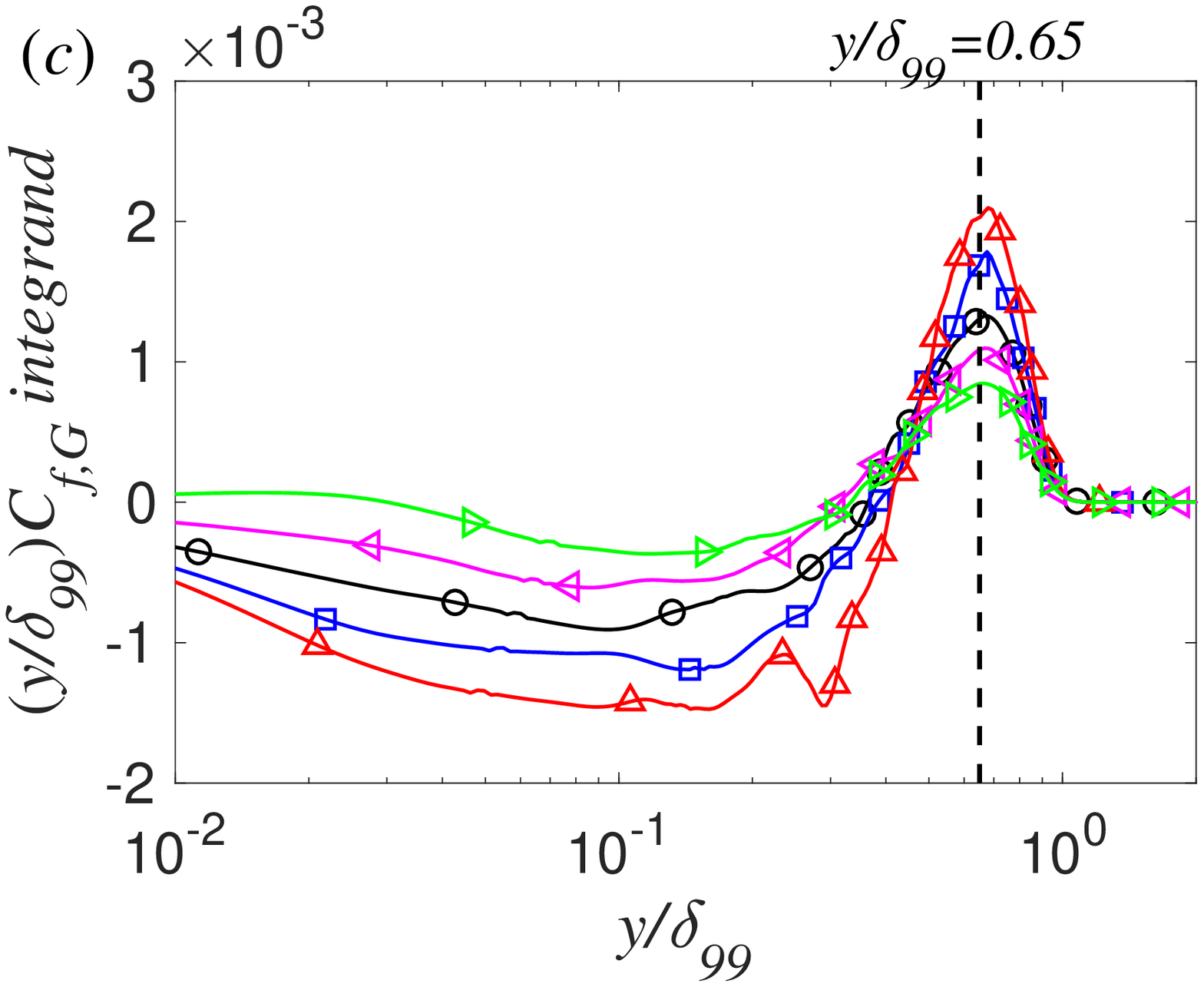}\label{topcfd2:c}}
\subfigure{\includegraphics[width = 5.5cm]{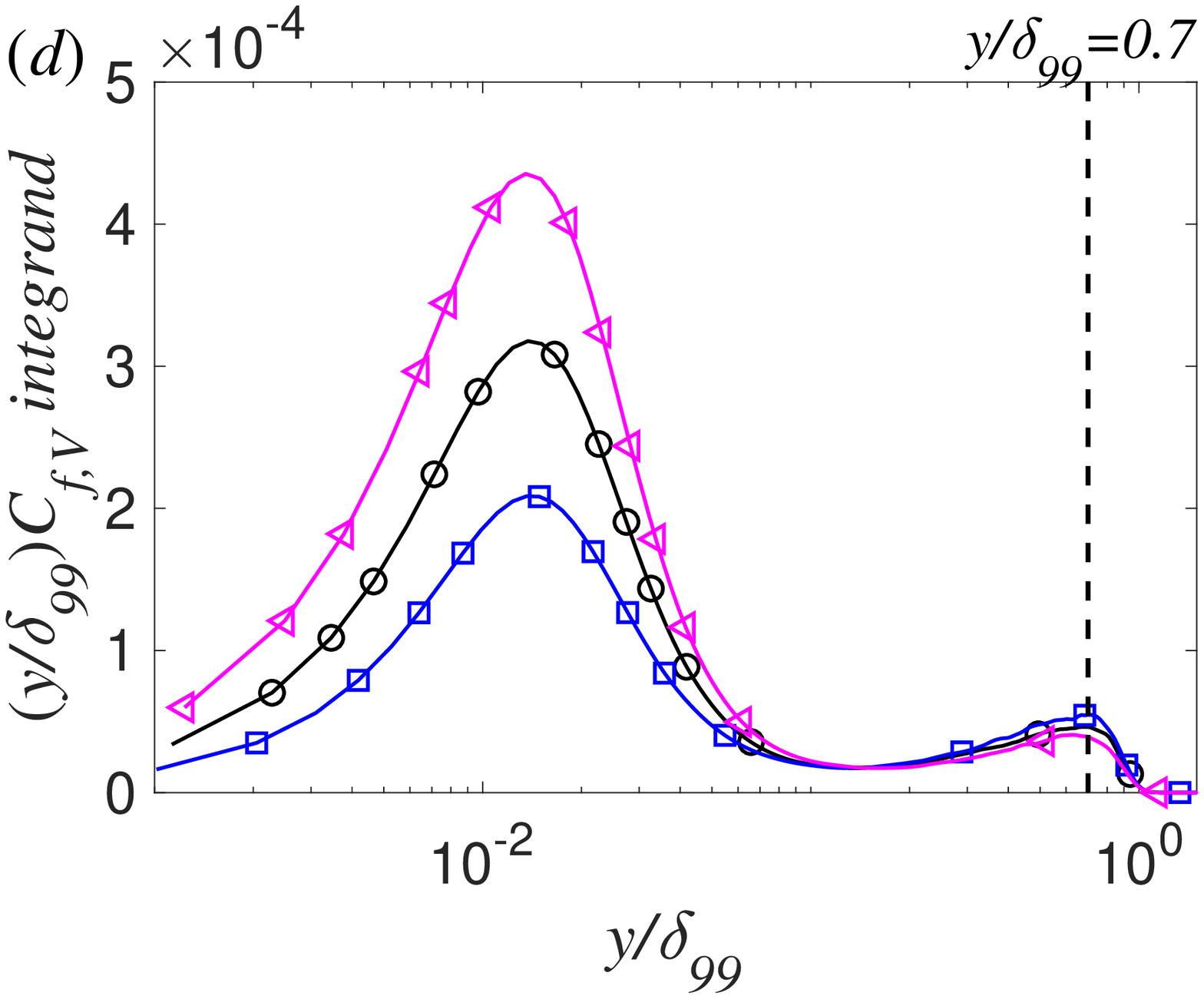}\label{topcfd2:d}}
\subfigure{\includegraphics[width = 5.5cm]{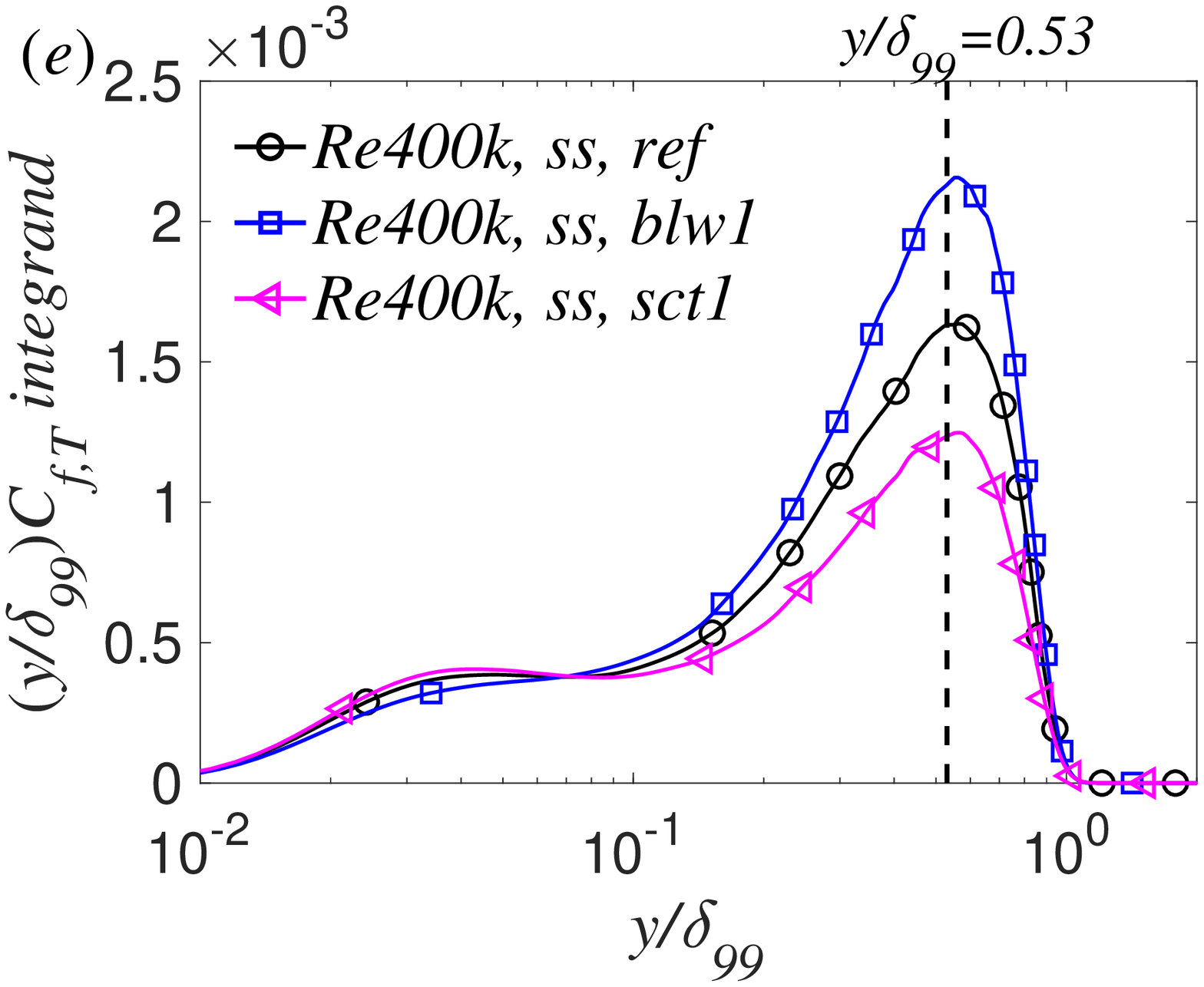}\label{topcfd2:e}}
\subfigure{\includegraphics[width = 5.5cm]{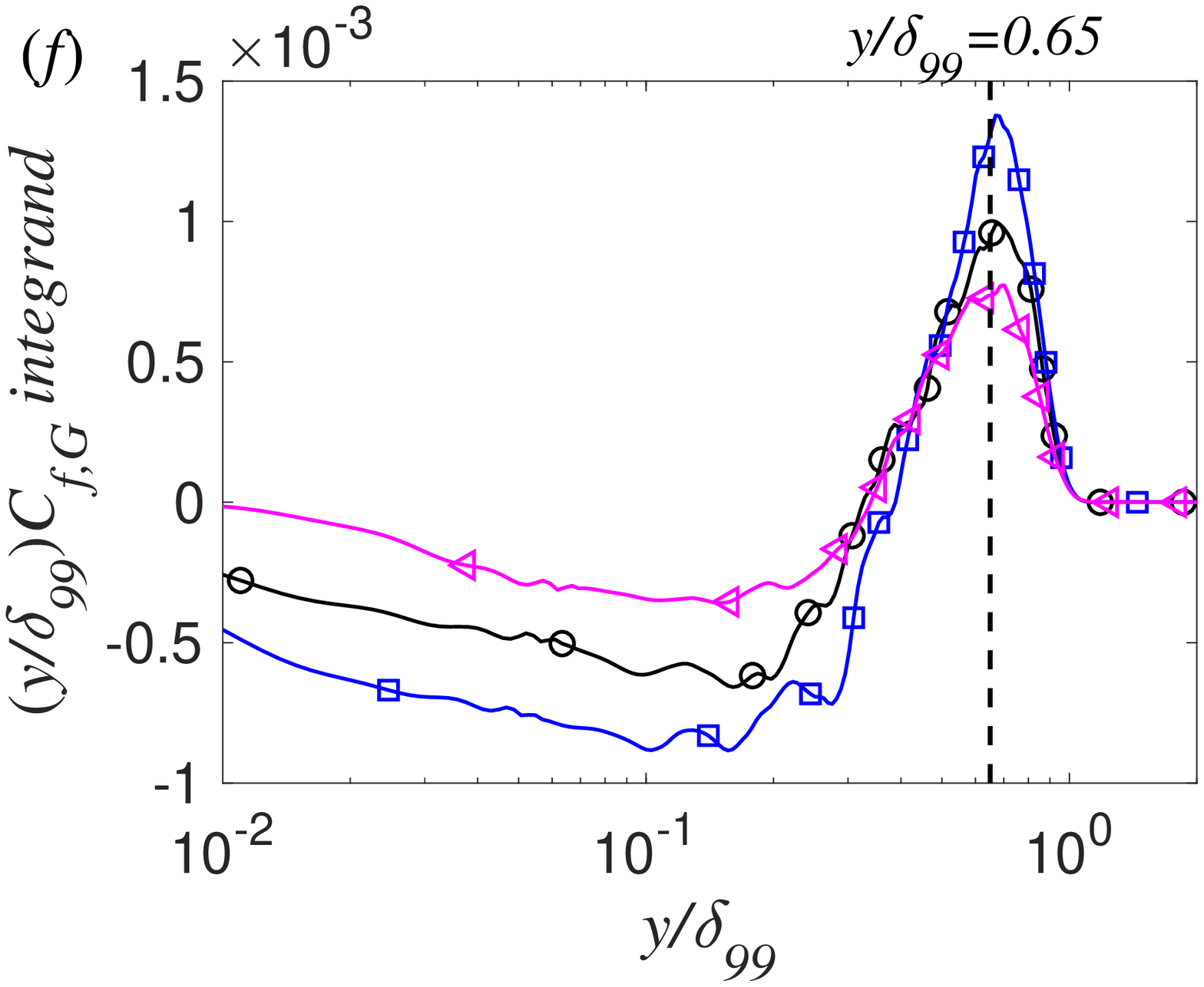}\label{topcfd2:f}}
\caption{Pre-multiplied integrands of ($a,d$) $C_{f,V}$, ($b,e$) $C_{f,T}$, and ($c,f$) $C_{f,G}$, as a function of $y/\delta_{99}$, on the suction side of a NACA4412 wing section at ($a-c$) $Re_c=200,000$ and ($d-f$) $Re_c=400,000$.}
\label{topcfd2}
\end{figure}

In contrast to the wall-normal distributions as a function of $y^+$ shown in figure \ref{topcfd},  figure \ref{topcfd2} plots their profiles as a function of $y/\delta_{99}$. The outer-peak locations of  $C_{f,V}-$, $C_{f,T}-$, and $C_{f,G}-$contributions normalized by the outer scale are well-collapsed at $y/\delta_{99}\approx 0.7$, $0.53$, and $ 0.65$, respectively, as marked with vertical dashed lines in figure \ref{topcfd2}, regardless of the control scheme and Reynolds number. 
These phenomena are consistent with our previous finding \citep{Fan2020}, \textit{i.e.} the inner-peak locations (in $C_{f,V}-$ and $C_{f,T}-$contribution) exhibit good scaling in the inner unit ($\delta_\nu$), and the outer-peak locations in the outer unit ($\delta_{99}$), regardless of the friction Reynolds number, the magnitude of APG and its development history. This finding suggests that self similarity is exhibited in inner or outer scales for the turbulence statistics associated with the friction-drag generation.

\begin{figure}[h]
\centering
\subfigure{\includegraphics[width = 5.5cm]{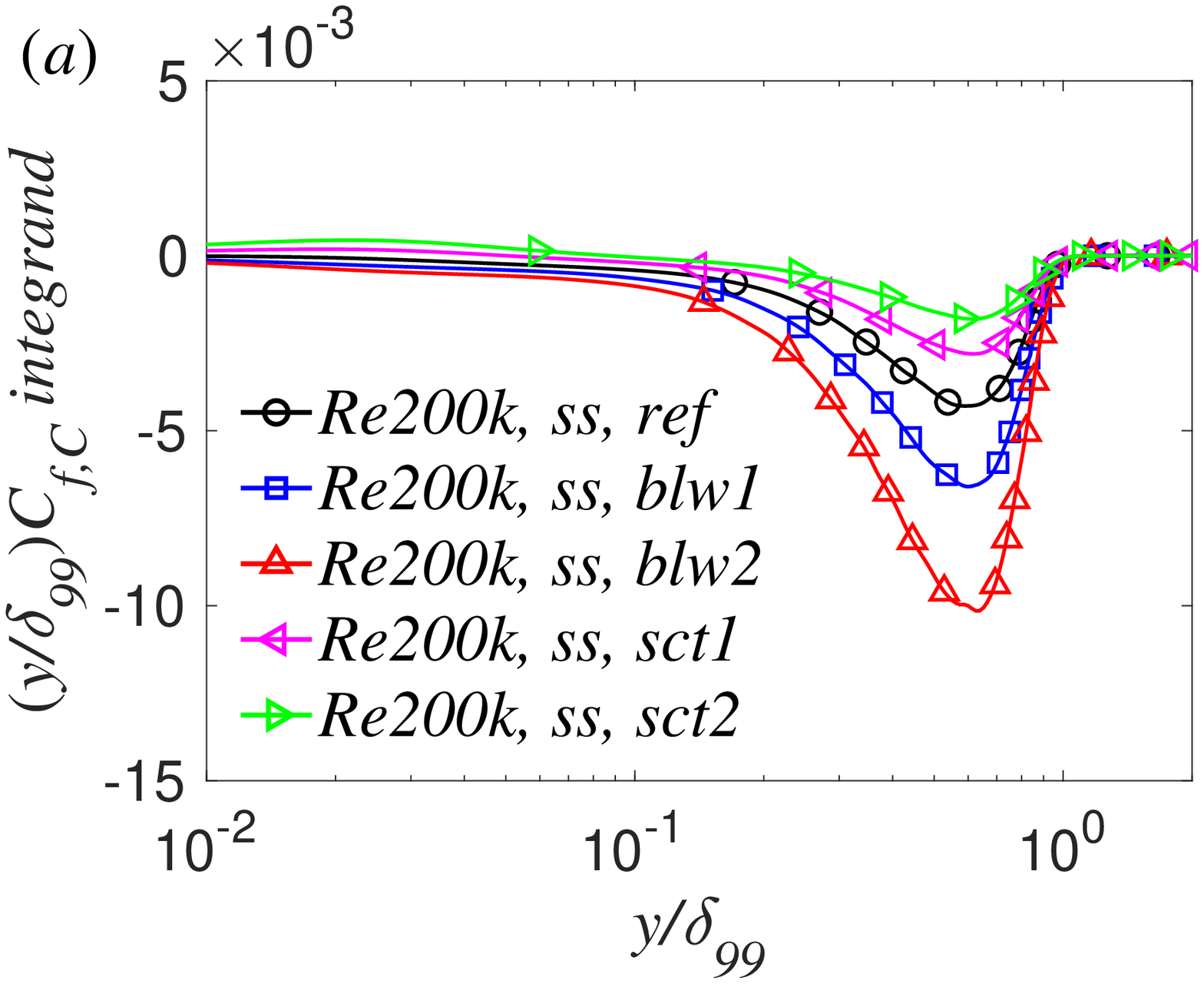}\label{topcf3d:a}}
\subfigure{\includegraphics[width = 5.5cm]{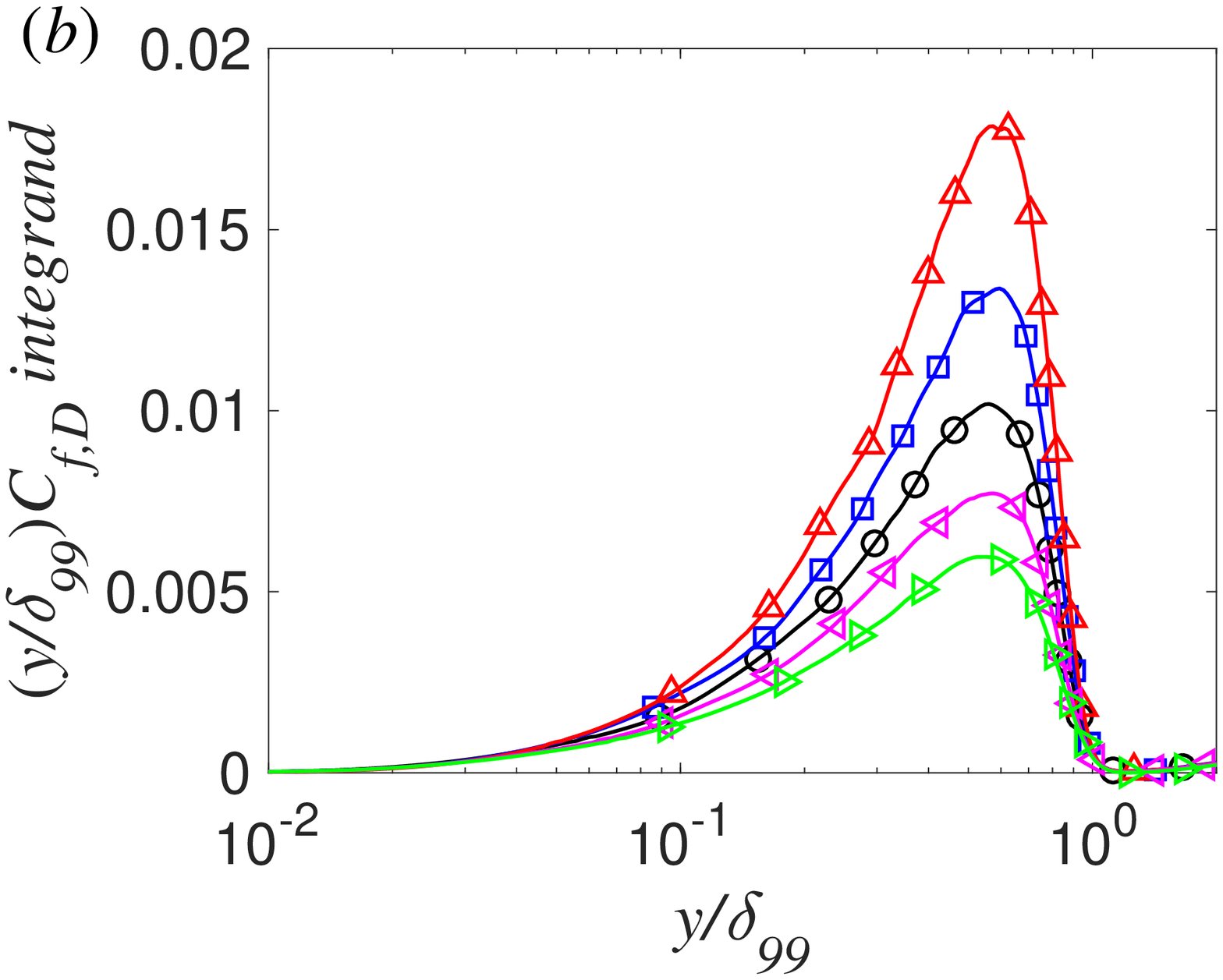}\label{topcf3d:b}}
\subfigure{\includegraphics[width = 5.5cm]{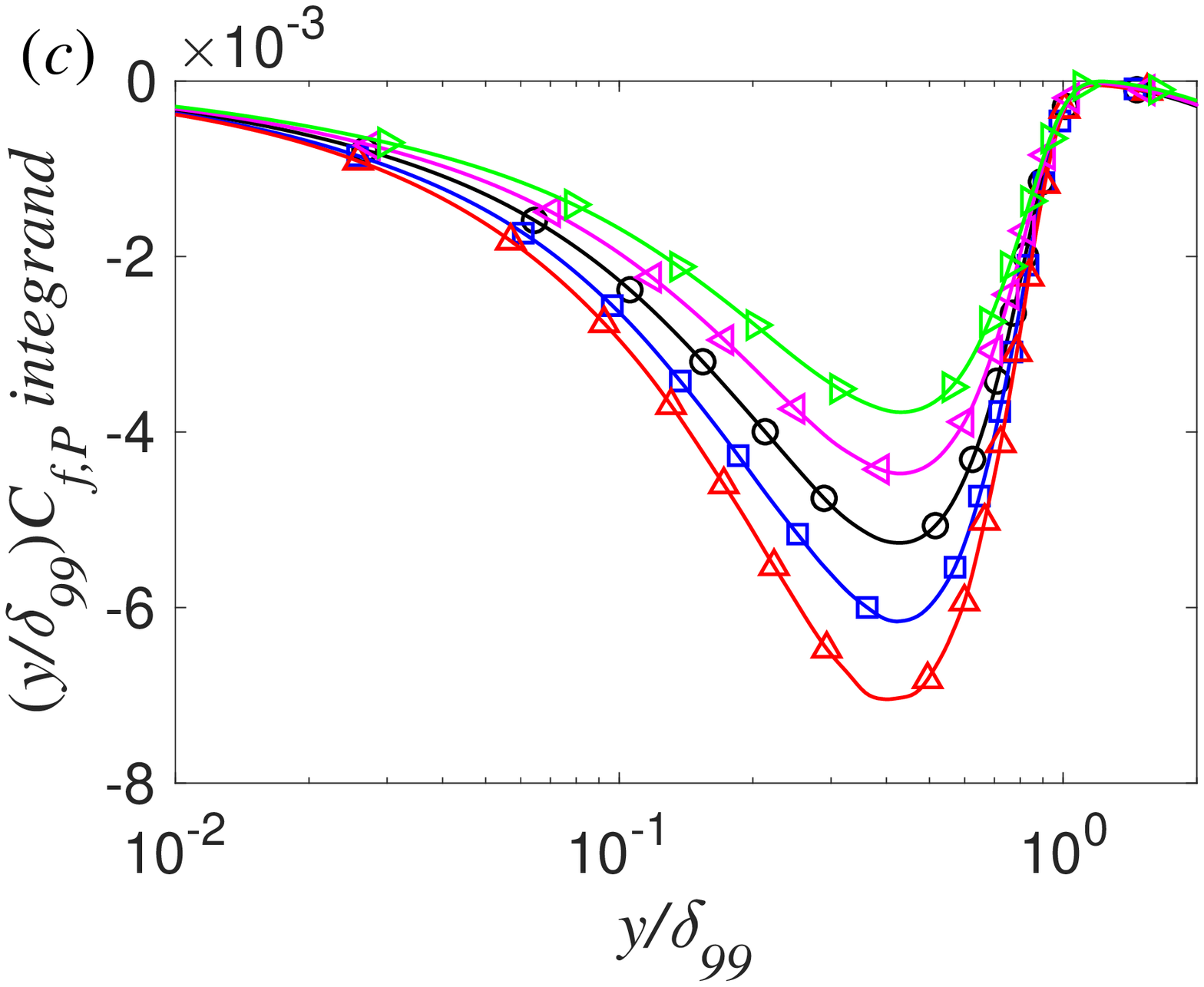}\label{topcf3d:c}}
\subfigure{\includegraphics[width = 5.5cm]{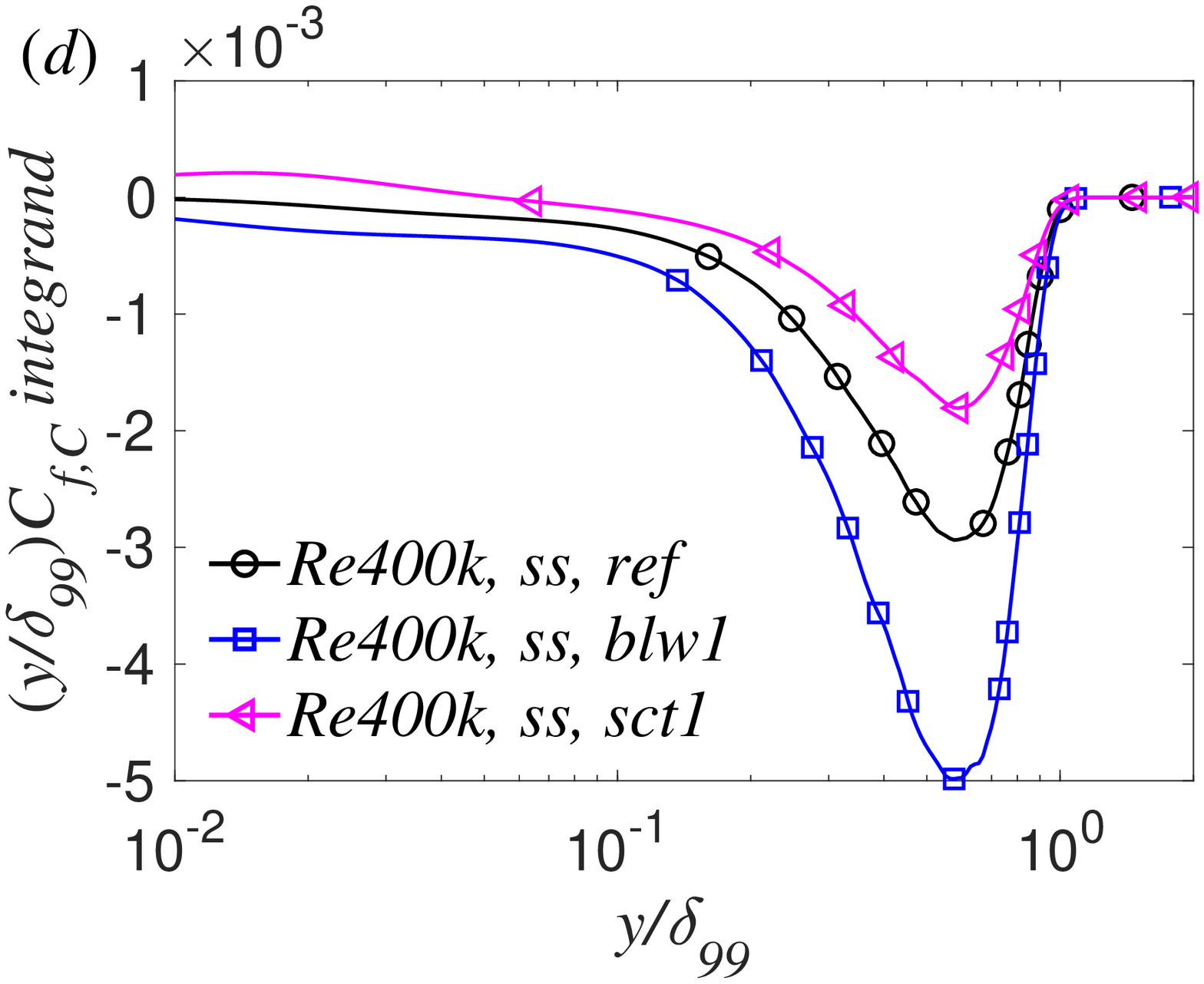}\label{topcf3d:d}}
\subfigure{\includegraphics[width = 5.5cm]{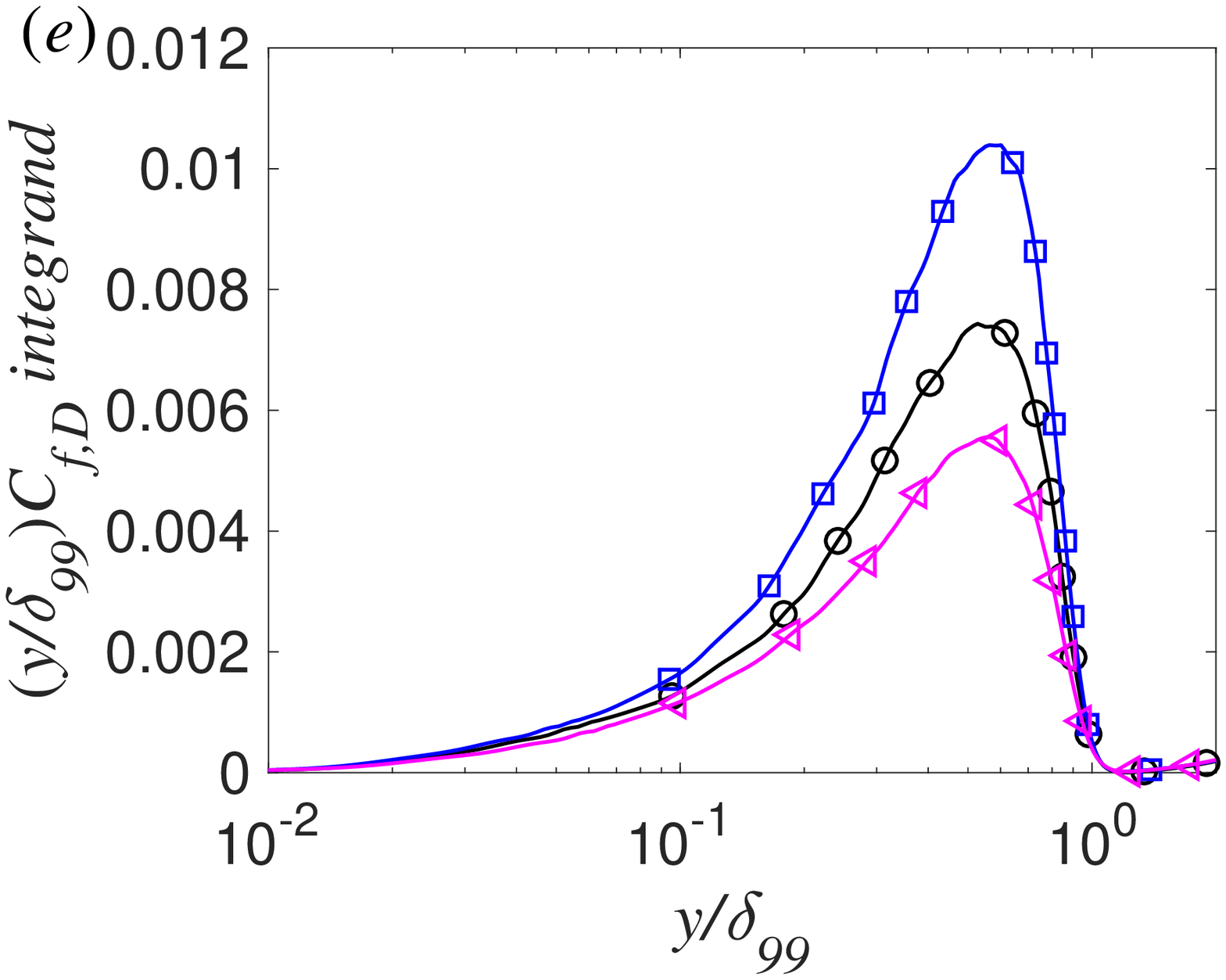}\label{topcf3d:e}}
\subfigure{\includegraphics[width = 5.5cm]{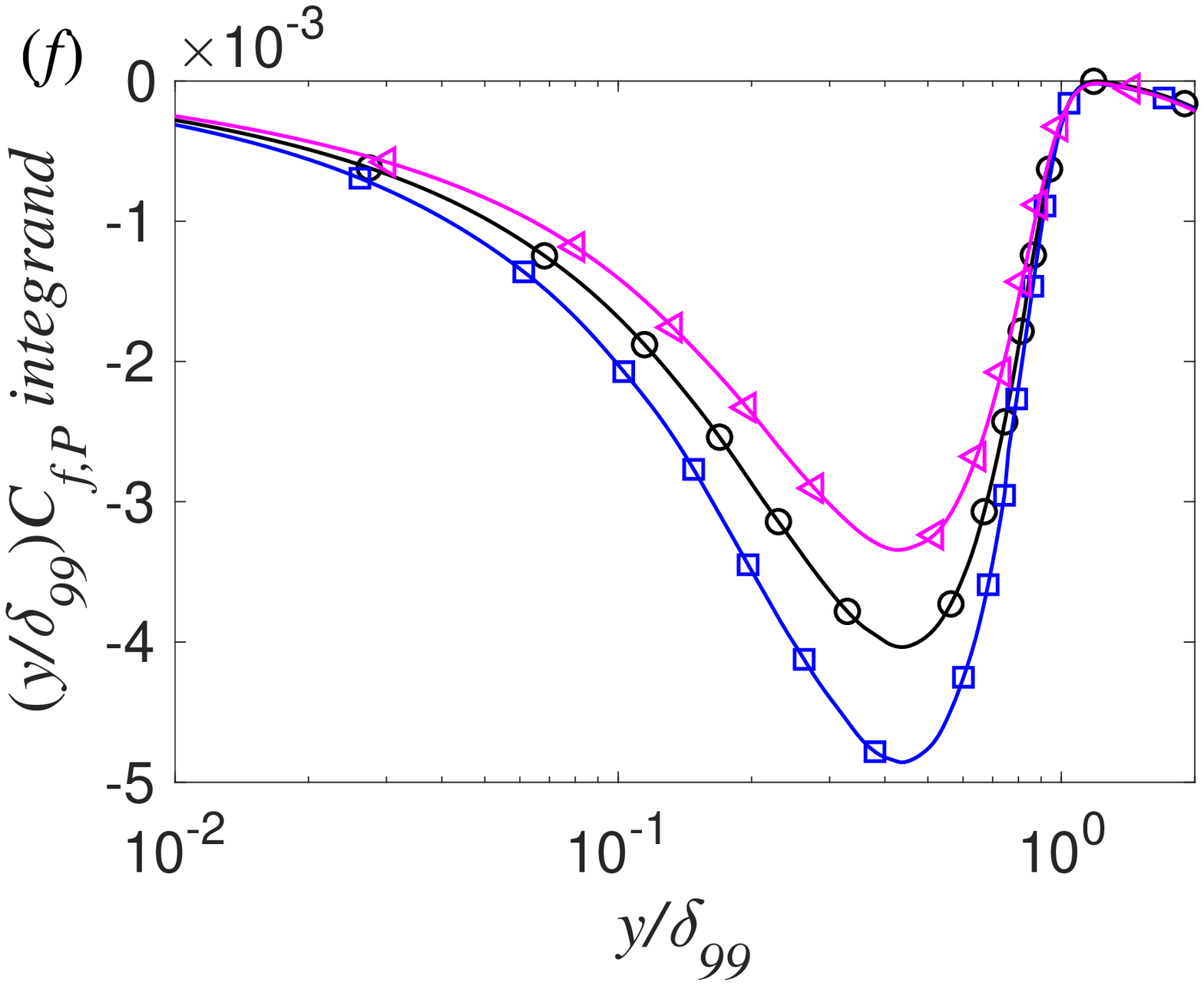}\label{topcf3d:f}}
\caption{Pre-multiplied integrands of ($a,d$) $C_{f,C}$, ($b,e$) $C_{f,D}$, and ($c,f$) $C_{f,P}$, as a function of $y/\delta_{99}$, on the suction side of a NACA4412 wing section at ($a-c$) $Re_c=200,000$ and ($d-f$) $Re_c=400,000$.}
\label{topcf3d}
\end{figure}

The generation of $C_{f,G}$ results from a counterbalance between the negative work done by $C_{f,C}$ and $C_{f,P}$ and the positive work by $C_{f,D}$. Figure \ref{topcf3d} quantifies their wall-normal distributions, and only the outer scaling by $\delta_{99}$ is applied herein. 
Good collapses of the peak locations are also observed at $y/\delta_{99}\approx 0.59$, $0.56$, and $ 0.43$ for $C_{f,C}-$, $C_{f,D}-$, and $C_{f,P}-$contributions, respectively.
In the blowing cases, the APG effects, a fact that promotes the population/energization of outer-layer structures \citep{Harun2013}, are strengthened. Meanwhile, the wall-normal convection and the streamwise boundary-layer growth are intensified in the outer region \citep{Vinuesa2018}.
Therefore, in an absolute sense, the generations of the components ($C_{f,C}$, $C_{f,D}$, and $C_{f,P}$) are all enhanced.
Thereinto, as the convection and pressure gradient do negative work for the friction-drag generation, blowing acts to reduce the $C_{f,C}-$ and $C_{f,P}-$contributions. On the other hand, the positive contribution of $C_{f,D}$ is  increased, as shown in figures \ref{topcf3d:b} and \ref{topcf3d:e}.
These three components counterbalance each other. The negative $C_{f,C}-$ and $C_{f,P}-$contributions are responsible for the $C_{f,G}$ reduction in the near-wall region, whereas the positive $C_{f,D}-$contributions is responsible for the increase of $C_{f,G}$ in the outer region, as seen in figures \ref{topcfd:c} and \ref{topcfd:f}.

\subsection{Contributions of small- and large-scale structures to the friction-drag generation}
Firstly, we use empirical mode decomposition (EMD) \citep{Huang1998} to identity the small- and large-scale turbulence structures.
EMD is an adaptive mode-decomposition technique, which extracts characteristic wavelengths of non-stationary signals automatically without {\it a-priori} basis functions. It has been applied for wall-bounded turbulence \citep{Agostini2014,Agostini2019,Cheng2019,doga19}, and details of the methodology of EMD can be found in \cite{Huang1998}. Here we just describe EMD very briefly. 

With EMD, a raw temporal or spatial signal $f(t)$ is decomposed into a sum of multiple intrinsic mode functions (IMFs) with a residual $R(t)$:
\begin{equation}
    f(t)=\sum_{i=1}^{m}{\rm IMF_i(t)}+R(t),
\end{equation}
where $m$ is the number of IMFs.  
The IMFs are data-driven functions, representing components with different wavelengths or scales in the full field.
In this study, the velocity fluctuations ($u'$ and $v'$) at the streamwise location $x/a\approx0.75$ are decomposed into four modes (three IMFs with a final residual). The first two modes represent the small-scale structures and the others characterize the large scales, which is justified based on a preliminary analysis (similar to the studies of \cite{Agostini2014,Agostini2019}).

\begin{figure}[h]
\centering
\subfigure{\includegraphics[width = 5.5cm]{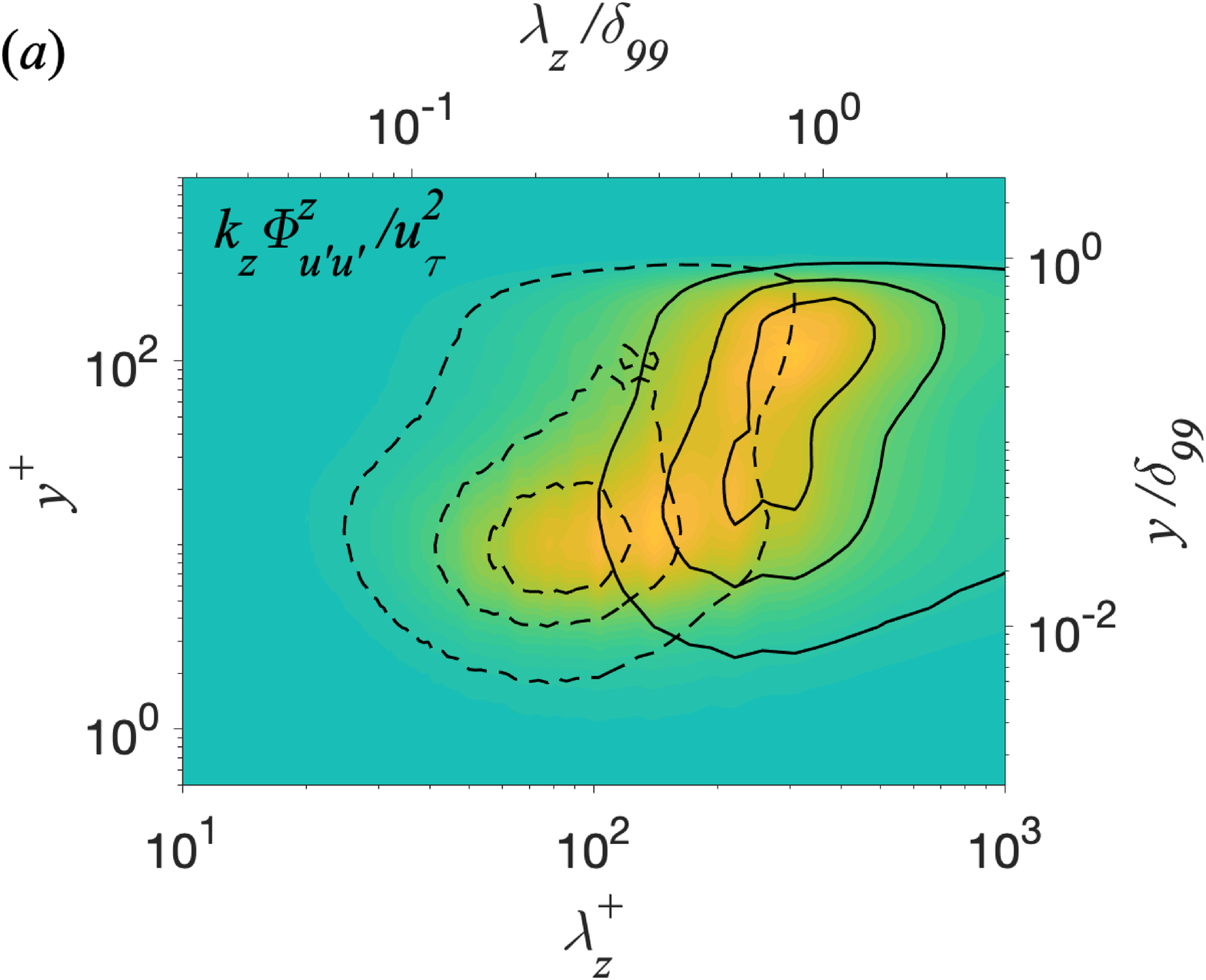}\label{top4:a}}
\subfigure{\includegraphics[width = 5.5cm]{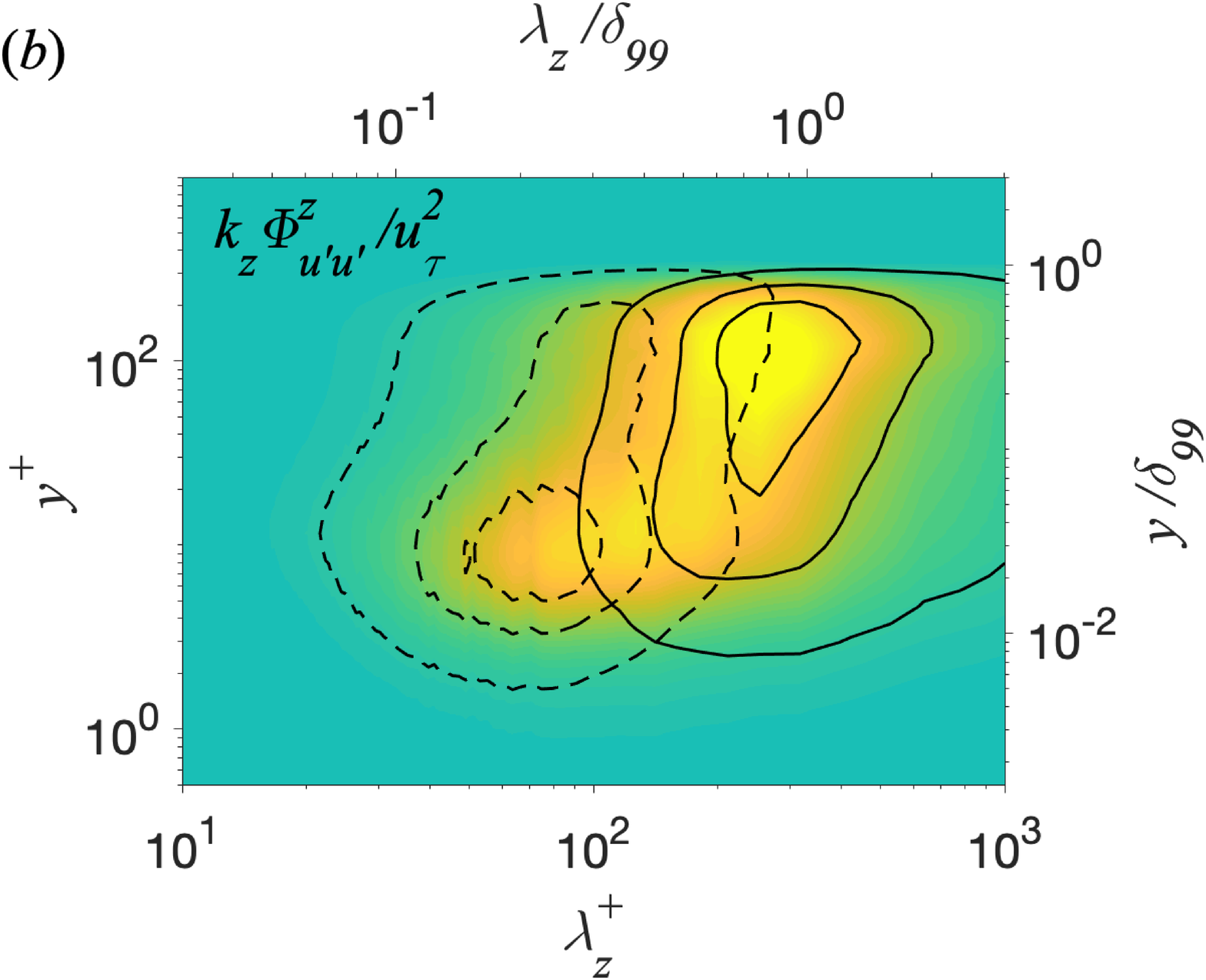}\label{top4:b}}
\subfigure{\includegraphics[width = 5.5cm]{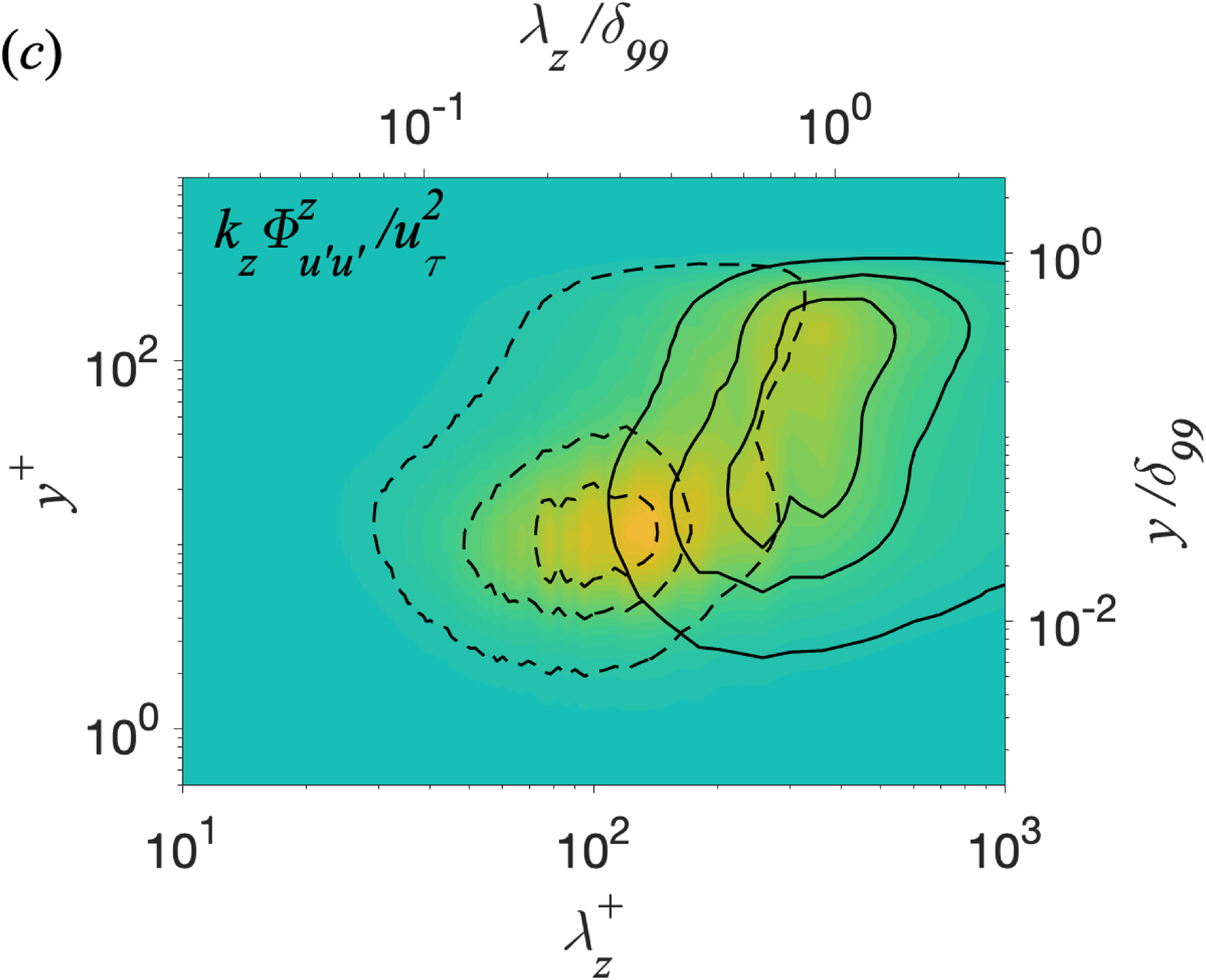}\label{top4:c}}
\subfigure{\includegraphics[width = 5.5cm]{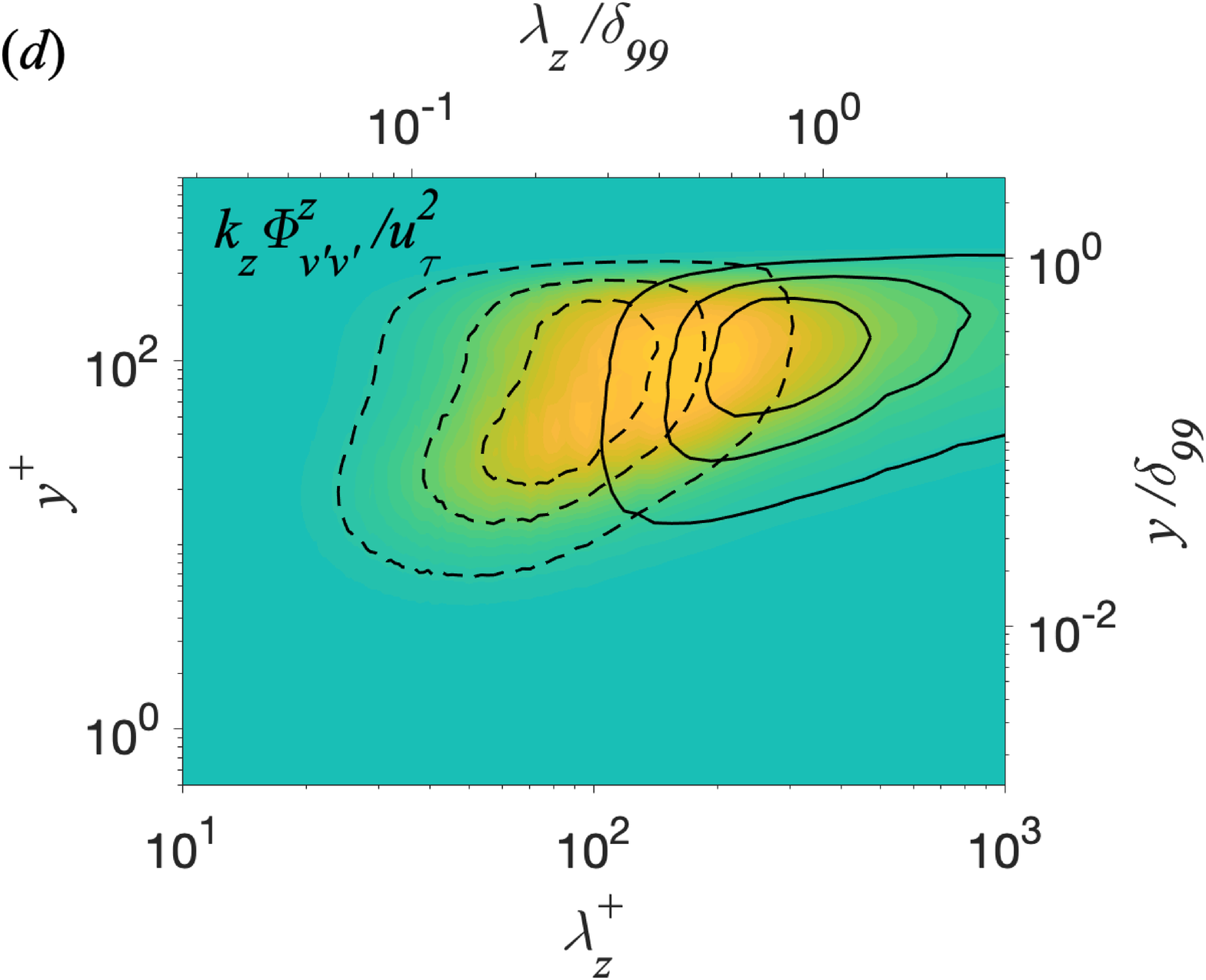}\label{top4:d}}
\subfigure{\includegraphics[width = 5.5cm]{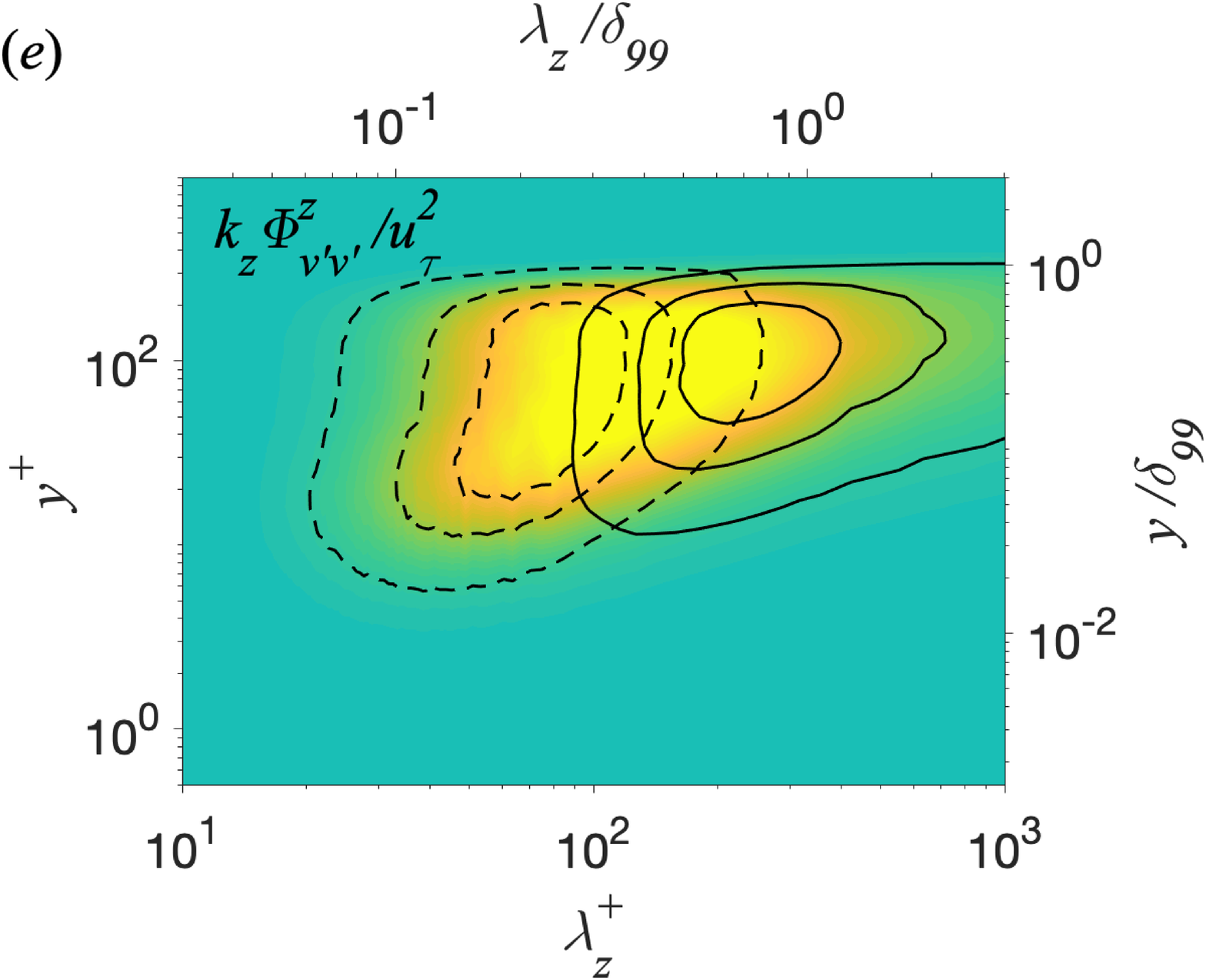}\label{top4:e}}
\subfigure{\includegraphics[width = 5.5cm]{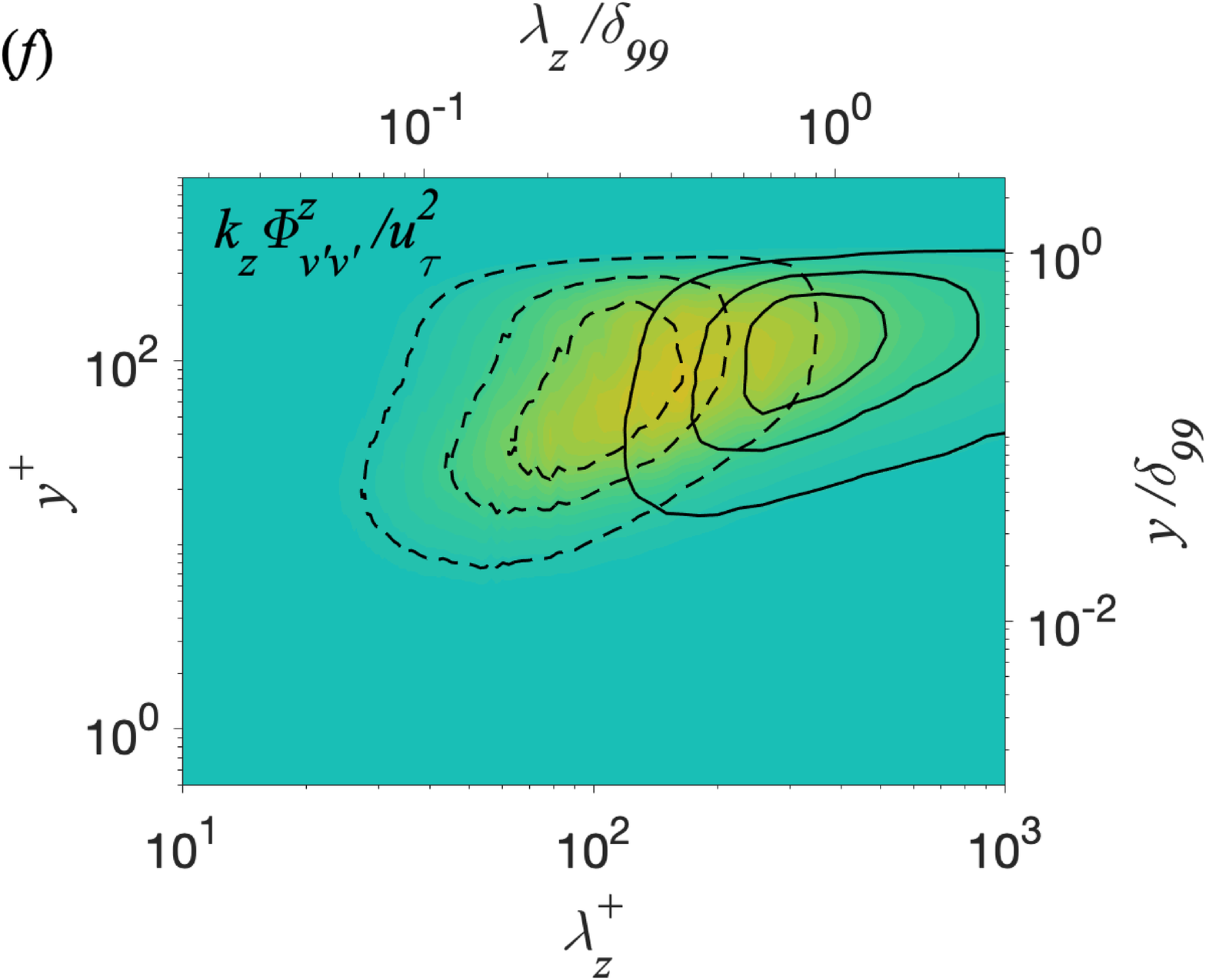}\label{top4:f}}
\caption{Spanwise pre-multiplied spectra of the tangential and normal velocity fluctuations on the suction side of a NACA4412 wing section at $Re_c=400,000$ ($a,c$) without control, with ($b,e$) uniform blowing and ($c,f$) uniform suction. The nephogram represents the spectra of the full field; the dashed and solid contour lines represent those of small and large scales, respectively.}
\label{top4}
\end{figure}

The contours of spanwise pre-multiplied spectra of $u'$ and $v'$ scaled with friction velocities are plotted in figure \ref{top4} for the $Re_c$=$400,000$ cases. Similar features can also be found in the low-Reynolds-number ($Re_c$=$200,000$) cases, which are not shown here for brevity.
In figure \ref{top4}, the iso-contour levels marked with dashed and solid lines represent the spectra of small- and large-scale velocity fluctuations, respectively. These contour lines indicate 0.12, 0.42, and 0.72 of their maxima, from outside to inside.
For the reference case, as shown in figure \ref{top4:a}, the spectra of small-scale $u'-$structures peak at the wall-normal distance $y^+\approx10$ with a spanwise wavelength $\lambda_z^+\approx 80$, and those of large-scale $u'-$structures peak at $y^+\approx100$ ($y/\delta_{99}\approx0.28$) with $\lambda_z^+\approx 310$ ($\lambda_z / \delta_{99}\approx 0.85$). This observation is consistent with the study of \cite{Cheng2019}, that small-scale $u'-$structures identified by EMD are representative of the near-wall coherent motions, whereas the large-scale $u'-$structures characterize large-scale motions.

When blowing/suction is applied on the airfoil surface, such energy spectra are affected, as seen in figures \ref{top4:b} and \ref{top4:c}. 
In the case of blowing, the small-scale structures are enhanced, and penetrate deeper into the outer region.
The peak of spectra locates at $(\lambda_z^+,y^+)\approx (80,8)$. 
Whereas suction has the opposite influence on the amplitude of small scales, with the peak location faintly influenced at $(\lambda_z^+,y^+)\approx (90,10)$.
As for the large-scale structures, they are enhanced by blowing, due to the energized large-scale motions, which agrees well with the conclusion drawn by \citet{Kametani2015}. The peak of large-scale spectra locates at $(\lambda_z/\delta_{99},y/\delta_{99})\approx (0.96,0.33)$. As shown in figure \ref{top4:c}, suction trends to diminish the secondary peak in the outer region with the peak at $(\lambda_z/\delta_{99},y/\delta_{99})\approx (0.93,0.35)$.

As for the $v'-$structures shown in figures \ref{top4:d}--\ref{top4:f}, the small- and large-scale structures have approximately the same spanwise wavelengths as the $u'-$structures, which is consistent with the EMD results of channel flows \citep{Cheng2019}, suggesting that the decomposed two scales of $u'-$ and $v'-$structures can be characterized with the same spanwise wavelength. Both the small- and large-scale $v'-$structures are greatly enhanced by blowing while reduced by suction.
On the other hand, the wall-normal locations of the $v'-$structures are much different from those of $u'-$structures, because that the 
presence of wall prevents the normal velocity fluctuations from extending close to the near-wall region, in contrast to the wall-parallel component $u'$. 
As shown in figures \ref{top4:d}--\ref{top4:f}, the small-scale spectra of $v'$ are more intense beyond the buffer layer at $y^+\approx50$, with the spanwise wavelength scale $\lambda_z^+$ peaking at around $80$--$100$ for all of the three cases, and the large-scale $v'-$spectra peak around $(\lambda_z/\delta_{99},y/\delta_{99})\approx (0.9,0.35)$. 

Hereafter, we denote the small- and large-scale tangential and wall-normal velocity fluctuations as $u_s'$, $u_l'$, $v_s'$, and $v_l'$, respectively. Then the Reynolds stress is decomposed as:
\begin{equation}
-\left<u'v'\right>=-\left<u_s'v_s'\right>-\left<u_s'v_l'\right>-\left<u_l'v_s'\right>-\left<u_l'v_l'\right>,
\end{equation}
where $-\left<u_s'v_s'\right>$ represents the Reynolds stress carried by small-scale structures, $-\left<u_l'v_l'\right>$ represents the Reynolds stress associated with large-scale structures, and $-\left<u_s'v_l'\right>$ and $-\left<u_l'v_s'\right>$ denote the scale interactions from small- to large- and from large- to small-scale structures, respectively. 
Substituting these Reynolds-stress components into equation \eqref{cf2_cf}, the term of TKE production ($C_{f,T}$) can be further divided into four parts, \textit{viz.}
\begin{eqnarray}
    {C_{f,T,ss}} &=& 2/U_e^{+3}\int_{0}^{\infty}\left<-u_s'v_s'\right>^+\frac{\partial \left<u\right>^+}{\partial y^+}{\rm d}y^+,\label{cfss}\\
    {C_{f,T,sl}} &=& 2/U_e^{+3}\int_{0}^{\infty} \left<-u_s'v_l'\right>^+\frac{\partial \left<u\right>^+}{\partial y^+}{\rm d}y^+,\label{cfsl}\\
    {C_{f,T,ls}} &=& 2/U_e^{+3}\int_{0}^{\infty} \left<-u_l'v_s'\right>^+\frac{\partial \left<u\right>^+}{\partial y^+}{\rm d}y^+,\label{cfls}\\
    {C_{f,T,ll}} &=& 2/U_e^{+3}\int_{0}^{\infty} \left<-u_l'v_l'\right>^+\frac{\partial \left<u\right>^+}{\partial y^+}{\rm d}y^+.\label{cfll}
\end{eqnarray}

\begin{figure}[h]
\centering
\subfigure{\includegraphics[width = 5.5cm]{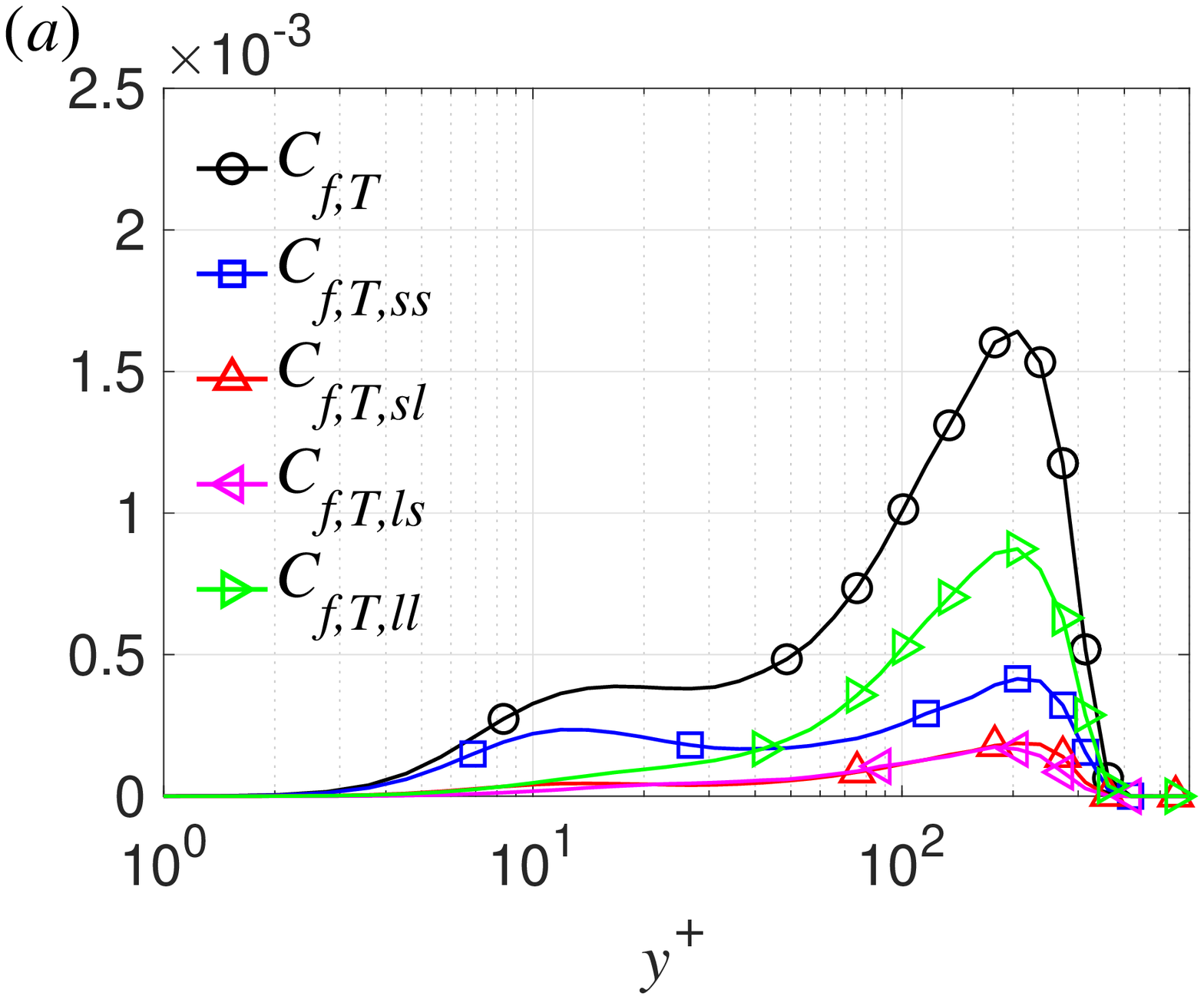}\label{top4dist:a}}
\subfigure{\includegraphics[width = 5.5cm]{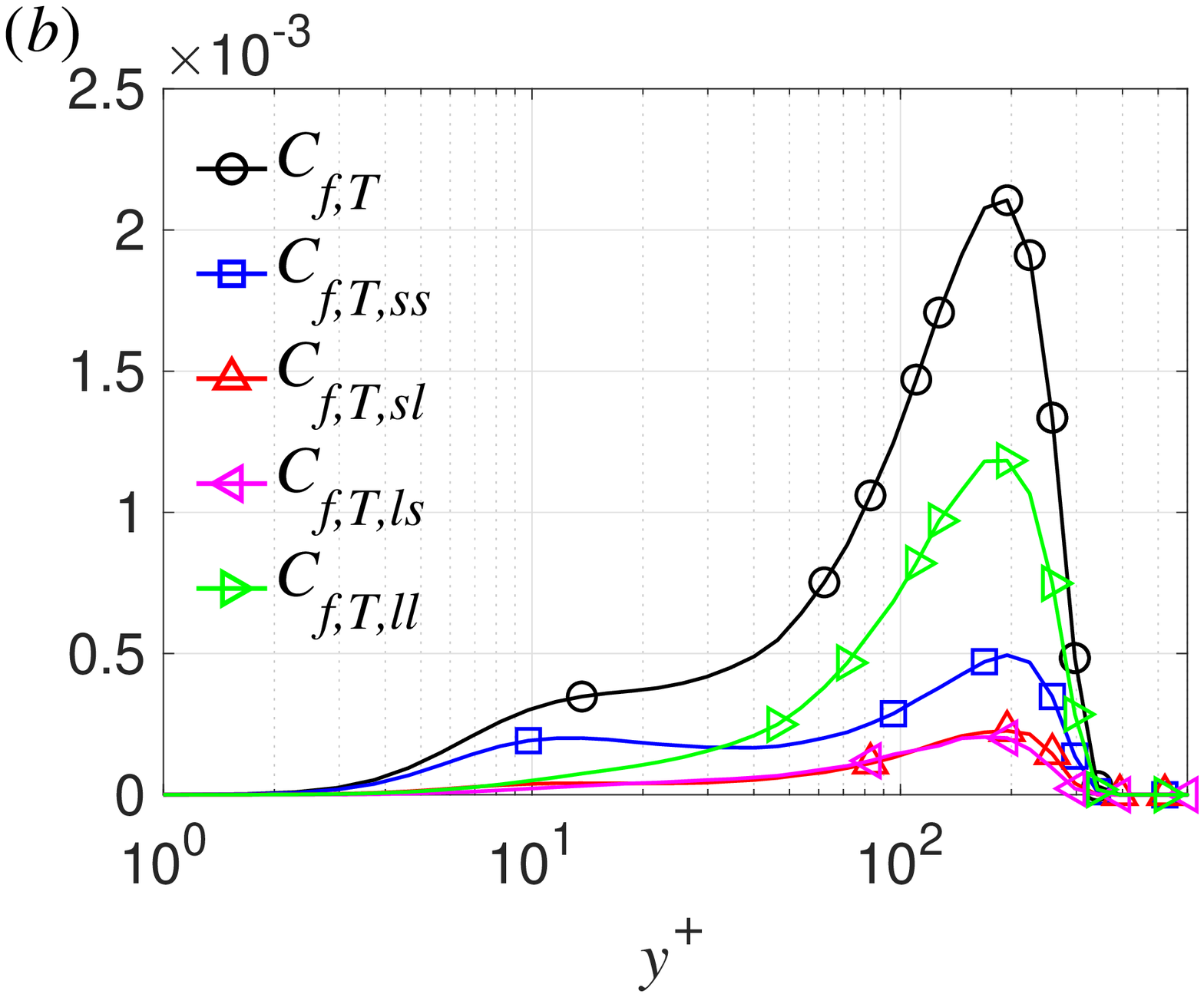}\label{top4dist:b}}
\subfigure{\includegraphics[width = 5.5cm]{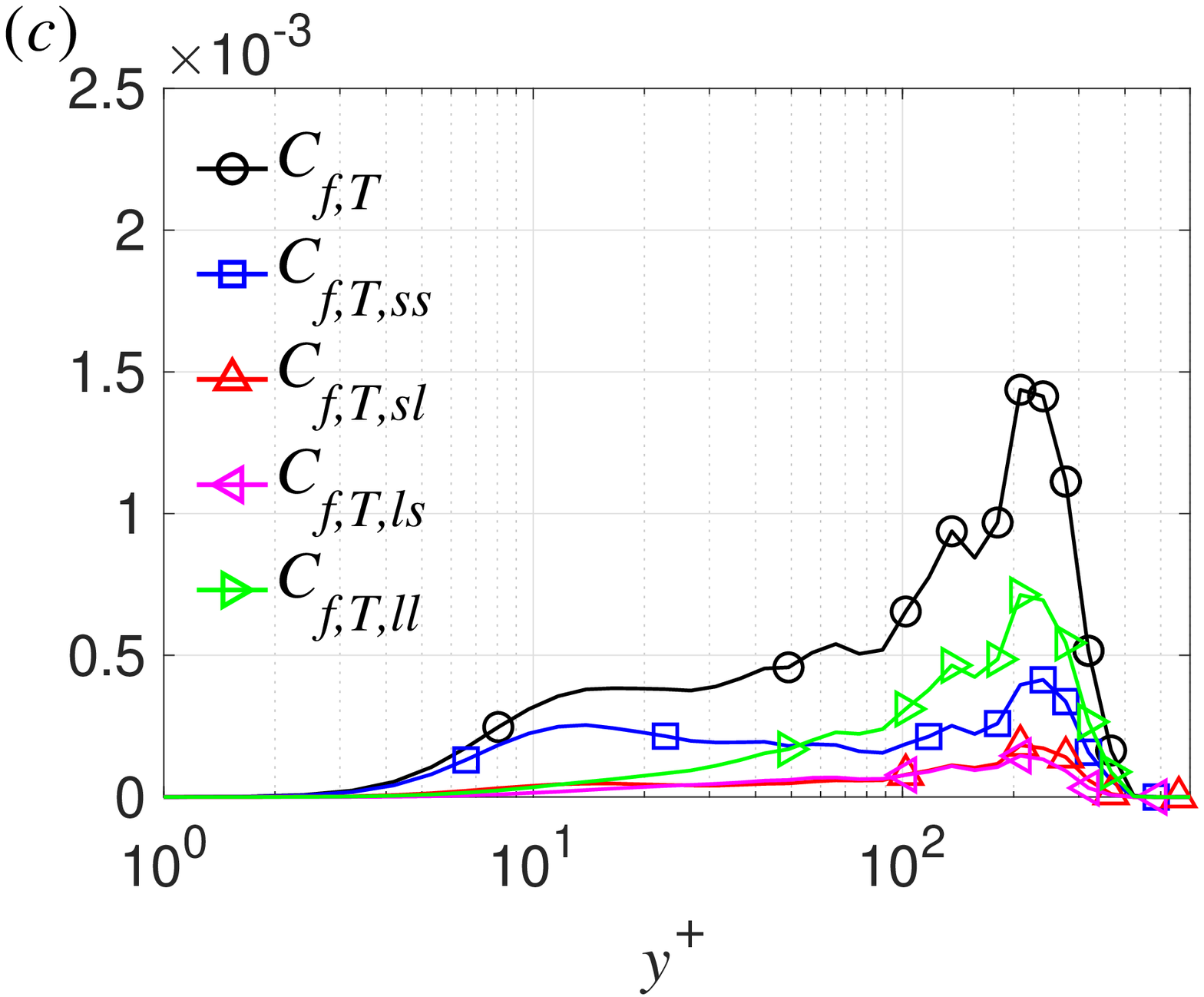}\label{top4dist:c}}
\caption{Pre-multiplied integrands of $C_{f,T,i}$ as a function of $y^+$ on the suction side of a NACA4412 wing section at $Re_c=400,000$ ($a$) without control, with ($b$) uniform blowing and ($c$) uniform suction. Note that the premultiplication factor is $y^+$.}
\label{top4dist}
\end{figure}

\begin{figure}[h]
\centering
\subfigure{\includegraphics[width = 5.5cm]{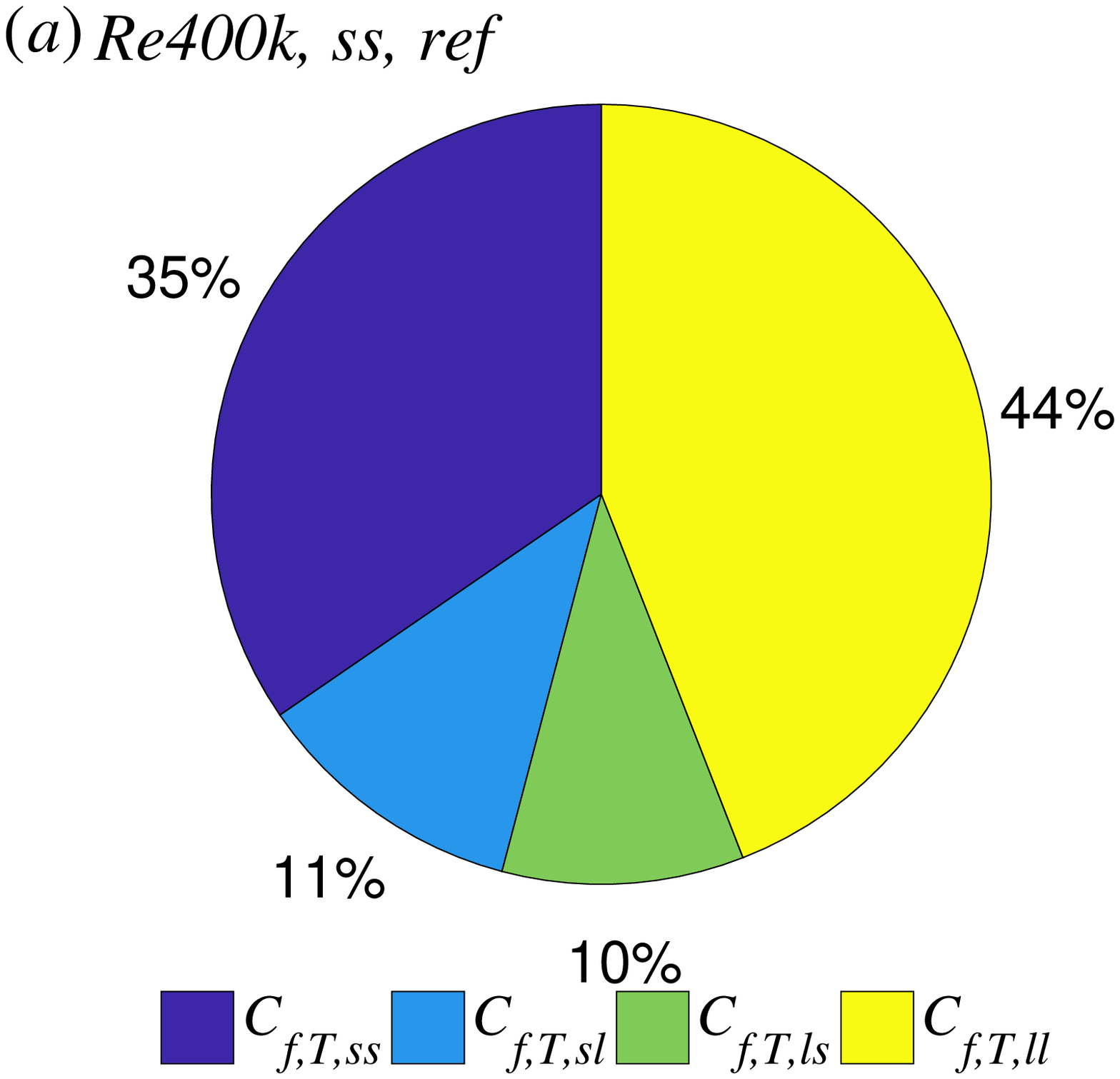}\label{top4cont:a}}
\subfigure{\includegraphics[width = 5.5cm]{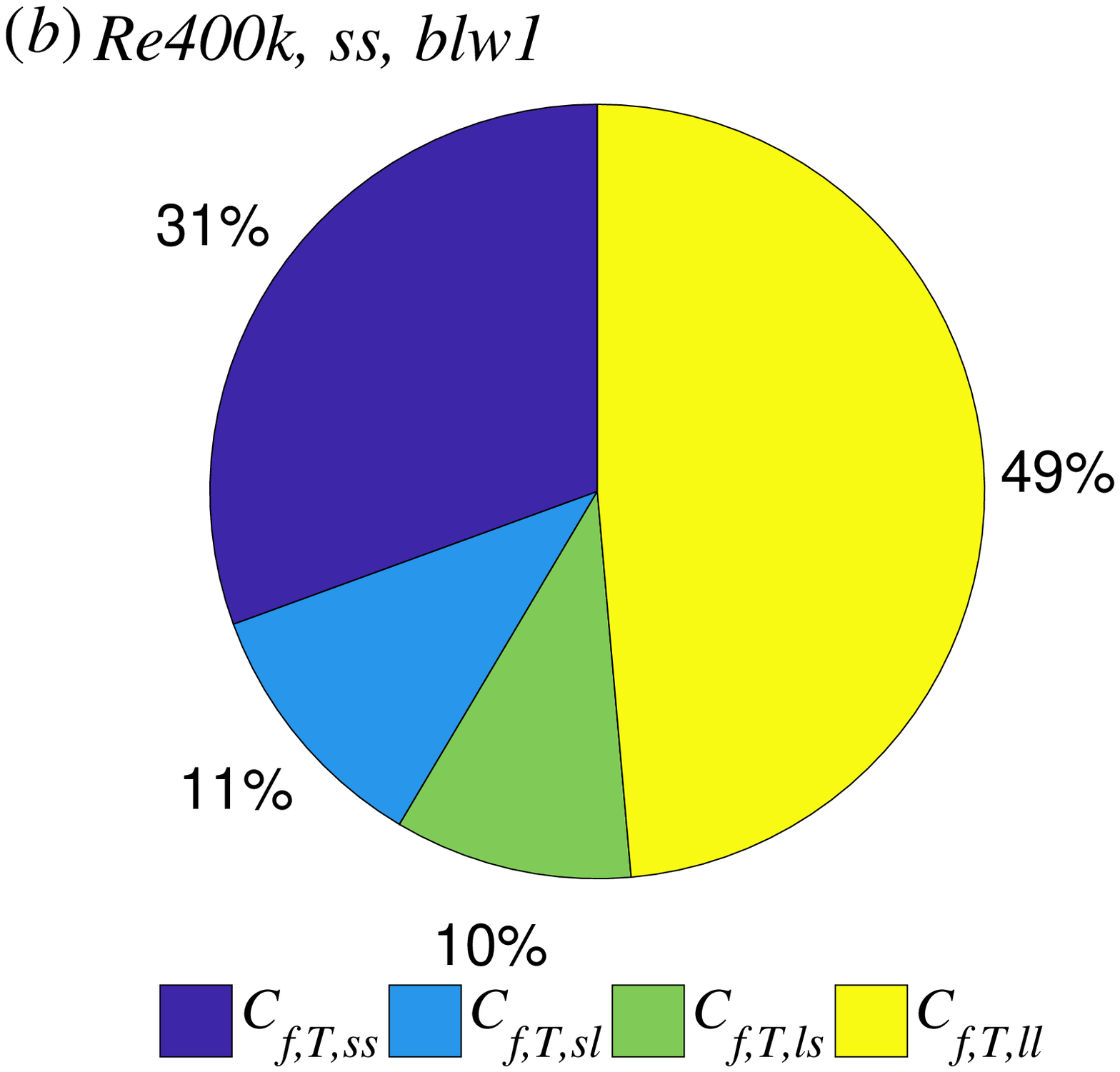}\label{top4cont:b}}
\subfigure{\includegraphics[width = 5.5cm]{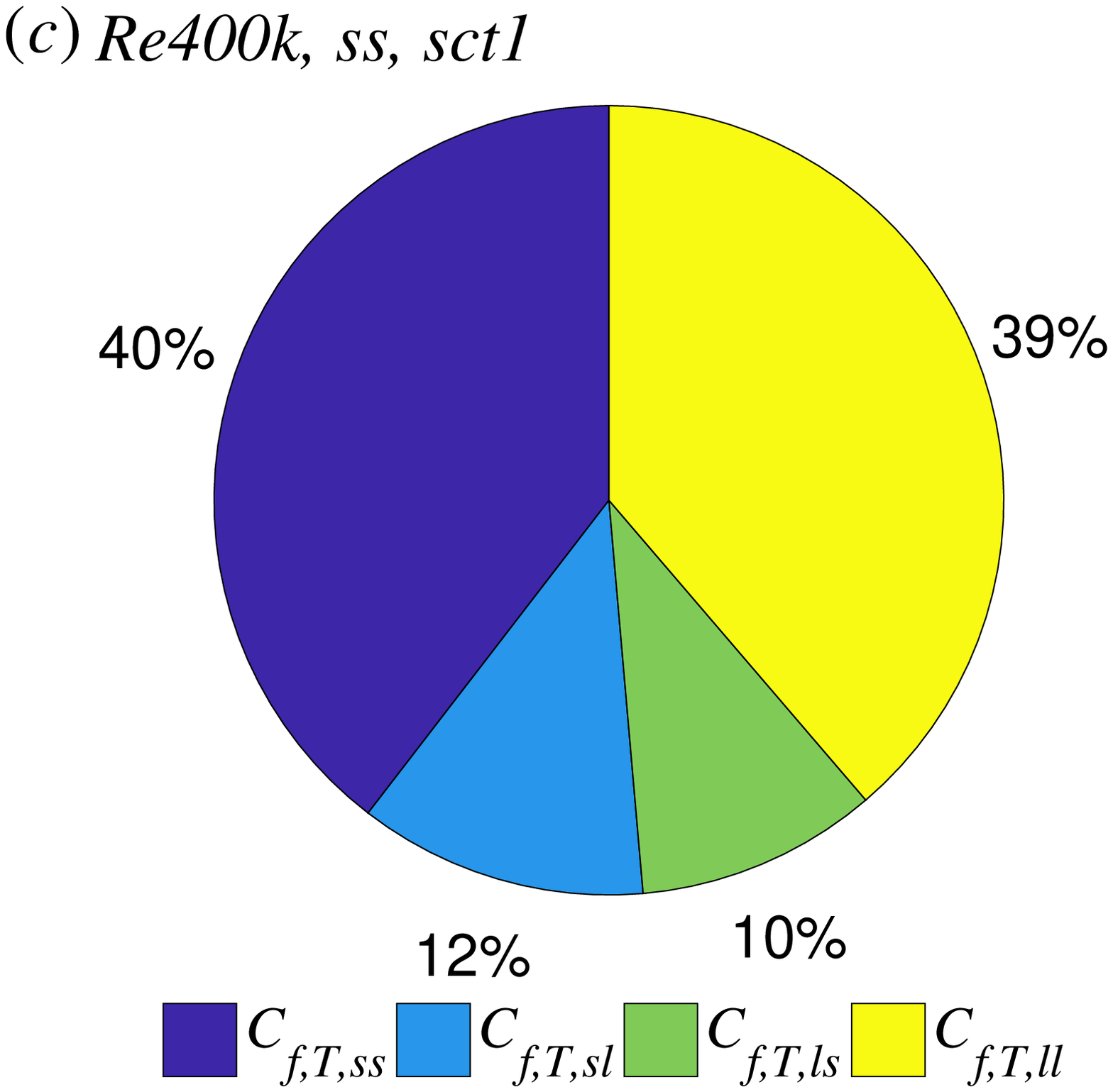}\label{top4cont:c}}
\caption{Scale-specific contributions ($C_{f,T,ss}$, $C_{f,T,sl}$, $C_{f,T,ls}$, and $C_{f,T,ll}$ normalized by $C_{f,T}$ itself) on the suction side of the NACA4412 wing section at $Re_c=400,000$ ($a$) without control, with ($b$) uniform blowing and ($c$) uniform suction.}
\label{top4cont}
\end{figure}

Similar to the analysis in Sec. \ref{sec:2}, we show the wall-normal distributions of the $y^+$-pre-multiplied integrands in equations \eqref{cfss}--\eqref{cfll} in figure \ref{top4dist}. For the reference case without control, the small-scale motions play important roles both in the inner and outer region, as two comparable peaks are respectively observed there, as seen in figure \ref{top4dist:a}. This is different from the phenomenon in channel flows \citep{Agostini2019} and ZPG-TBLs, 
since the APG strengthens the energy of small-scale structures in the outer region \citep{Tanarro2020}.
Whereas the large-scale structures dominate the production of $C_{f,T}$ in the outer region, which is associated with the enhanced generation of large-scale motions in APG-TBLs \citep{Lee2008,Harun2013,Vinuesa2018}.
When blowing is applied, the distributions of $C_{f,T,ss}-$contribution and $C_{f,T,ll}-$contribution in the outer region are shifted upwards, suggesting that the wall-normal mass flux enhances the outer fluctuations of both small- and large-scale structures. Comparison between $C_{f,T,ss}$ and $C_{f,T,ll}$ indicates that the blowing raises the relative importance of large scales with respect to the small ones.
In addition, the scale interactions ($C_{f,T,sl}$ and $C_{f,T,ls}$) mainly exhibited in the outer region are not very sensitive to the control schemes.

In order to further check the role of small- and large-scale motions in the generation of turbulence-kinetic-energy production, figure \ref{top4cont} quantifies their integrations normalized by $C_{f,T}$ itself. We can find that the blowing and suction have opposite influences on the contributions of small- and large-scale motions, that is, blowing is able to enhance the contribution of large-scales and to suppress the contribution of small-scales, whereas suction behaves contrarily. As for the scale-interactions, \textit{i.e.} $C_{f,T,sl}$ and $C_{f,T,ls}$, they account for approximately 20\% of the total $C_{f,T}$ and remain almost unchanged with different control strategies.

\section{Friction-drag decomposition on the pressure side}\label{pressure}

In this section, we pay attention to the pressure side of a NACA4412 wing section, where the TBLs are subjected to favorable pressure gradients.
Figure \ref{botcf} shows the skin-friction coefficients on the pressure side of the airfoil. In spite of the favorable pressure gradients, blowing is still able to reduce the friction drag. Whereas, the skin-friction coefficients are no longer decreased monotonously with regard to $x/c$ if compared with those on the suction side of the airfoil, which may be caused by the coupling influences of $Re_\tau$ and FPGs \citep{Atzori2020}. 
The relative errors $([C_{f,V}+C_{f,T}+C_{f,G}-C_f]/C_f)$ of the friction-drag decomposition are limited within $\pm0.03\%$.

Key points of the decomposition results on the pressure side are listed below:

\begin{figure}[h]
\centering
\includegraphics[width = 5.5cm]{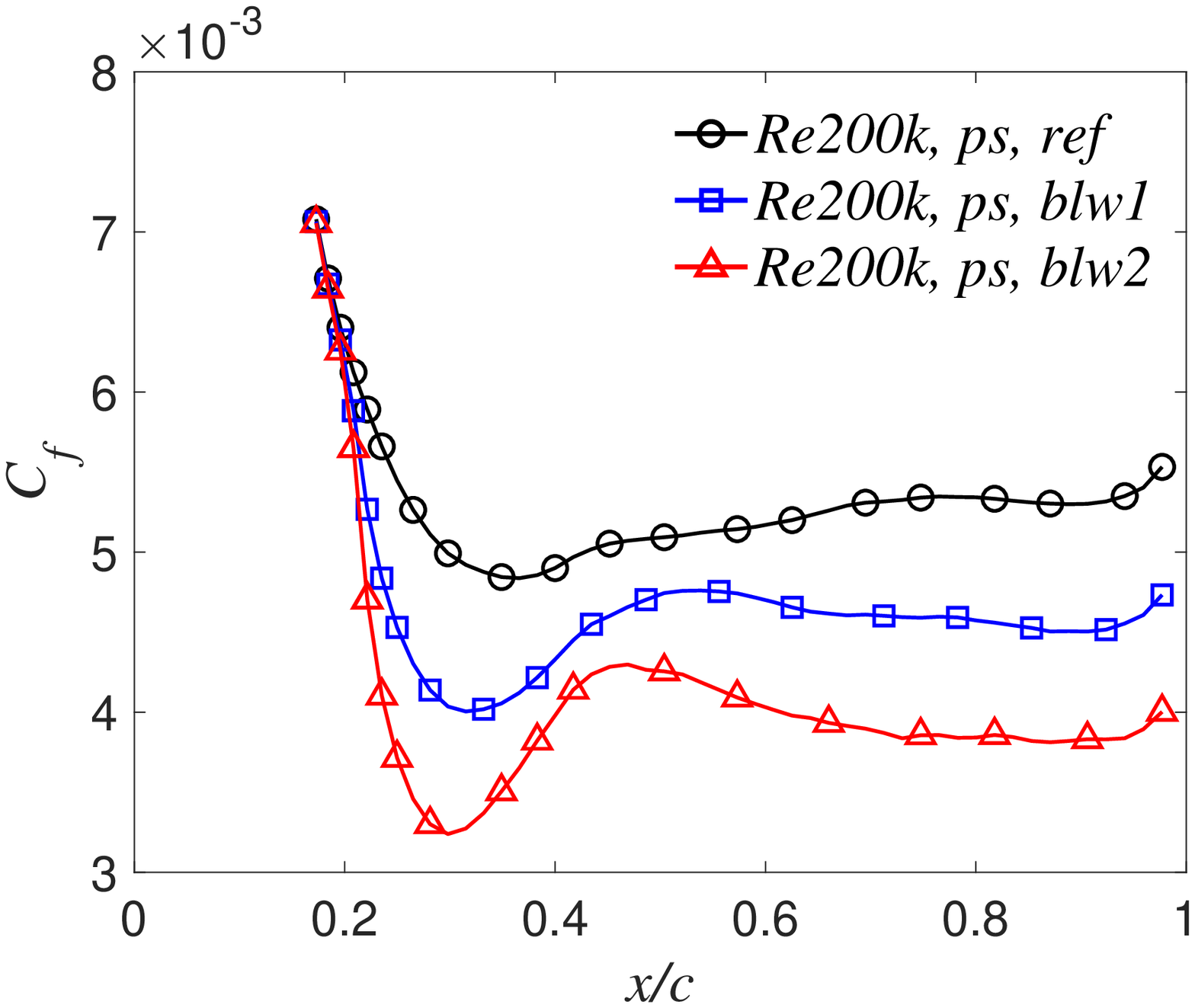}
\caption{Variation of skin-friction coefficients on the pressure side of the NACA4412 wing section at $Re_c=200,000$. }
\label{botcf}
\end{figure}

\begin{figure}[h]
\centering
\subfigure{\includegraphics[width = 5.5cm]{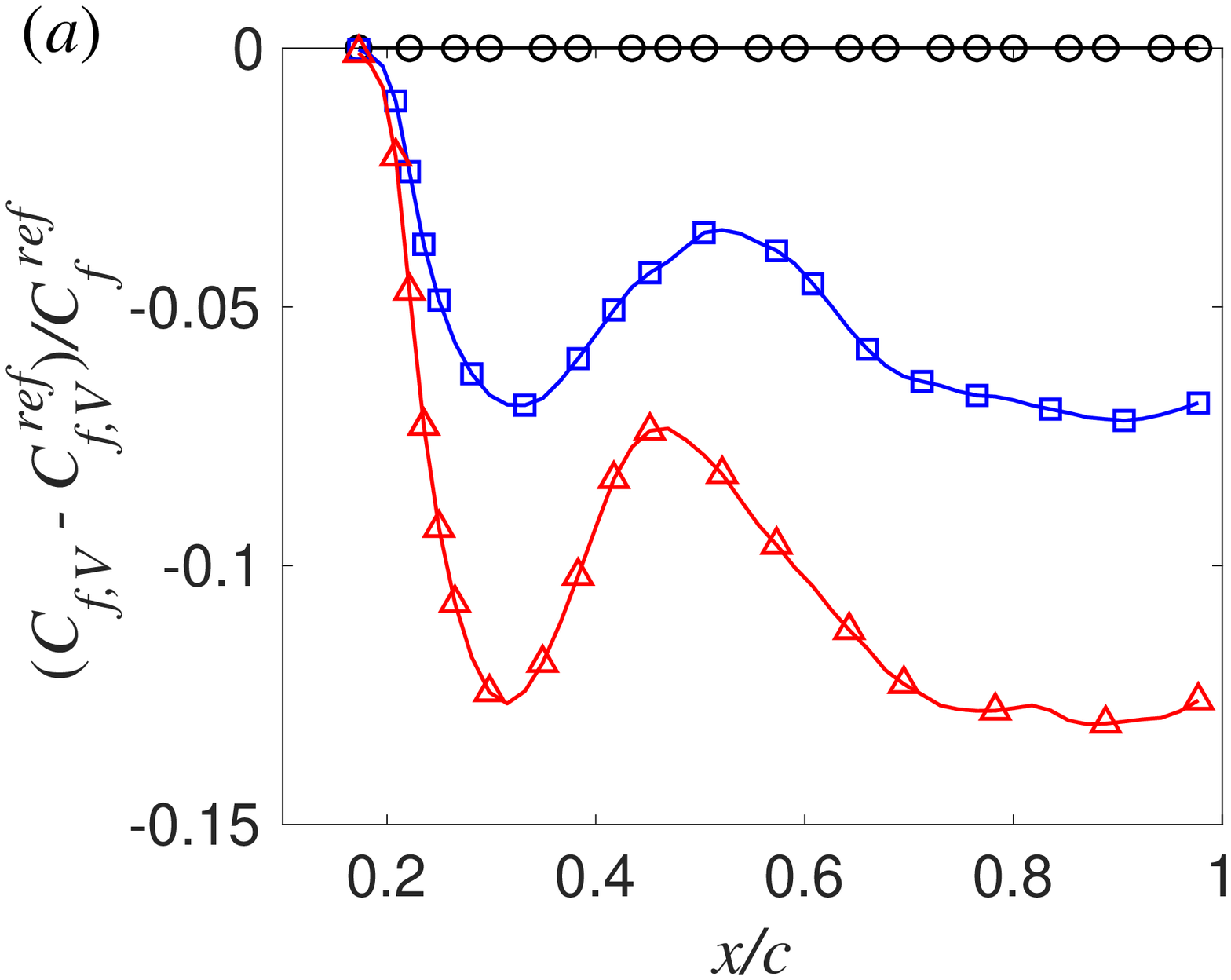}\label{bot2cf:a}}
\subfigure{\includegraphics[width = 5.5cm]{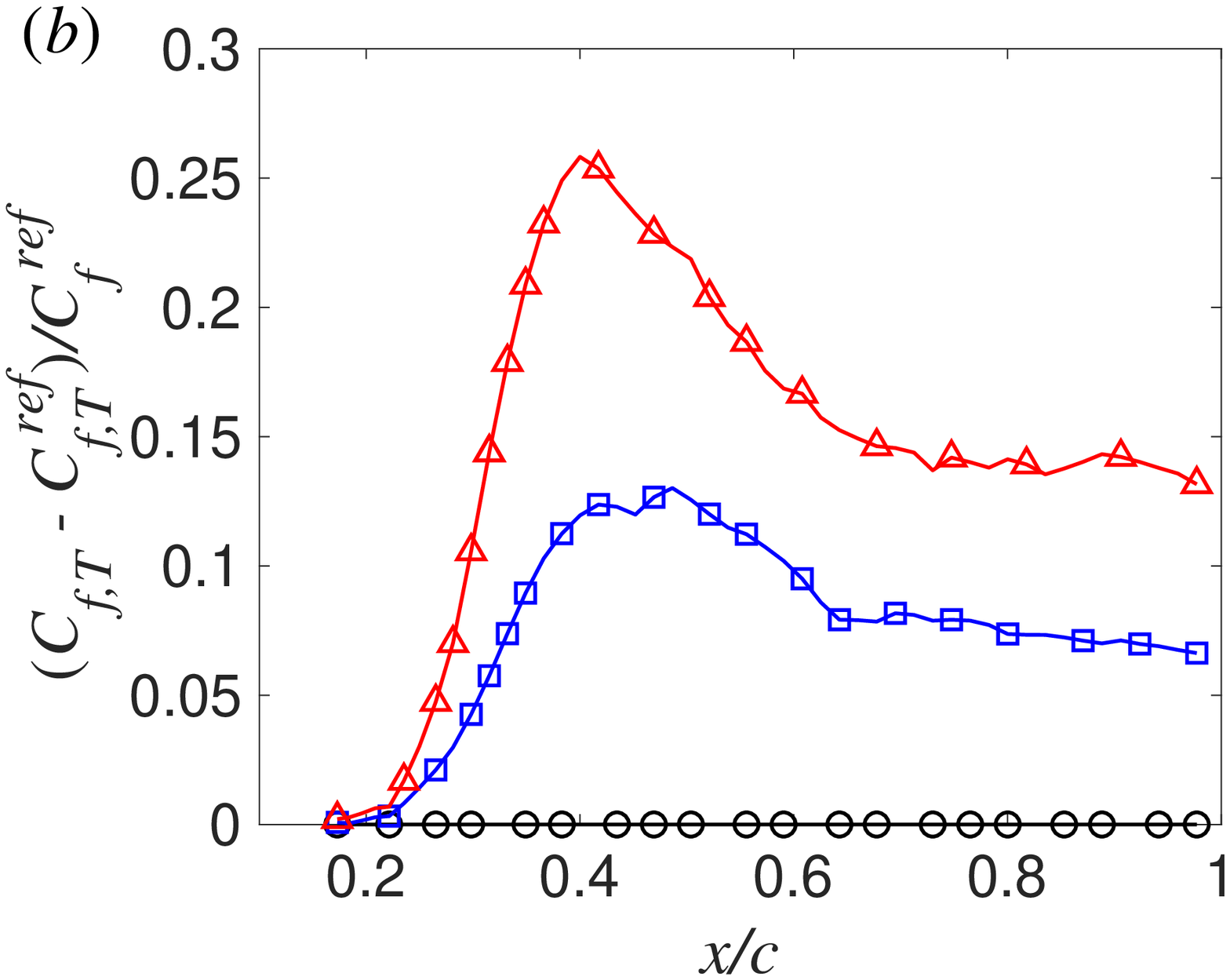}\label{bot2cf:b}}
\subfigure{\includegraphics[width = 5.5cm]{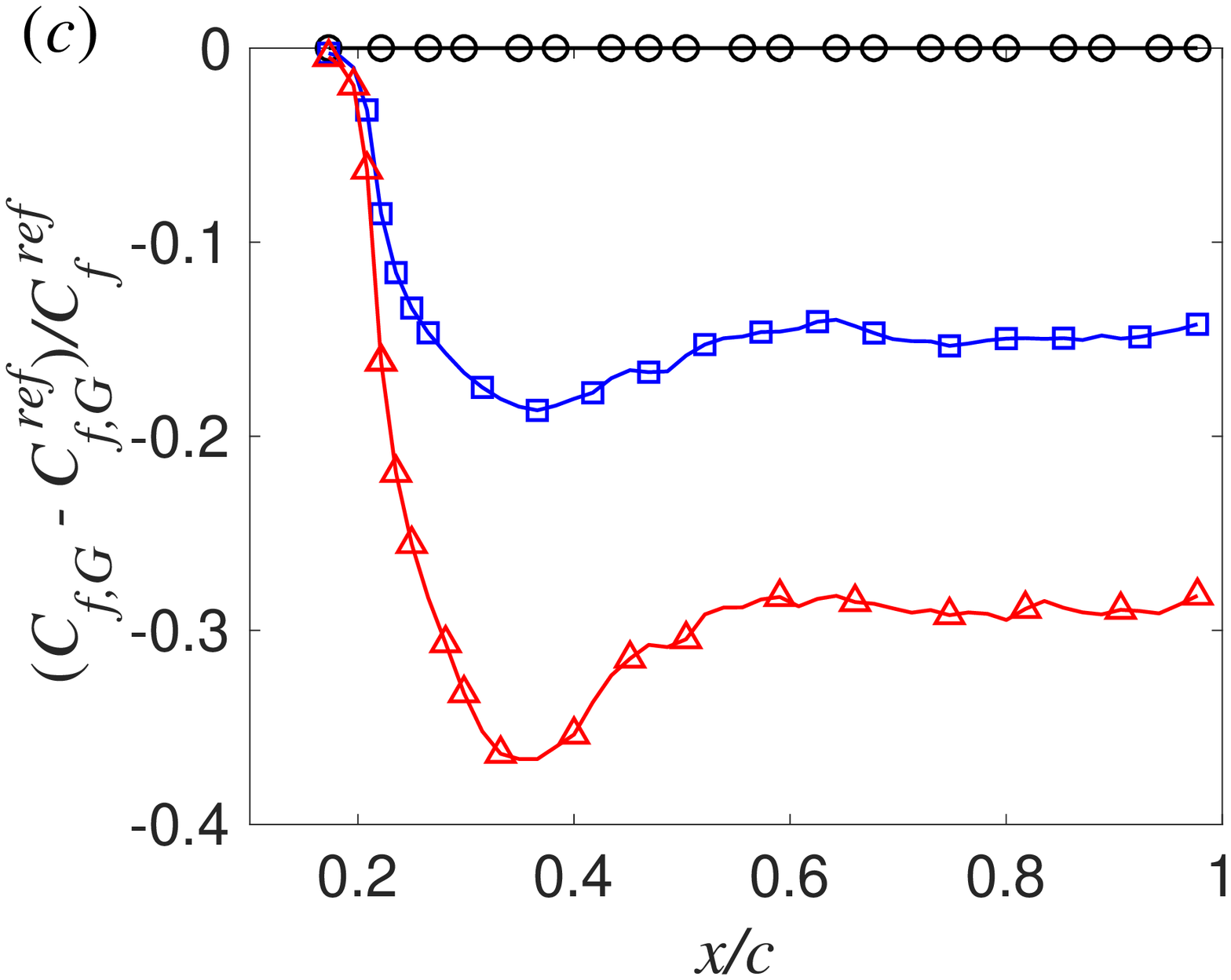}\label{bot2cf:c}}
\subfigure{\includegraphics[width = 5.5cm]{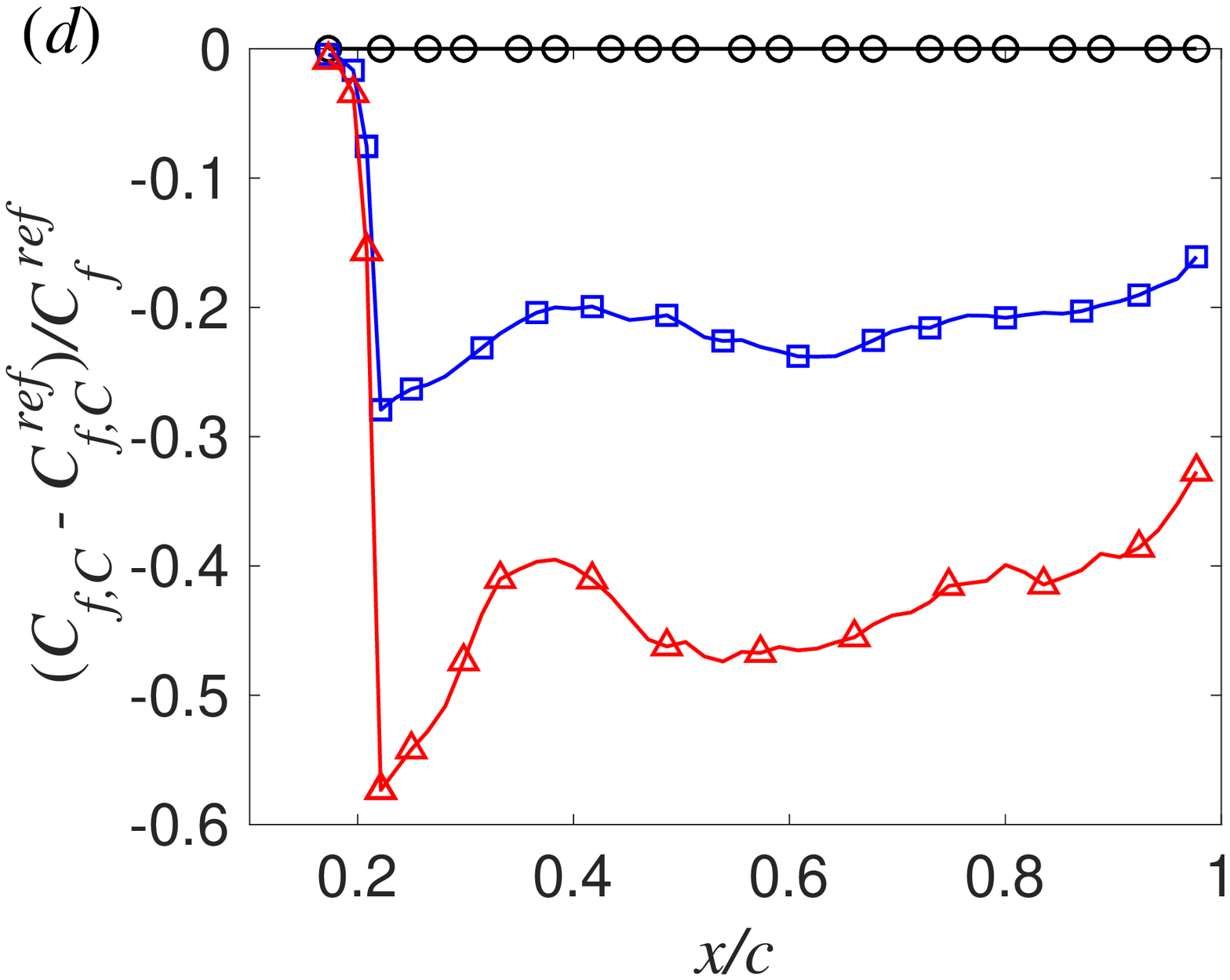}\label{bot2cf:d}}
\subfigure{\includegraphics[width = 5.5cm]{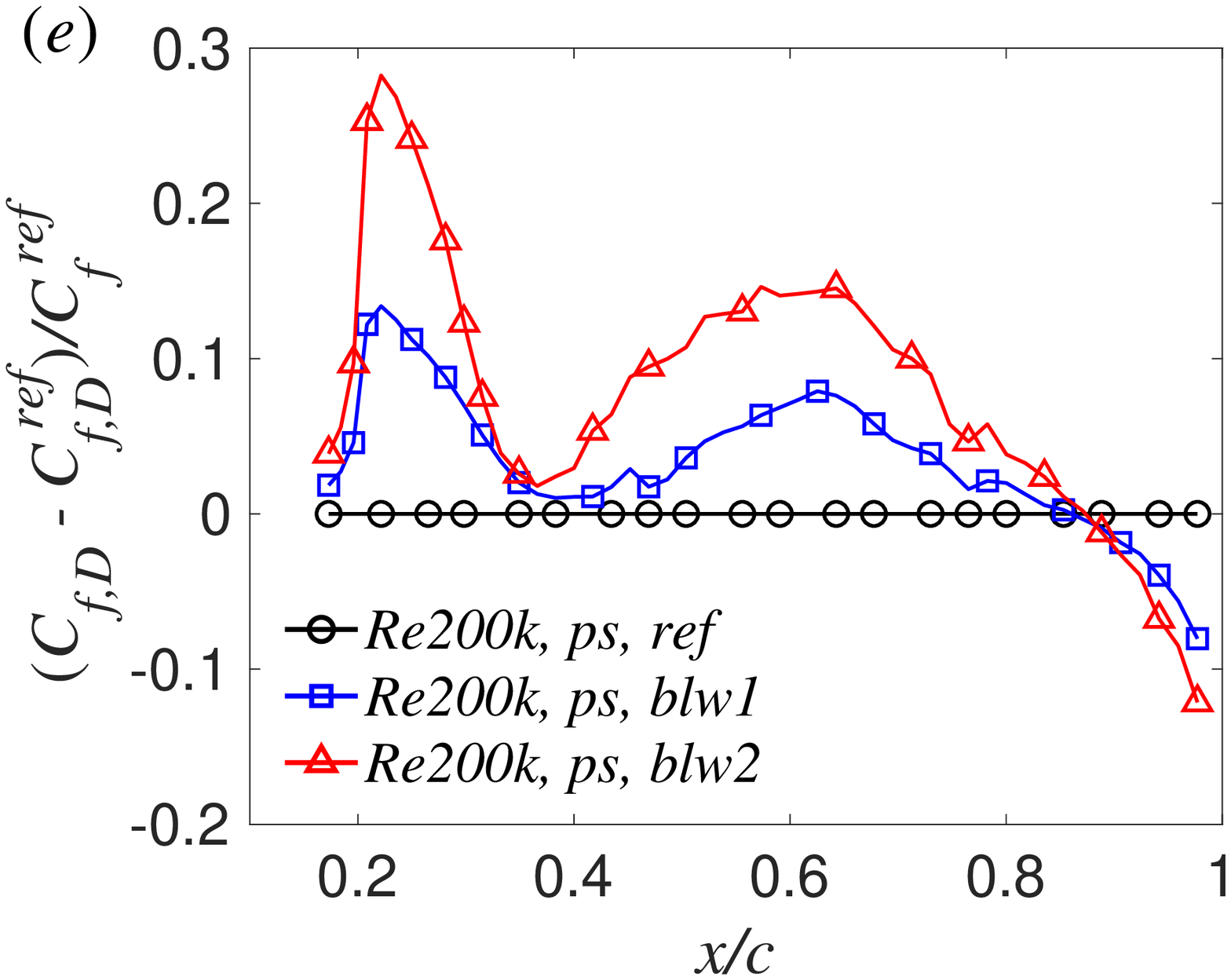}\label{bot2cf:e}}
\subfigure{\includegraphics[width = 5.5cm]{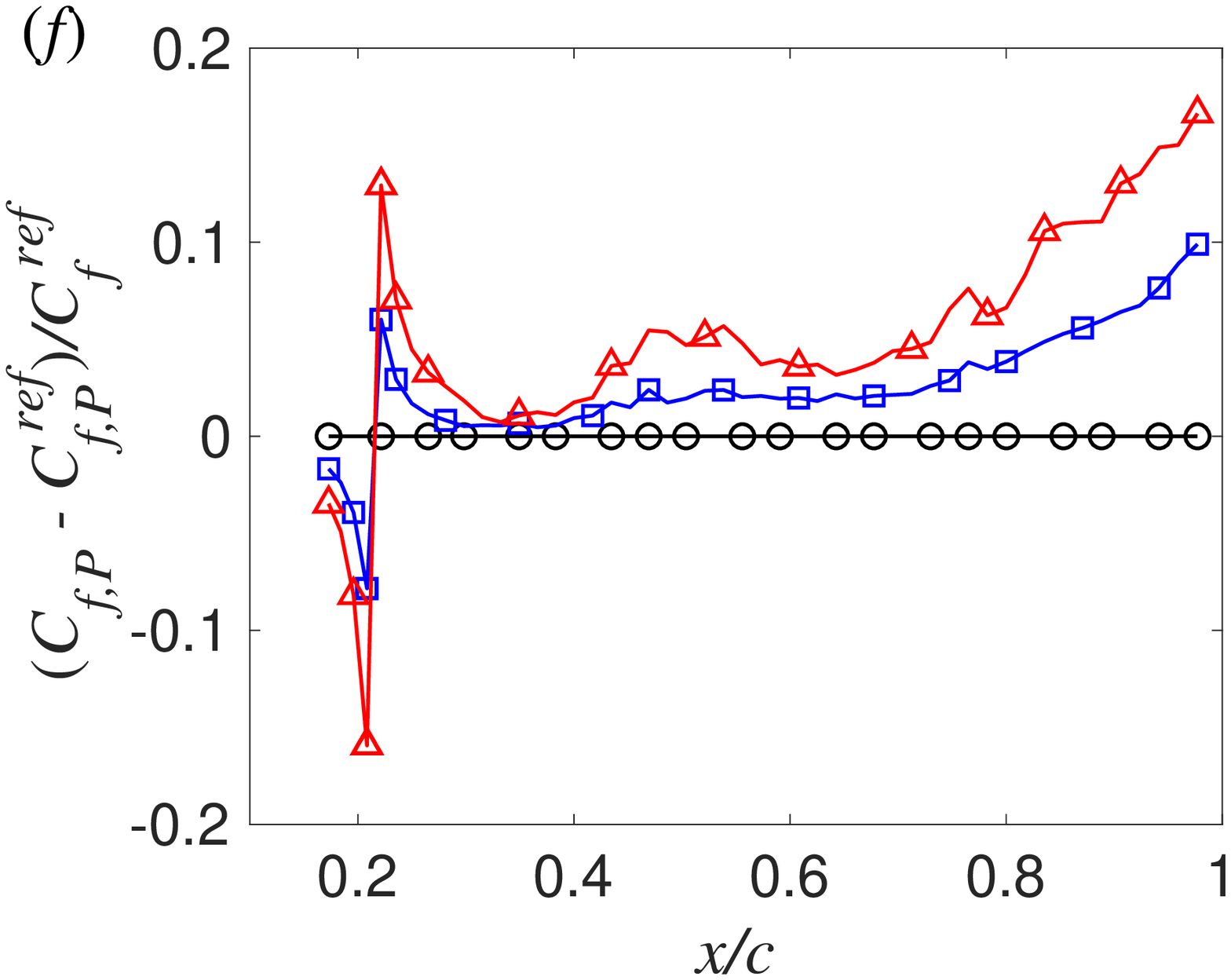}\label{bot2cf:f}}
\caption{Variation of ($a$) $C_{f,V}$, ($b$) $C_{f,T}$, ($c$) $C_{f,G}$, ($d$) $C_{f,C}$, ($e$) $C_{f,D}$, and ($f$) $C_{f,P}$ with regard to the reference case on the pressure side of NACA4412 at $Re_c=200,000$.}
\label{bot2cf}
\end{figure}

$\bullet$ The variations of $C_{f,V}$, $C_{f,T}$, and $C_{f,G}$ (see figures \ref{bot2cf:a}--\ref{bot2cf:c}) are similar to the results on the suction side, that is, by blowing, $C_{f,V}$ and $C_{f,T}$ are decreased, and $C_{f,G}$ is reduced. 
As for the sub-constituents of $C_{f,G}$ in figures \ref{bot2cf:d}--\ref{bot2cf:f}, blowing reduces $C_{f,C}$, similar to the observations on the suction side; however, differently, blowing just increases $C_{f,D}$ until $x/c \approx 0.85$ and then decreases; in contrary to the control effects on the suction side, $C_{f,P}$ is increased, suggesting that blowing strengthens the FPG on the pressure side of NACA4412. Stronger intensity of blowing leads to larger increases or decreases of the constituents.

\begin{figure}[h]
\centering
\subfigure{\includegraphics[width = 5.5cm]{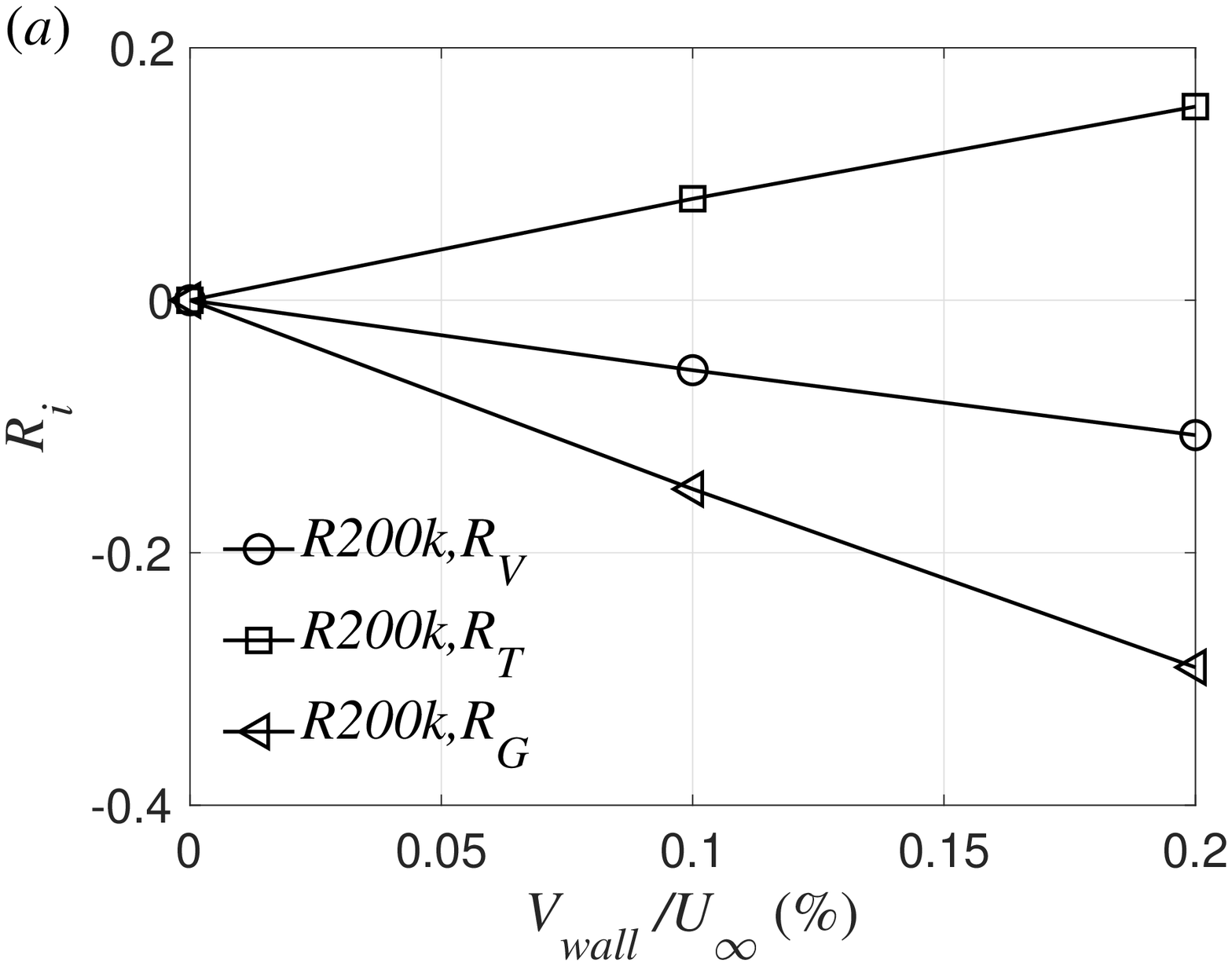}\label{bot_rr:a}}		
\subfigure{\includegraphics[width = 5.5cm]{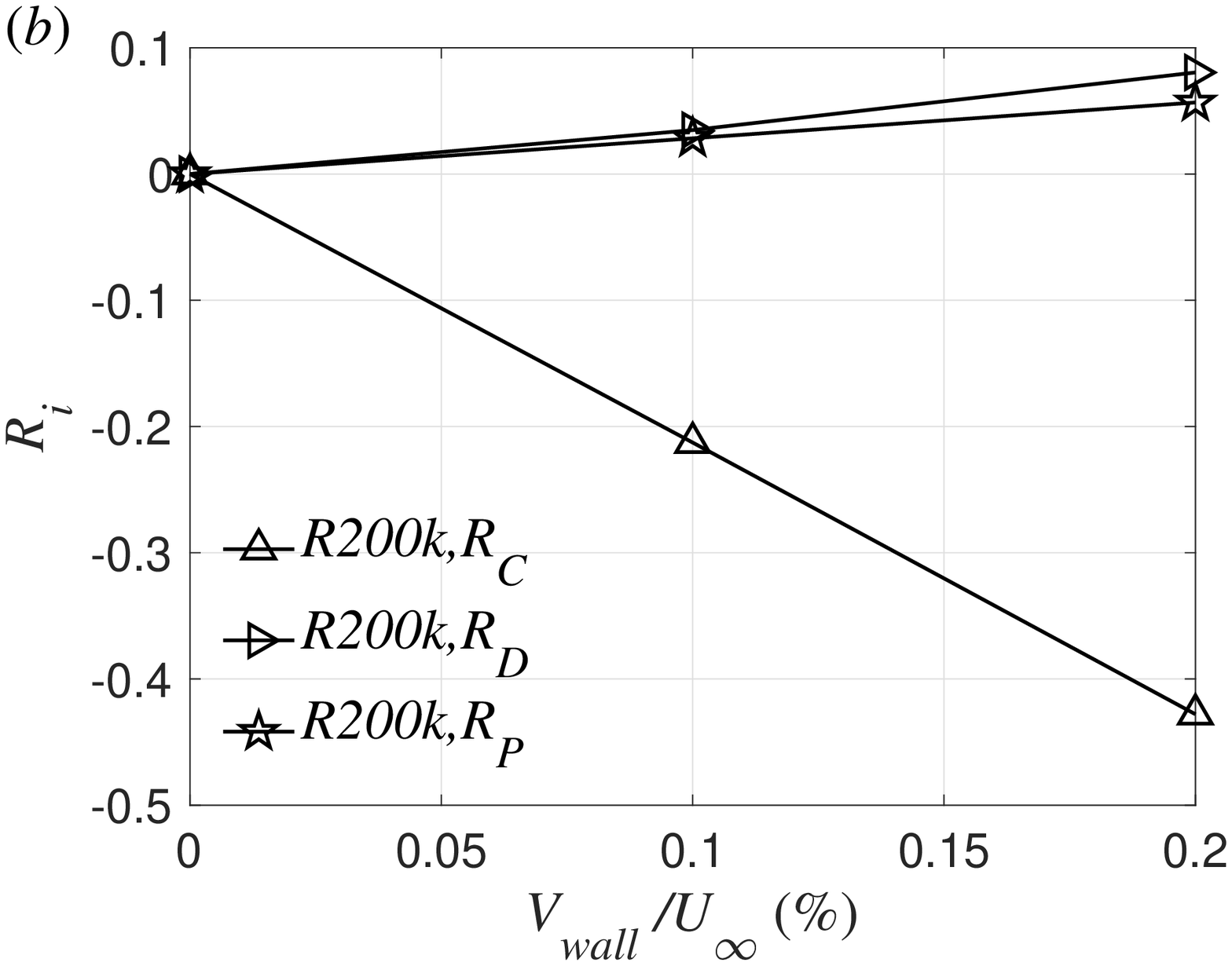}\label{bot_rr:b}}
\caption{Total friction-drag change rate as a function of control intensity ($V_{wall}/U_{\infty}$) on the pressure side of NACA4412.}
\label{bot_rr}
\end{figure}

$\bullet$ Again, near-linear dependence of $R_i$  on the intensity of blowing is still observed within the range of $V_{wall}/U_\infty$ under scrutiny, as seen in figure \ref{bot_rr}.

\begin{figure}[h]
\centering
\subfigure{\includegraphics[width = 5.5cm]{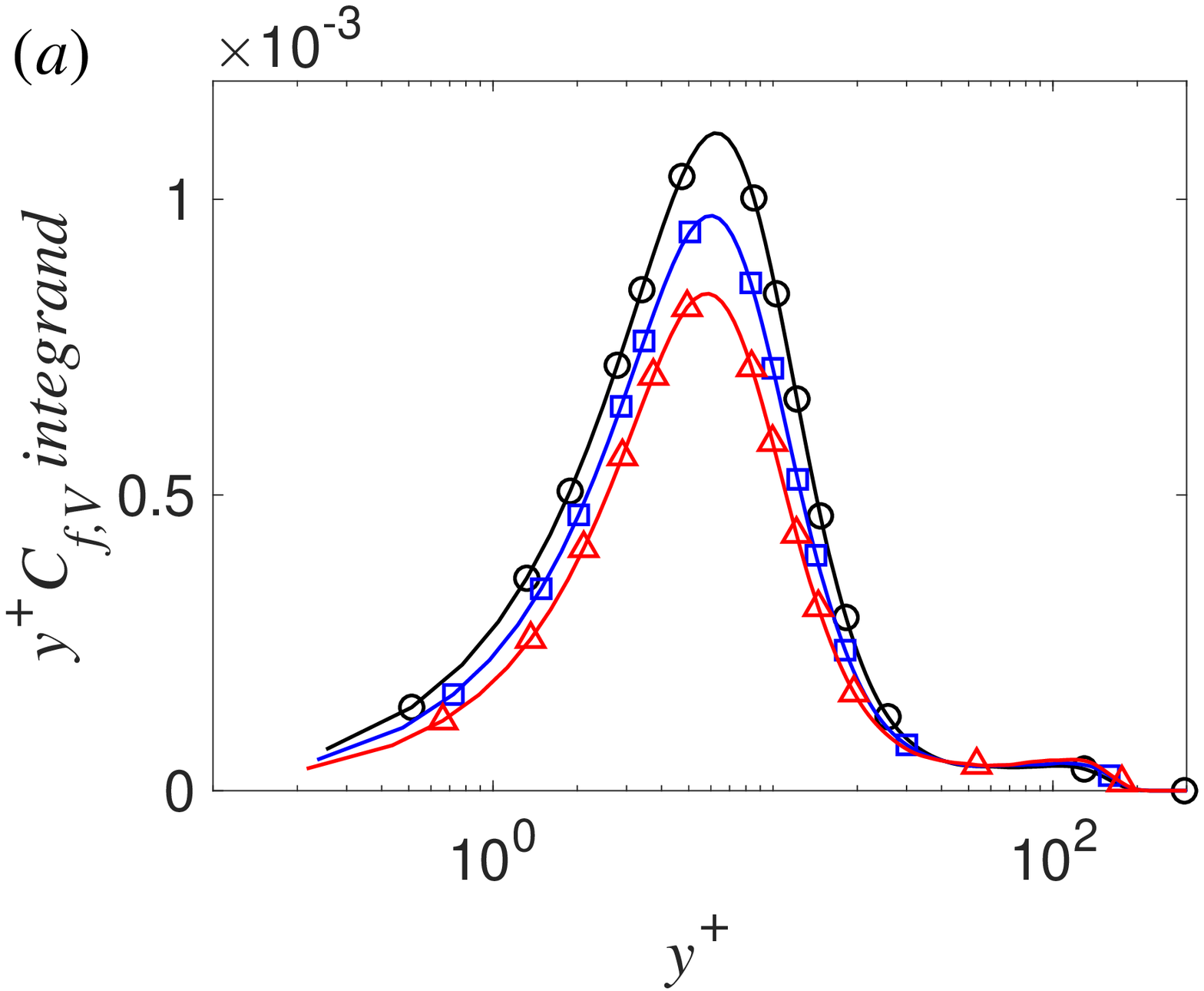}\label{bot2cfd:a}}
\subfigure{\includegraphics[width = 5.5cm]{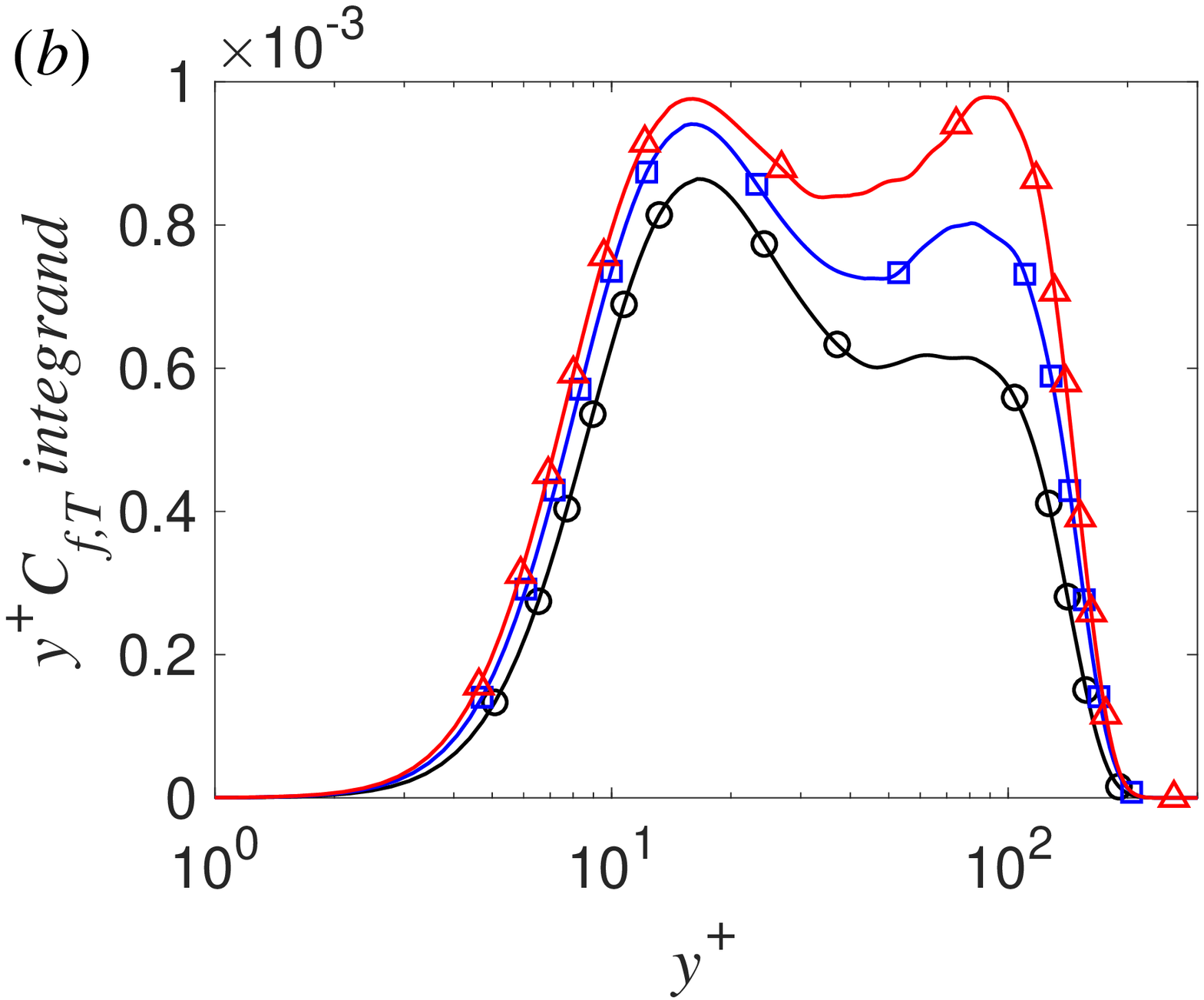}\label{bot2cfd:b}}
\subfigure{\includegraphics[width = 5.5cm]{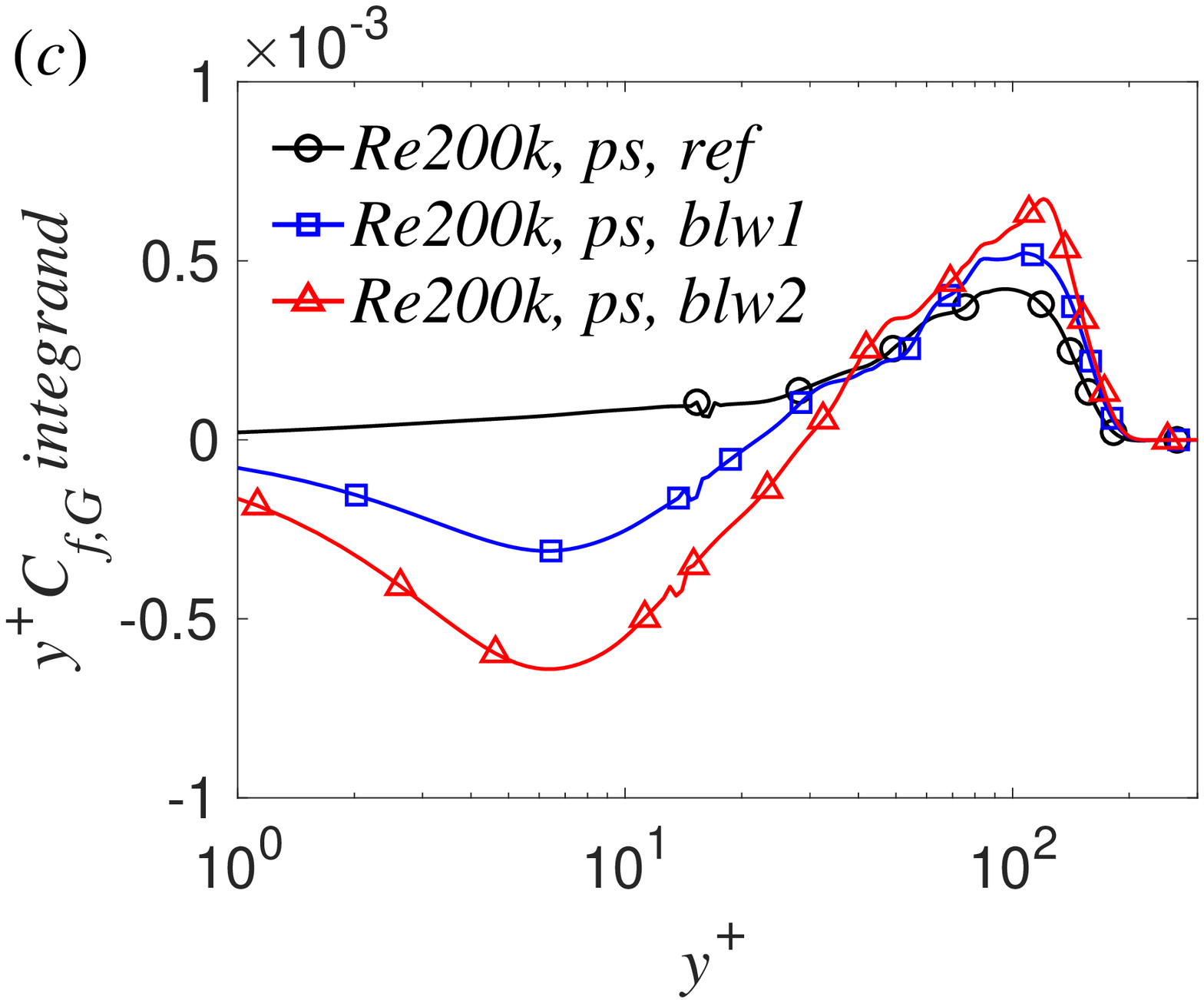}\label{bot2cfd:c}}
\caption{Pre-multiplied integrands of ($a$) $C_{f,V}$, ($b$) $C_{f,T}$, and ($c$) $C_{f,G}$, as a function of $y^+$, on the pressure side of NACA4412 at $Re_c=200,000$.}
\label{bot2cfd}
\end{figure}

\begin{figure}[h]
\centering
\subfigure{\includegraphics[width = 5.5cm]{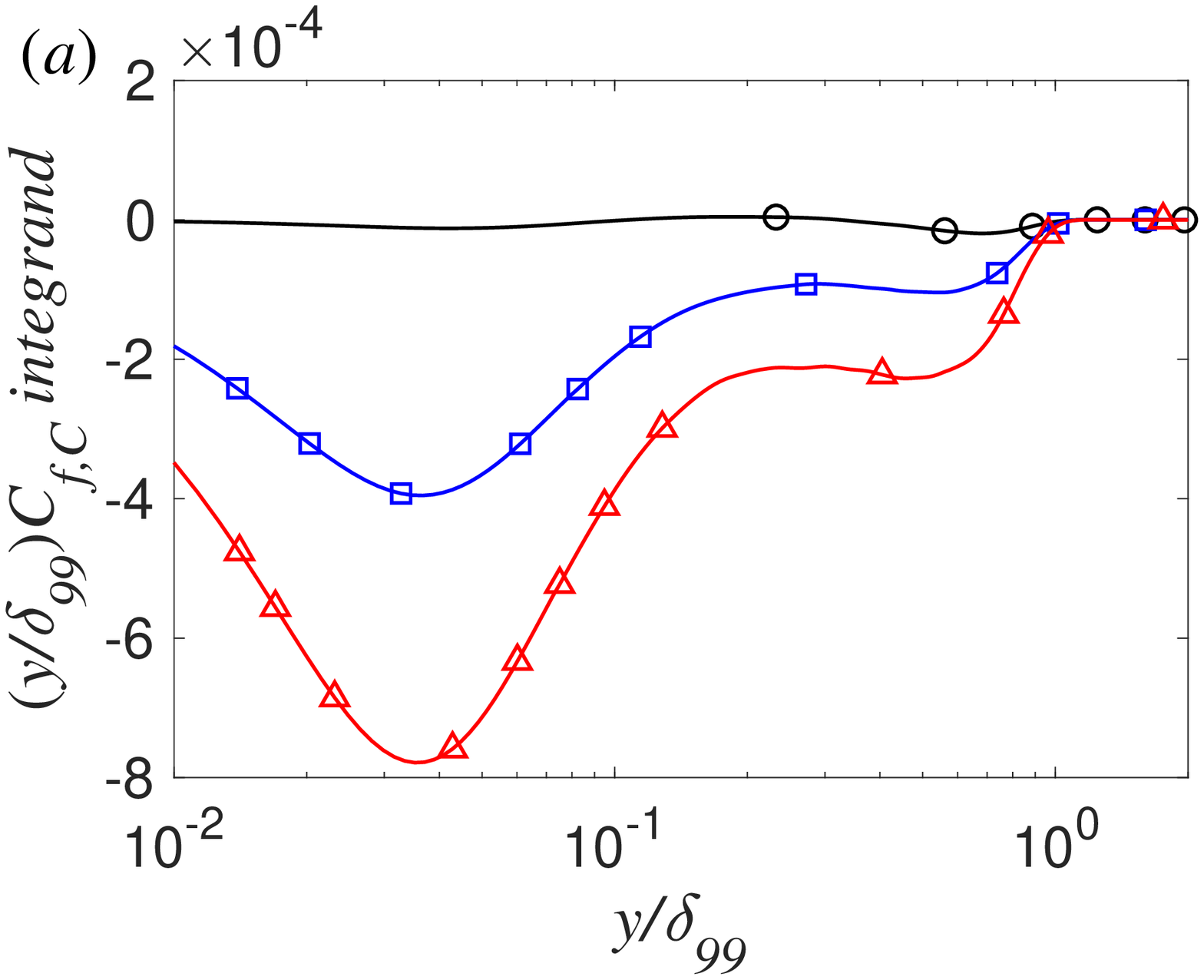}\label{bot2cf3d:a}}
\subfigure{\includegraphics[width = 5.5cm]{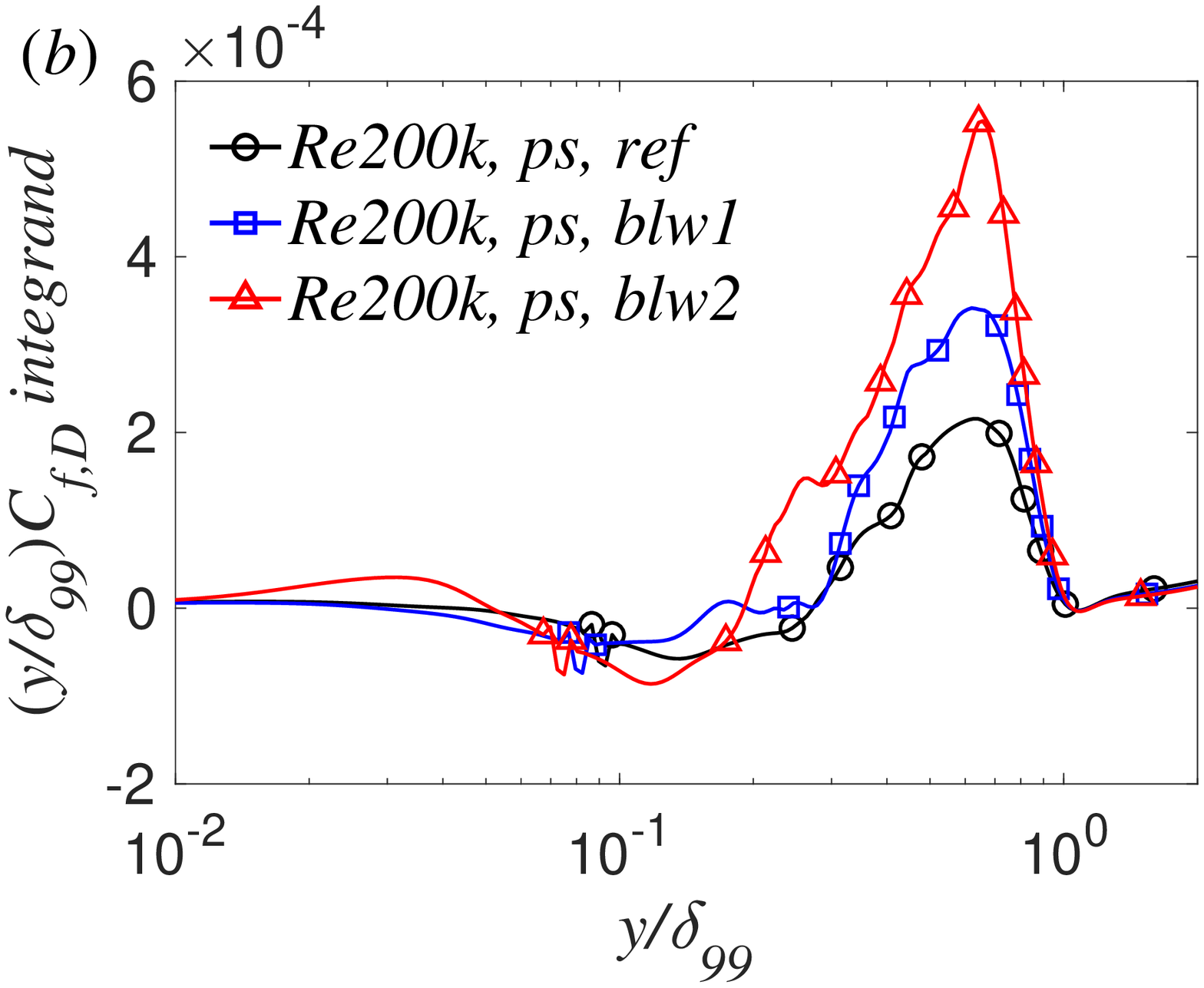}\label{bot2cf3d:b}}
\subfigure{\includegraphics[width = 5.5cm]{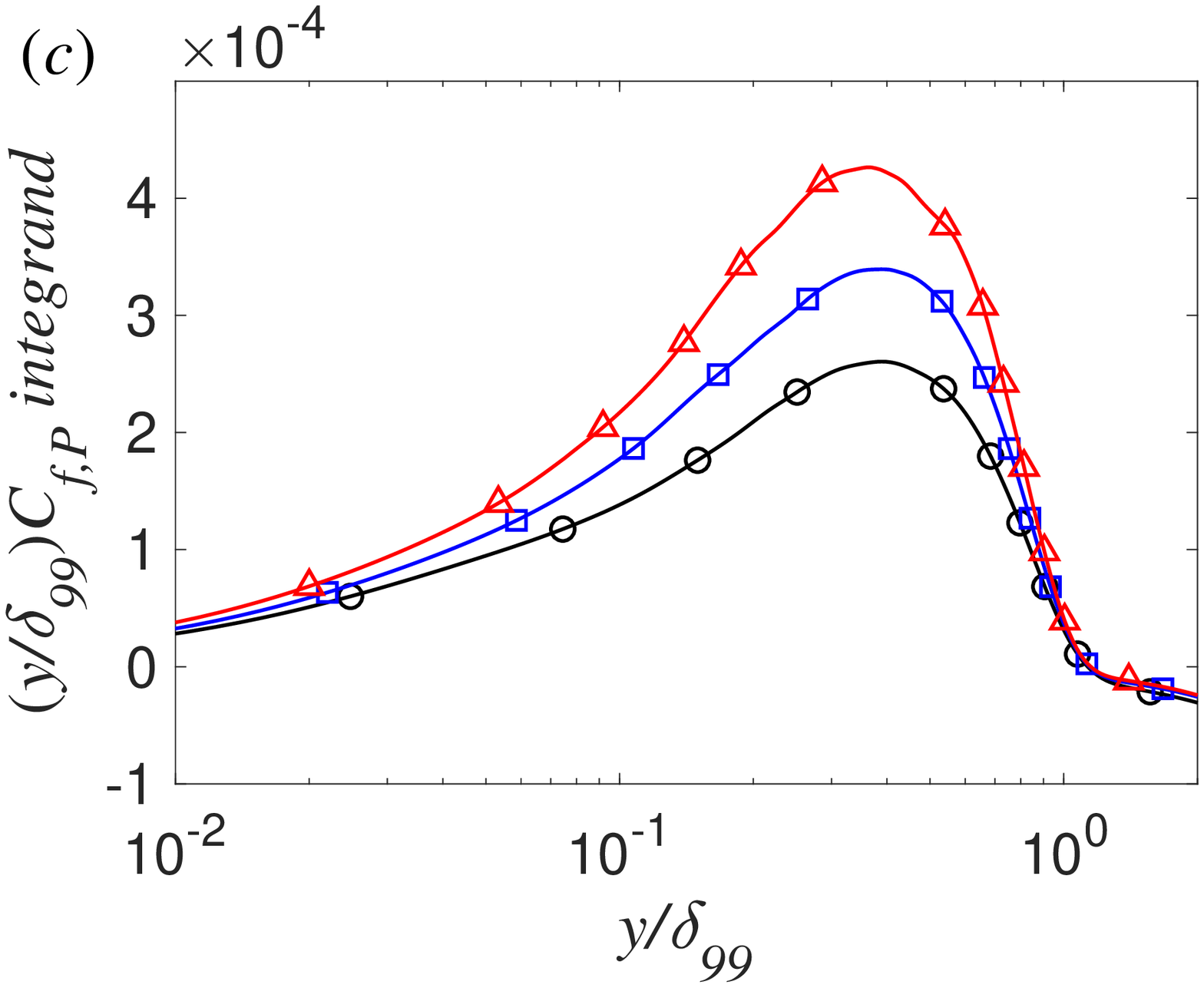}\label{bot2cf3d:c}}
\caption{Pre-multiplied integrands of ($a$) $C_{f,C}$, ($b$) $C_{f,D}$, and ($c$) $C_{f,P}$, as a function of $y/\delta_{99}$, on the pressure side of NACA4412 at $Re_c=200,000$.}
\label{bot2cf3d}
\end{figure}

$\bullet$ 
At $x/c\approx0.75$ (similar features are observed at other positions), the wall-normal distributions of the $C_f-$constituents are plotted in figure \ref{bot2cfd}.
With the presence of FPG, the outer peak of $C_{f,V}$ is barely legible, and the outer peak of $C_{f,T}$ is comparable to the inner one, which are different from the results on the suction side. With blowing, the outer peak of $C_{f,T}$ is also increased however by a lesser amount, probably because that the FPG attenuates the outer-layer Reynolds shear stress and the production of turbulence kinetic energy \citep{Harun2013}. 
In figure \ref{bot2cf3d}, different from the observation in figure \ref{topcf3d}, the generation of $C_{f,C}$ mainly arises from the inner-layer dynamics on the pressure side of the airfoil. 

$\bullet$ The peak locations in figures \ref{bot2cfd} and \ref{bot2cf3d} are well collapsed when normalized by inner or outer units. It suggests that the self-similar features of turbulence statistics will not be affected by the presence of FPG.
Note that the distributions in figures \ref{bot2cfd:a}-\ref{bot2cfd:c} are not plotted as a function of $y / \delta_{99}$ since the friction Reynolds numbers of these three cases at $x/c\approx0.75$ are very similar, being $Re_\tau\approx177$, $185$, and $187$ in the case of ``$Re200k, ps, ref$'', ``$Re200k, ps, blw1$'', and ``$Re200k, ps, blw2$'', respectively.

$\bullet$ At last, EMD was not conducted, since the inner-outer scale separation is less evident in the FPG-TBLs at such low $Re_\tau$ than those on the suction side of the airfoil.

\section{Conclusion} \label{conclusion}
We employed the RD identity in conjunction with empirical mode decomposition (EMD) to study the control effects of uniform blowing and suction on the generation of mean friction drag on a NACA4412 airfoil at chord Reynolds numbers $Re_c$=$200,000$ and $400,000$.
In general, blowing reduces the mean friction drag, and suction increases the friction drag, both on the suction and pressure sides of the airfoil. With the RD identity, the mean friction drag on the airfoil is decomposed into three components, associated with viscous dissipation ($C_{f,V}$), turbulence-kinetic-energy production ($C_{f,T}$), and spatial growth of the flow ($C_{f,G}$), respectively. The $C_{f,G}$ component is further decomposed into three terms related to the mean wall-normal convection ($C_{f,C}$), streamwise development ($C_{f,D}$), and the pressure gradient ($C_{f,P}$). 
The effects of suction on these $C_f-$constituents are quite opposite to those of blowing, thus we just summarize the key conclusions in the blowing cases. 

For the adverse-pressure-gradient turbulent boundary layers on the suction side of the airfoil, concluding remarks  are listed as below:

$\bullet$ Blowing reduces the generation of $C_{f,V}$ and $C_{f,G}$, while increases that of $C_{f,T}$.
The drag reduction with blowing is mostly attributed to $C_{f,G}$, which results from the decreased $C_{f,C}$ and $C_{f,P}$, which overwhelm the increase in  $C_{f,D}$.
The integrated $C_f-$constituents over the control surface are observed to be linearly dependent on the intensity of blowing, and weakly influenced by the chord Reynolds number (at least within the parameters we considered).

$\bullet$ Wall-normal distributions of the $C_f-$constituents at $x/c \approx 0.75$ are checked to clarify how the control schemes impact the sources of friction-drag generation. 
With blowing, the generation of $C_{f,V}$, which is mainly related to the inner-layer dynamics, is suppressed in the inner region ($y^+<30$), while it is enhanced in the outer region. 
Contrarily, the generation of $C_{f,T}$, mainly originats from the outer-layer motions, is amplified by the blowing.
These phenomena are linked to the variations of the wall-normal velocity gradients and Reynolds shear stresses in the wall-normal direction.
The generation of $C_{f,G}$ results from a counterbalance between the negative work done by $C_{f,C}$ and $C_{f,P}$ and the positive work by $C_{f,D}$. 
Blowing is able to enhance the generation of all sub-components, as the strengthened adverse pressure gradient promotes a more pronounced growth of the boundary layer and a more prominent outer region \citep{Vinuesa2018}.

$\bullet$ We observed that, in the wall-normal direction, the inner-peak locations of $C_{f,V}-$ and $C_{f,T}-$contributions scale well in the inner unit ($\delta_\nu$), and the outer-peak locations of $C_{f,V}-$, $C_{f,T}-$, $C_{f,G}-$ as well as its sub-contributions, collapse well in the outer unit ($\delta_{99}$), regardless of the friction Reynolds number, control scheme, and the intensity of blowing/suction. This reveals that self similarity is exhibited in inner or outer scales for the turbulence statistics associated with the friction-drag generation.

$\bullet$ The small- and large-scale structures are separated with empirical mode decomposition (EMD), aiming to analyze the scale-specific contribution of turbulent motions to friction-drag generation.
Results unveil that, normalized by $C_{f,T}$ itself,  blowing is able to enhance the contribution of large-scale motions and to suppress that of small scales; note that suction behaves contrarily. The contributions related to cross-scale interactions remain almost unchanged with different control strategies.

For the favorable-pressure-gradient turbulent boundary layers on the pressure side of the airfoil, the most significant observation is that the outer-layer motions are of less importance for the generation of $C_f-$constituents.
In the case of blowing,  the generation of $C_{f,P}$ is increased, which is the opposite behavior to that on the suction side of the airfoil.

\section*{Acknowledgments}
The funding support of the National Natural Science Foundation of China (under the grant No. 91952302 and 92052101) is acknowledged. Davide Gatti acknowledges support by the state of Baden-W\"urttemberg through bwHPC. Marco Atzori, Ricardo Vinuesa and Philipp Schlatter also acknowledge support from the Swedish Foundation for Strategic Research, project ``In-Situ Big Data Analysis for Flow and Climate Simulations'' (ref. number BD15-0082), from the Knut and Alice Wallenberg Foundation and from the Swedish Research Council (VR). The simulations were performed on resources provided by the Swedish National Infrastructure for Computing (SNIC) and within the project CWING on the national supercomputer Cray XC40 Hazel Hen at the High Performance Computing Center Stuttgart (HLRS).

\bibliographystyle{plainnat}
\bibliography{ref_control}

\begin{thebibliography}{64}
\providecommand{\natexlab}[1]{#1}
\providecommand{\url}[1]{\texttt{#1}}
\expandafter\ifx\csname urlstyle\endcsname\relax
  \providecommand{\doi}[1]{doi: #1}\else
  \providecommand{\doi}{doi: \begingroup \urlstyle{rm}\Url}\fi

\bibitem[Agostini and Leschziner(2014)]{Agostini2014}
L.~Agostini and M.~A. Leschziner.
\newblock On the influence of outer large-scale structures on near-wall
  turbulence in channel flow.
\newblock \emph{Phys. Fluids}, 26\penalty0 (7):\penalty0 075107, 2014.

\bibitem[Agostini and Leschziner(2016)]{Agostini2016}
L.~Agostini and M.~A. Leschziner.
\newblock Predicting the response of small-scale near-wall turbulence to
  large-scale outer motions.
\newblock \emph{Phys. Fluids}, 28\penalty0 (1):\penalty0 015107, 2016.

\bibitem[Agostini and Leschziner(2019)]{Agostini2019}
L.~Agostini and M.~A. Leschziner.
\newblock The connection between the spectrum of turbulent scales and the
  skin-friction statistics in channel flow at ${R}e_\tau \approx 1000$.
\newblock \emph{J. Fluid Mech.}, 871:\penalty0 22–51, 2019.

\bibitem[Ansell and Balajewicz(2017)]{Ansell2017}
P.~J. Ansell and M.~J. Balajewicz.
\newblock Separation of unsteady scales in a mixing layer using empirical mode
  decomposition.
\newblock \emph{AIAA J.}, 55\penalty0 (2):\penalty0 419--434, 2017.

\bibitem[Atzori et~al.(2020)Atzori, Vinuesa, Fahland, Stroh, Gatti, Frohnapfel,
  and Schlatter]{Atzori2020}
M.~Atzori, R.~Vinuesa, G.~Fahland, A.~Stroh, D.~Gatti, B.~Frohnapfel, and
  P.~Schlatter.
\newblock Aerodynamic effects of uniform blowing and suction on a {NACA4412}
  airfoil.
\newblock \emph{Flow, Turbul. Combust.}, 105:\penalty0 735--759, 2020.

\bibitem[Bannier et~al.(2015)Bannier, Garnier, and Sagaut]{Bannier2015}
A.~Bannier, {\'E}.~Garnier, and P.~Sagaut.
\newblock Riblet flow model based on an extended {FIK} identity.
\newblock \emph{Flow, Turbul. Combust.}, 95\penalty0 (2-3):\penalty0 351--376,
  2015.

\bibitem[Bobke et~al.(2016)Bobke, {\"O}rl{\"u}, and Schlatter]{bobk2016}
A.~Bobke, R.~{\"O}rl{\"u}, and P.~Schlatter.
\newblock Simulations of turbulent asymptotic suction boundary layers.
\newblock \emph{J. Turbul.}, 17\penalty0 (2):\penalty0 157--180, 2016.

\bibitem[Cheng et~al.(2019)Cheng, Li, Lozano-Dur{\'a}n, and Liu]{Cheng2019}
C.~Cheng, W.-P. Li, A.~Lozano-Dur{\'a}n, and H.~Liu.
\newblock Identity of attached eddies in turbulent channel flows with
  bidimensional empirical mode decomposition.
\newblock \emph{J. Fluid Mech.}, 870:\penalty0 1037–1071, 2019.

\bibitem[Clauser(1954)]{Clauser1954}
F.~H. Clauser.
\newblock The turbulent boundary layer in adverse pressure gradients.
\newblock \emph{J. Aero. Sci.}, 21:\penalty0 91--108, 1954.

\bibitem[Clauser(1956)]{Clauser1956}
F.~H. Clauser.
\newblock The turbulent boundary layer.
\newblock \emph{Adv. Appl. Mech.}, 4:\penalty0 1--51, 1956.

\bibitem[Deck et~al.(2014)Deck, Renard, Laraufie, and Weiss]{Deck2014}
S.~Deck, N.~Renard, R.~Laraufie, and P.~{\'E}. Weiss.
\newblock Large-scale contribution to mean wall shear stress in
  high-{R}eynolds-number flat-plate boundary layers up to ${Re}_{{\theta}}{=}$
  13650.
\newblock \emph{J. Fluid Mech.}, 743:\penalty0 202--248, 2014.

\bibitem[Dogan et~al.(2019)Dogan, \"Orl\"u, Gatti, Vinuesa, and
  Schlatter]{doga19}
E.~Dogan, R.~\"Orl\"u, D.~Gatti, R.~Vinuesa, and P.~Schlatter.
\newblock Quantification of amplitude modulation in wall-bounded turbulence.
\newblock \emph{Fluid Dyn. Res.}, 51:\penalty0 011408, 2019.

\bibitem[Eto et~al.(2019)Eto, Kondo, Fukagata, and Tokugawa]{Eto2019}
K.~Eto, Y.~Kondo, K.~Fukagata, and N.~Tokugawa.
\newblock Assessment of friction drag reduction on a {Clark-Y} airfoil by
  uniform blowing.
\newblock \emph{AIAA J.}, 57\penalty0 (7):\penalty0 2774--2782, 2019.

\bibitem[Fahland et~al.(2021)Fahland, Stroh, Frohnapfel, Atzori, Vinuesa,
  Schlatter, and Gatti]{fahl21}
G.~Fahland, A.~Stroh, B.~Frohnapfel, M.~Atzori, R.~Vinuesa, P.~Schlatter, and
  D.~Gatti.
\newblock Investigation of blowing and suction for turbulent flow control on
  airfoils.
\newblock \emph{AIAA J.}, \penalty0 ({To Appear}), 03 2021.

\bibitem[Fan et~al.(2019{\natexlab{a}})Fan, Cheng, and Li]{Fan2019a}
Y.-T. Fan, C.~Cheng, and W.-P. Li.
\newblock Effects of the {R}eynolds number on the mean skin friction
  decomposition in turbulent channel flows.
\newblock \emph{Appl. Math. Mech. (English Ed.)}, 40\penalty0 (3):\penalty0
  331--342, 2019{\natexlab{a}}.

\bibitem[Fan et~al.(2019{\natexlab{b}})Fan, Li, and Pirozzoli]{Fan2019}
Y.-T. Fan, W.-P. Li, and S.~Pirozzoli.
\newblock Decomposition of the mean friction drag in zero-pressure-gradient
  turbulent boundary layers.
\newblock \emph{Phys. Fluids}, 31\penalty0 (8):\penalty0 086105,
  2019{\natexlab{b}}.

\bibitem[Fan et~al.(2020{\natexlab{a}})Fan, Li, Atzori, Pozuelo, Schlatter, and
  Vinuesa]{Fan2020}
Y.-T. Fan, W.-P. Li, M.~Atzori, R.~Pozuelo, P.~Schlatter, and R.~Vinuesa.
\newblock Decomposition of the mean friction drag in adverse-pressure-gradient
  turbulent boundary layers.
\newblock \emph{Phys. Rev. Fluids}, 5:\penalty0 114608, 2020{\natexlab{a}}.

\bibitem[Fan et~al.(2020{\natexlab{b}})Fan, Li, and Pirozzoli]{Fan2020a}
Y.-T. Fan, W.-P. Li, and S.~Pirozzoli.
\newblock Energy-based decomposition of friction drag in turbulent square-duct
  flows.
\newblock \emph{Int. J. Heat Fluid Flow}, 86:\penalty0 108731,
  2020{\natexlab{b}}.
\newblock ISSN 0142-727X.

\bibitem[Fischer et~al.(2008)Fischer, Lottes, and Kerkemeier]{fisc08}
P.~F. Fischer, J.~W. Lottes, and S.~G. Kerkemeier.
\newblock Nek5000: Open source spectral element {CFD} solver.
\newblock Available at: \url{http://nek5000.mcs.anl.gov}, 2008.

\bibitem[Fukagata et~al.(2002)Fukagata, Iwamoto, and Kasagi]{Fukagata2002}
K.~Fukagata, K.~Iwamoto, and N.~Kasagi.
\newblock Contribution of reynolds stress distribution to the skin friction in
  wall-bounded flows.
\newblock \emph{Phys. Fluids}, 14\penalty0 (11):\penalty0 L73--L76, 2002.

\bibitem[Gad-el Hak(1994)]{Gad-el-Hak1994}
M.~Gad-el Hak.
\newblock Interactive control of turbulent boundary layers - {A} futuristic
  overview.
\newblock \emph{AIAA J.}, 32\penalty0 (9):\penalty0 1753--1765, 1994.

\bibitem[Harun et~al.(2013)Harun, Monty, Mathis, and Marusic]{Harun2013}
Z.~Harun, J.~P. Monty, R.~Mathis, and I.~Marusic.
\newblock Pressure gradient effects on the large-scale structure of turbulent
  boundary layers.
\newblock \emph{J. Fluid Mech.}, 715:\penalty0 477--498, 2013.

\bibitem[Huang et~al.(1998)Huang, Shen, Long, Wu, Shih, Zheng, Yen, Tung, and
  Liu]{Huang1998}
N.~E. Huang, Z.~Shen, S.~R. Long, M.~C. Wu, H.~H. Shih, Q.~Zheng, N.-C. Yen,
  C.~C. Tung, and H.~H. Liu.
\newblock {The empirical mode decomposition and the Hilbert spectrum for
  nonlinear and non-stationary time series analysis}.
\newblock \emph{Proc. R. Soc. Lond. A}, 454:\penalty0 903--995, 1998.

\bibitem[Huang et~al.(2008)Huang, Schmitt, Lu, and Liu]{Huang2008}
Y.~X. Huang, F.~G. Schmitt, Z.~M. Lu, and Y.~L. Liu.
\newblock An amplitude-frequency study of turbulent scaling intermittency using
  empirical mode decomposition and {H}ilbert spectral analysis.
\newblock \emph{Europhys. Lett.}, 84\penalty0 (4):\penalty0 40010, nov 2008.

\bibitem[Hwang(1996)]{hwan97}
D.~Hwang.
\newblock A proof of concept experiment for reducing skin friction by using a
  micro-blowing technique.
\newblock \emph{NASA}, pages TM--107315, 1996.

\bibitem[Hwang(2004)]{hwan04}
D.~Hwang.
\newblock Review of research into the concept of the microblowing technique for
  turbulent skin friction reduction.
\newblock \emph{Prog. Aerosp. Sci.}, 40:\penalty0 559--575, 2004.

\bibitem[Kametani and Fukagata(2011)]{Kametani2011}
Y.~Kametani and K.~Fukagata.
\newblock Direct numerical simulation of spatially developing turbulent
  boundary layers with uniform blowing or suction.
\newblock \emph{J. Fluid Mech.}, 681:\penalty0 154–172, 2011.

\bibitem[Kametani et~al.(2015)Kametani, Fukagata, {\"O}rl{\"u}, and
  Schlatter]{Kametani2015}
Y.~Kametani, K.~Fukagata, R.~{\"O}rl{\"u}, and P.~Schlatter.
\newblock {Effect of uniform blowing/suction in a turbulent boundary layer at
  moderate Reynolds number}.
\newblock \emph{Int. J. Heat Fluid Flow}, 55:\penalty0 132--142, 2015.

\bibitem[Kim et~al.(2002)Kim, Sung, and Chung]{Kim2002}
K.~Kim, H-.J. Sung, and M.-K. Chung.
\newblock Assessment of local blowing and suction in a turbulent boundary
  layer.
\newblock \emph{AIAA J.}, 40\penalty0 (1):\penalty0 175--177, 2002.

\bibitem[Kornilov(2015)]{Kornilov2015}
V.~I. Kornilov.
\newblock Current state and prospects of researches on the control of turbulent
  boundary layer by air blowing.
\newblock \emph{Prog. Aerosp. Sci.}, 76:\penalty0 1--23, 2015.

\bibitem[Kornilov(2017)]{korn17}
V.~I. Kornilov.
\newblock Implementation of air injection into the turbulent boundary layer of
  aircraft wing using external pressurized flow.
\newblock \emph{Thermophys. Aeromech.}, 24:\penalty0 175--185, 2017.

\bibitem[Kornilov(2021)]{korn20}
V.~I. Kornilov.
\newblock Combined blowing/suction flow control on low-speed airfoils.
\newblock \emph{Flow Turbul. Combust.}, 106:\penalty0 81--108, 2021.

\bibitem[Kornilov et~al.(2019)Kornilov, Kavun, and Popkov]{korn19}
V.~I. Kornilov, I.~N. Kavun, and A.~N. Popkov.
\newblock Modification of turbulent airfoil section flow using a combined
  control action.
\newblock \emph{Thermophys. Aeromech.}, 26:\penalty0 165--178, 2019.

\bibitem[Lee and Sung(2008)]{Lee2008}
J.-H. Lee and H.-J. Sung.
\newblock Effects of an adverse pressure gradient on a turbulent boundary
  layer.
\newblock \emph{Int. J. Heat Fluid Flow}, 29\penalty0 (3):\penalty0 568--578,
  2008.

\bibitem[Li(2020)]{Li2020}
W.-P. Li.
\newblock Turbulence statistics of flow over a drag-reducing and a
  drag-increasing riblet-mounted surface.
\newblock \emph{Aerosp. Sci. Technol.}, 104:\penalty0 106003, 2020.
\newblock ISSN 1270-9638.

\bibitem[Li et~al.(2019)Li, Fan, Modesti, and Cheng]{Li2019}
W.-P. Li, Y.-T. Fan, D.~Modesti, and C.~Cheng.
\newblock Decomposition of the mean skin-friction drag in compressible
  turbulent channel flows.
\newblock \emph{J. Fluid Mech.}, 875:\penalty0 101--123, 2019.

\bibitem[Lumley(1967)]{Lumley1967}
J.~L. Lumley.
\newblock The structure of inhomogeneous turbulent flows.
\newblock In A.~M. Yaglom and V.~I. Tartarsky, editors, \emph{Atmospheric
  Turbulence and Radio Wave Propagation}, pages 166--177. 1967.

\bibitem[Mahfoze et~al.(2019)Mahfoze, Moody, Wynn, Whalley, and
  Laizet]{Mahfoze2019}
O.~A. Mahfoze, A.~Moody, A.~Wynn, R.~D. Whalley, and S.~Laizet.
\newblock Reducing the skin-friction drag of a turbulent boundary-layer flow
  with low-amplitude wall-normal blowing within a {B}ayesian optimization
  framework.
\newblock \emph{Phys. Rev. Fluids}, 4:\penalty0 094601, 2019.

\bibitem[Mehdi and White(2011)]{Mehdi2011}
F.~Mehdi and C.~M. White.
\newblock Integral form of the skin friction coefficient suitable for
  experimental data.
\newblock \emph{Exp. Fluids}, 50\penalty0 (1):\penalty0 43--51, 2011.

\bibitem[Mehdi et~al.(2014)Mehdi, Johansson, White, and Naughton]{Mehdi2014}
F.~Mehdi, T.~G. Johansson, C.~M. White, and J.~W. Naughton.
\newblock On determining wall shear stress in spatially developing
  two-dimensional wall-bounded flows.
\newblock \emph{Exp. Fluids}, 55\penalty0 (1):\penalty0 1656, 2014.

\bibitem[Modesti et~al.(2018)Modesti, Pirozzoli, Orlandi, and
  Grasso]{Modesti2018}
D.~Modesti, S.~Pirozzoli, P.~Orlandi, and F.~Grasso.
\newblock On the role of secondary motions in turbulent square duct flow.
\newblock \emph{J. Fluid Mech.}, 847:\penalty0 R1, 2018.
\newblock \doi{10.1017/jfm.2018.391}.

\bibitem[Park and Choi(1999)]{Park1999}
J.~Park and H.~Choi.
\newblock Effects of uniform blowing or suction from a spanwise slot on a
  turbulent boundary layer flow.
\newblock \emph{Phys. Fluids}, 11\penalty0 (10):\penalty0 3095--3105, 1999.

\bibitem[Patera(1984)]{pate84}
A.~T. Patera.
\newblock A spectral element method for fluid dynamics: laminar flow in a
  channel expansion.
\newblock \emph{J. Comput. Phys.}, 54:\penalty0 468--488, 1984.

\bibitem[Peet and Sagaut(2009)]{Peet2009}
Y.~Peet and P.~Sagaut.
\newblock Theoretical prediction of turbulent skin friction on geometrically
  complex surfaces.
\newblock \emph{Phys. Fluids}, 21\penalty0 (10):\penalty0 105105, 2009.

\bibitem[Ran et~al.(2021)Ran, Zare, and Jovanovi{\'c}]{Ran2021}
W.~Ran, A.~Zare, and M.~R. Jovanovi{\'c}.
\newblock Model-based design of riblets for turbulent drag reduction.
\newblock \emph{J. Fluid Mech.}, 906:\penalty0 A7, 2021.
\newblock \doi{10.1017/jfm.2020.722}.

\bibitem[Rastegari and Akhavan(2015)]{Rastegari2015}
A.~Rastegari and R.~Akhavan.
\newblock On the mechanism of turbulent drag reduction with super-hydrophobic
  surfaces.
\newblock \emph{J. Fluid Mech.}, 773:\penalty0 R4, 2015.
\newblock \doi{10.1017/jfm.2015.266}.

\bibitem[Renard and Deck(2016)]{Renard2016}
N.~Renard and S.~Deck.
\newblock A theoretical decomposition of mean skin friction generation into
  physical phenomena across the boundary layer.
\newblock \emph{J. Fluid Mech.}, 790:\penalty0 339--367, 2016.

\bibitem[Rotta(1950)]{Rotta1950}
J.~C. Rotta.
\newblock {\"U}ber die theorie der turbulenten grenzschichten.
\newblock \emph{Mitt. Max Planck Inst. Str{\"o}mungsforsch., G{\"o}ttingen},
  1950.

\bibitem[Sanmiguel~Vila et~al.(2020)Sanmiguel~Vila, Vinuesa, Discetti, Ianiro,
  Schlatter, and {\"O}rl{\"u}]{Vila2020}
C.~Sanmiguel~Vila, R.~Vinuesa, S.~Discetti, A.~Ianiro, P.~Schlatter, and
  R.~{\"O}rl{\"u}.
\newblock Separating adverse-pressure-gradient and {R}eynolds-number effects in
  turbulent boundary layers.
\newblock \emph{Phys. Rev. Fluids}, 5:\penalty0 064609, 2020.

\bibitem[Schlatter and \"Orl\"u(2012)]{schl12}
P.~Schlatter and R.~\"Orl\"u.
\newblock Turbulent boundary layers at moderate {R}eynolds numbers: inflow
  length and tripping effects.
\newblock \emph{J. Fluid Mech.}, 710:\penalty0 5--34, 2012.

\bibitem[Stroh et~al.(2015)Stroh, Frohnapfel, Schlatter, and Hasegawa]{stro15}
A.~Stroh, B.~Frohnapfel, P.~Schlatter, and Y.~Hasegawa.
\newblock A comparison of opposition control in turbulent boundary layer and
  turbulent channel flow.
\newblock \emph{Phys. Fluids}, 27:\penalty0 075101, 2015.

\bibitem[Stroh et~al.(2016)Stroh, Hasegawa, Schlatter, and
  Frohnapfel]{Stroh2016}
A.~Stroh, Y.~Hasegawa, P.~Schlatter, and B.~Frohnapfel.
\newblock Global effect of local skin friction drag reduction in spatially
  developing turbulent boundary layer.
\newblock \emph{J. Fluid Mech.}, 805:\penalty0 303–321, 2016.
\newblock \doi{10.1017/jfm.2016.545}.

\bibitem[Tanarro et~al.(2020)Tanarro, Vinuesa, and Schlatter]{Tanarro2020}
{\'A}.~Tanarro, R.~Vinuesa, and P.~Schlatter.
\newblock Effect of adverse pressure gradients on turbulent wing boundary
  layers.
\newblock \emph{J. Fluid Mech.}, 883:\penalty0 A8, 2020.

\bibitem[Touber and Leschziner(2012)]{Touber2012}
E.~Touber and M.~A. Leschziner.
\newblock Near-wall streak modification by spanwise oscillatory wall motion and
  drag-reduction mechanisms.
\newblock \emph{J. Fluid Mech.}, 693:\penalty0 150–200, 2012.
\newblock \doi{10.1017/jfm.2011.507}.

\bibitem[Vinuesa and Schlatter(2017)]{Vinuesa2017}
R.~Vinuesa and P.~Schlatter.
\newblock Skin-friction control of the flow around a wing section through
  uniform blowing.
\newblock In \emph{Proceedings of European Drag Reduction and Flow Control
  Meeting (EDRFCM)}, 2017.

\bibitem[Vinuesa et~al.(2018)Vinuesa, Negi, Atzori, Hanifi, Henningson, and
  Schlatter]{Vinuesa2018}
R.~Vinuesa, P.~S. Negi, M.~Atzori, A.~Hanifi, D.~S. Henningson, and
  P.~Schlatter.
\newblock Turbulent boundary layers around wing sections up to
  ${R}e_c$=1,000,000.
\newblock \emph{Int. J. Heat Fluid Flow}, 72:\penalty0 86 -- 99, 2018.

\bibitem[Wang et~al.(2018)Wang, Pan, and Wang]{Wang2018}
W.-K. Wang, C.~Pan, and J.-J. Wang.
\newblock Quasi-bivariate variational mode decomposition as a tool of scale
  analysis in wall-bounded turbulence.
\newblock \emph{Exp. Fluids}, 59:\penalty0 1, 2018.

\bibitem[Wang et~al.(2019)Wang, Pan, and Wang]{Wang2019}
W.-K. Wang, C.~Pan, and J.-J. Wang.
\newblock Multi-component variational mode decomposition and its application on
  wall-bounded turbulence.
\newblock \emph{Exp. Fluids}, 60:\penalty0 95, 2019.

\bibitem[Wei(2018)]{Wei2018}
T.~Wei.
\newblock Integral properties of turbulent-kinetic-energy production and
  dissipation in turbulent wall-bounded flows.
\newblock \emph{J. Fluid Mech.}, 854:\penalty0 449--473, 2018.

\bibitem[Welch et~al.(2001)Welch, Larosiliere, Hwang, and Wood]{welc01}
G.~Welch, L.~Larosiliere, D.~Hwang, and J.~Wood.
\newblock \emph{Effectiveness of micro-blowing technique in adverse pressure
  gradients}.
\newblock 2001.

\bibitem[White and Mungal(2008)]{White2008}
C.~M. White and M.~G. Mungal.
\newblock Mechanics and prediction of turbulent drag reduction with polymer
  additives.
\newblock \emph{Annu. Rev. Fluid Mech.}, 40\penalty0 (1):\penalty0 235--256,
  2008.

\bibitem[Wu and Christensen(2010)]{Wu2010}
Y.~Wu and K.~T. Christensen.
\newblock Spatial structure of a turbulent boundary layer with irregular
  surface roughness.
\newblock \emph{J. Fluid Mech.}, 655:\penalty0 380–418, 2010.

\bibitem[Yoon et~al.(2016)Yoon, Ahn, Hwang, and Sung]{Yoon2016}
M.~Yoon, J.~Ahn, J.~Hwang, and H.~J. Sung.
\newblock Contribution of velocity-vorticity correlations to the frictional
  drag in wall-bounded turbulent flows.
\newblock \emph{Phys. Fluids}, 28\penalty0 (8):\penalty0 081702, 2016.

\bibitem[Yoon et~al.(2018)Yoon, Hwang, and Sung]{Yoon2018}
M.~Yoon, J.~Hwang, and H.~J. Sung.
\newblock Contribution of large-scale motions to the skin friction in a
  moderate adverse pressure gradient turbulent boundary layer.
\newblock \emph{J. Fluid Mech.}, 848:\penalty0 288--311, 2018.

\end{thebibliography}

\end{document}